\documentclass[aps, prd,  onecolumn, preprintnumbers, nofootinbib,11pt,eqsecnum]{revtex4-1}
\pdfoutput=1
\usepackage{myoptions}

\usepackage{bbm}
\usepackage[caption=false]{subfig}

\begin{document}

\title{On entanglement spreading from holography}

\author{M\'ark Mezei}
\affiliation{
Princeton Center for Theoretical Science,\\ Princeton University, Princeton, NJ 08544}

\begin{abstract}

\noindent A global quench is an interesting setting where we can study thermalization of subsystems in a  pure state. We investigate entanglement entropy (EE) growth in global quenches in holographic field theories and relate some of its aspects to quantities characterizing chaos. More specifically we obtain four key results:\vspace{0.1cm}\\
\makebox[1.5em][l]{1.} We prove holographic bounds on the entanglement velocity $v_E$ and the butterfly effect speed $v_B$ that arises in the study of chaos.\vspace{0.1cm}\\
\makebox[1.5em][l]{2.} We obtain the EE as a function of time for large spherical entangling surfaces analytically. We show that the EE is insensitive to the details of the initial state or quench protocol. \vspace{0.1cm}\\
\makebox[1.5em][l]{3.} In a thermofield double state we determine analytically the two-sided mutual information between two large concentric spheres separated in time.\vspace{0.1cm}\\
\makebox[1.5em][l]{4.} We derive a bound on the rate of growth of EE for arbitrary shapes, and develop an expansion for EE at early times.\vspace{0.1cm}\\
In a companion paper~\cite{Mezei:2016wfz}, these results are put in the broader context of EE growth in chaotic systems: we relate EE growth to the chaotic spreading of operators, derive bounds on EE at a given time, and compare the holographic results to spin chain numerics and toy models. In this paper, we perform holographic calculations that provide the basis of arguments presented in that paper.

\end{abstract}

\maketitle
\tableofcontents

\newpage

\section{Introduction and summary of results}

A global quantum quench -- unitary time evolution from a translation invariant short-range entangled initial state --  is an interesting setting in which we can study thermalization in closed systems and probe their chaotic dynamics. Studying the real time dynamics of strongly interacting many-body systems or field theories is a formidable task. Holography has the potential to provide invaluable insight into the dynamics of these systems, by mapping the quantum problem into a higher dimensional classical geometric one~\cite{Maldacena:1997re,Gubser:1998bc,Witten:1998qj,Ryu:2006bv,Ryu:2006ef,Hubeny:2007xt}. In conformal field theories in two spacetime dimensions, global quenches have been thoroughly explored~\cite{Calabrese:2005in,Asplund:2015eha,Cardy:2015xaa}. In two spacetime dimensions one can also study the problem in integrable lattice models~\cite{Calabrese:2005in,DeChiara:2005wb,Calabrese:2007rg,2008PhRvA..78a0306F}, and using numerics in non-integrable chains~\cite{2008JSMTE..05..018L,Kim:2013bc,Mezei:2016wfz}. One can also study free theories in higher dimensions~\cite{2016PhRvB..94b4306L,Cotler:2016acd}. In this paper, we will concentrate on higher dimensional  chaotic systems.

Many aspects of the questions we ask in this paper have been already understood in prior work~\cite{Hartman:2013qma,Liu:2013iza,Liu:2013qca}. See also~\cite{AbajoArrastia:2010yt,Albash:2010mv,Balasubramanian:2010ce,Balasubramanian:2011ur} for early work on quenches in holography. The importance of the problem however warrants further scrutiny, and the results of this paper should provide valuable data for efforts into understanding quenches in strongly interacting chaotic systems. In a companion paper~\cite{Mezei:2016wfz}, we use the results of this paper along with numerical results from spin chains, to propose that the entanglement entropy in a quench in a chaotic systems is close to saturating a combination of two constraints: one that follows from recent insight into quantum chaos~\cite{Shenker:2013pqa,Roberts:2014isa,Maldacena:2015waa,Gu:2016oyy} and the positivity of relative entropy~\cite{Hartman:2015apr}, and another bounding the rate of growth of entropy. \cite{Mezei:2016wfz} can be read as putting the results of this paper into context, relating them to the picture of the chaotic growth of operators, and analyzing them from the point of view of toy models and bounds. 

The outline of the paper and the summary of the results is as follows. In Sec.~\ref{sec:Review} we first review the gravity duals of global quenches.
Second, we present a new computation of how close Ryu-Takayanagi (RT) surfaces~\cite{Ryu:2006bv,Ryu:2006ef} approach  the horizon of a static black brane with metric
\es{BH}{
ds^2&= {1\ov z^2}\le[-f(z)dt^2+ d\vec{x}^2+{dz^2\ov f(z)}\ri]\,,
}
with horizon at $z_h=1$. We work in the limit where the characteristic size $R$ of an arbitrary shaped boundary entangling surface $\Sig$ (on which the RT surface is anchored) is large compared to inverse temperature $\beta$:
\es{Limit}{
R\gg \beta\,.
}
The closest approach to the horizon is
\es{epdetIntro}{
\de z_\text{approach}\propto e^{-2\mu \,R_\text{insc}}\,,
}
where $\de z\equiv 1- z$, $R_\text{insc}$ is the radius of the largest  ball inscribable in  $\Sig$, and $\mu$ is defined in~\eqref{seq}. 
 It is explained in~\cite{Mezei:2016wfz} how this result can be used to compute the butterfly effect speed $v_B$ that characterizes the growth of operators in a thermal state~\cite{Shenker:2013pqa,Roberts:2014isa}.  
 Third, we turn to the time evolution of entanglement entropy in a global quench. It was derived for holographic theories in~\cite{Hartman:2013qma,Liu:2013iza,Liu:2013qca} that at early times the entropy grows linearly\footnote{See~\cite{Kundu:2016cgh} for a recent study of small regions.}
 \es{LinGrowth}{
\hat{S}_\Sig(t)=v_E\, s_\text{th}\, A_\Sig\, t\,, \qquad \beta\ll t\ll R\,,  
}
where $\beta$ is the effective inverse temperature associated to the system, $s_\text{th}$ is the thermal entropy density, $A_\Sig$ is the area of $\Sig$, and $\hat{S}(t)\equiv S(t) - S_\text{vacuum}$ is the entropy with the vacuum contribution subtracted.\footnote{The vacuum contribution is a divergent area law.} We will sometimes refer to $\hat{S}$ as the extensive piece of the entropy, as for $t\sim R$ it scales with the volume.  This equation is expected to hold in any chaotic system.\footnote{It is also obeyed in the quasiparticle model of entropy growth~\cite{Casini:2015zua} and in free scalar theory~\cite{Cotler:2016acd}.}
We review the holographic derivation of~\eqref{LinGrowth} for the (simplest) strip geometry in Sec.~\ref{sec:Review} using  Hubeny-Rangamani-Takayanagi (HRT) surfaces~\cite{Hubeny:2007xt}. Through this example, we introduce the entanglement velocity $v_E$ controlling the early time linear growth of entropy~\cite{Hartman:2013qma,Liu:2013iza,Liu:2013qca}.  We then move on to our main results, which are organized in an order of increasing of geometric complexity:

 
\makebox[1.5em][l]{\bf  1.} In Sec.~\ref{sec:Bounds} we discuss several inequalities involving $v_E$ and $v_B$ valid in holographic theories obeying the Null Energy Condition (NEC). We prove that in holographic theories
 \es{vEvB}{
 v_E\leq v_B\,.
 }
 This inequality is proven in~\cite{Mezei:2016wfz} for any unitary quantum system, so~\eqref{vEvB} is a consistency check for holographic theories in the dynamical regime.\footnote{That $v_E$ should be smaller than $v_B$ was also discussed recently in~\cite{Nahum:2016muy}. }
 The two velocities are controlled by different parts of the geometry, $v_E$ by the region behind the horizon and $v_B$ by the near horizon geometry, hence it is somewhat surprising that one can prove a relation between them. 
 
 We also prove several inequalities valid in holographic theories:
 \es{holoIneq}{
 v_E\leq v_E^\text{(S)}\,, \qquad  v_B\leq v_B^\text{(S)}\,, \qquad {v_E\ov v_B}\leq  {v_E^\text{(S)}\ov v_B^\text{(S)}}\,,
 }
 where the (S) superscript refers to the Schwarzschild black brane value given in~\eqref{vBRN} and~\eqref{vES}, which only depends on the number of dimensions. The first of these inequalities was conjectured in~\cite{Liu:2013iza,Liu:2013qca} based on the evaluation of $v_E$ in many examples. 
The Schwarzschild black brane is the holographic dual of a conformal field theory at finite temperature (or an eigenstate with energy density).  In a given dimension we can consider quenches that create not only energy density, but also charge density, or we can investigate non-conformal systems. These setups lead to final state black branes different from Schwarzschild. The inequalities~\eqref{holoIneq} imply that these complications slow down the spread of entanglement and operators, and also how close entanglement growth in the linear regime can come to  saturating~\eqref{vEvB}.

\begin{center}
\begin{figure}[!h]
\includegraphics[scale=0.7]{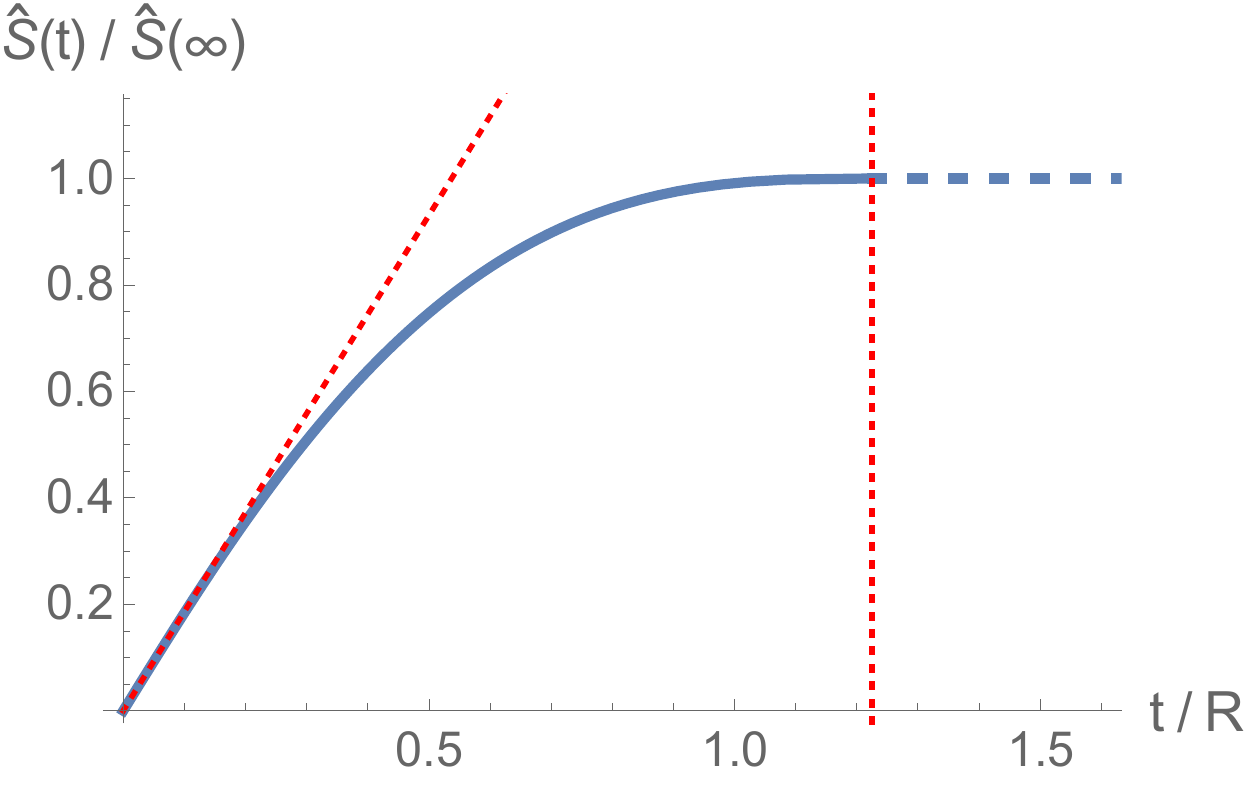} 
\caption{Entropy growth in a global quench for a sphere of radius $R$ in the limit~\eqref{Limit} in $d=4$. The solid blue curve is~\eqref{AnalyticCurve}, the dashed blue line is the saturation value of the entropy given by a static RT surface, and the dotted red lines indicate~\eqref{LinGrowth} and~\eqref{tsRes} respectively. 
\label{fig:4dSBH}}
\end{figure}
\end{center}

\makebox[1.5em][l]{\bf  2.}  In Sec.~\ref{sec:spherical} we discuss the entropy growth for spherical regions of radius $R$.   In the limit~\eqref{Limit},
we show that the extensive part of the entropy is the same irrespective of the details of the quench. While this finding conforms with the field theory intuition that all short-range entangled states should have the same dynamics in a strongly interacting chaotic system after the local thermalization time $t_\text{loc}\sim\beta$, the disparity in geometry between the end of the world brane quench model~\cite{Hartman:2013qma} and the infalling shell model~\cite{AbajoArrastia:2010yt,Albash:2010mv,Balasubramanian:2010ce,Balasubramanian:2011ur,Liu:2013iza,Liu:2013qca} makes this universality a nontrivial new result. We first use numerics, and motivated by the numerical results we take  a scaling limit of the extremal surface equations which leads to major simplification. We are able to solve these equations analytically. For the black brane with metric~\eqref{BH}
 the entropy as function of time is given by:
\es{AnalyticCurve}{
{t(z_f)\ov R}&=-{1\ov r(z_f)}\int_{z_f}^{z_\text{HM}} dz \ r'(z)\, {1\ov \sqrt{f(z)\le[1- {z\, {f'(z)}\ov 2(d-1)f(z) }\ri]}}\\
{\hat{S}(z_f)\ov \hat{S}_\text{thermal}}&=-{d-1\ov r(z_f)^{d-1}}\int_{z_f}^{z_\text{HM}} dz \ r'(z)\,  { r(z)^{d-2}\ov z^{d-1}}\, \sqrt{1-{1\ov 1- {z\, {f'(z)}\ov 2(d-1)f(z) }}}\\
r(z)&\equiv\le[{f(z_\text{HM}) \ov f(z) }\,\le({z\ov z_\text{HM}}\ri)^{2(d-1)}  {z\, f'(z)\ov 2(d-1)f(z) }\ri]^{1/2(d-2)}\,,
}
where $1\leq z_f \leq z_\text{HM}$, and $z_\text{HM}$ is the locus behind the horizon, where $-f(z)/z^{2(d-1)}$ is maximal. The equations are a bit complicated, but are completely explicit. On Fig.~\ref{fig:4dSBH} we show the evaluation of the integrals for the $d=4$ Schwarzschild black brane.

Let us discuss the results on Fig.~\ref{fig:4dSBH}. The entropy starts to grow linearly according to the law~\eqref{LinGrowth}
This behavior is indicated by a red dotted line.  After the initial linear growth the entropy saturates smoothly at some finite time $t_s=R/v_B$ indicated by a vertical red dotted line on Fig.~\ref{fig:4dSBH}.\footnote{In \cite{Liu:2013iza,Liu:2013qca} a saturation velocity $c_E=R/t_S$ was introduced to characterize the saturation time; in this language $c_E=v_B$. This relation is only true if the saturation is continuous. } This formula simply follows from taking $z_f\to1$ in~\eqref{AnalyticCurve}. Formally, this equation remains true for any black brane in any dimension. However, there is a subtlety: the parametric curve $\le(t(z_f),\hat{S}(z_f)\ri)$ determined by~\eqref{AnalyticCurve}  may be non-single valued, leading to the HRT surface changing discontinuously and  $t_S>R/v_B$. We will refer to this situation as discontinuous saturation; although the entropy as a function of time remains continuous, its derivative is discontinuous.

We analyze necessary conditions for discontinuous saturation in Sec.~\ref{sec:cusp} by analyzing~\eqref{AnalyticCurve}. We prove that in $d=3$ the entropy always saturates discontinuously, but in higher dimensions Schwarzschild black branes (e.g. in $d=4$ shown in Fig.~\ref{fig:4dSBH}) give continuous saturation and
\es{tsRes}{
t_S={R\ov v_B}\,.
} 
In~\cite{Mezei:2016wfz} we give an alternative proof of this equation for any black brane that gives continuous saturation based on  entanglement wedge subregion duality. The proof applies to any theory of gravity, and in~\cite{Mezei:2016wfz} we verify the relation~\eqref{tsRes} explicitly in four derivative gravity. 


\makebox[1.5em][l]{\bf  3.}  In~\cite{Liu:2013iza,Liu:2013qca} an ``entanglement tsunami" picture for entropy growth was suggested: there would be a sharp wave front propagating inward from the entangling surface $\Sig$. Behind this wave, degrees of freedom become fully entangled with the outside, while the degrees of freedom yet to be reached by the wave remain unentangled. At early times the wavefront moves with $v_E$ giving the relation~\eqref{LinGrowth}. 

In Sec.~\ref{sec:MutInf} we probe this proposal in the thermofield double state~\cite{Maldacena:2001kr}, where degrees of freedom in the right copy ($R$) are strongly entangled with their partner in the left copy ($L$).  If we time evolve only $R$, this high degree of local entanglement spreads in $R$.  The technology developed in Sec.~\ref{sec:spherical} can be used to compute the two-sided mutual information~\cite{Morrison:2012iz,Hartman:2013qma} between two (concentric) spherical regions $A_L(0)$ and $B_R(t)$, where the sphere $B$ in $R$  has been time evolved for time $t$ with respect to $A_L(0)$:
\es{TwoMut}{
I[B_R(t),A_L(0)] \equiv S[B_R(t)] + S[A_L(0)] - S[B_R(t)\cup A_L(0)]\,.
}
We regard  $L$ as auxiliary system that keeps track of the movement of information in $R$, and $I[B_R(t),A_L(0)]$ as the "mutual information in time"~\cite{Hosur:2015ylk} between $A_R(0)$ and $B_R(t)$. 

Let us fix $R_A$ and some time $t$. The "entanglement tsunami" picture would predict that  information in $R_A$ only gets scrambled in a region of size $R_A+v_E\,t$ at least for $t\ll R_A$. However, in Sec.~\ref{sec:MutInf} we find that as a function of $R_B$ the mutual information only saturates at
\es{TsunamiWave}{
R_B^\text{(saturates)}=R_A+v_B\,t\,,
} 
which is valid for any time $t$. More discussion of the two-sided mutual information and the interpretation of our results can be found in~\cite{Mezei:2016wfz}, where we advocate for this quantity as a useful probe of information spreading in any quantum system. In this paper we concentrate on the computational aspects of obtaining the two-sided mutual information for a holographic theory.

There is another result that may help in ameliorating the tsunami picture. In~\cite{Mezei:2016wfz} using ideas from~\cite{Hartman:2015apr}, we derive an upper bound on the entropy in quench
\es{TsunamiVolume}{
\hat{S}_\Sig(t)\leq s_{\text{th}}\, \text{vol}[\text{tsunami}(v_B,\,t)]\,,
}
where $\text{vol}[\text{tsunami}(v_B,\,t)]$ is the volume of the region behind covered by the tsunami wavefront propagating with $v_B$ in time $t$.
The results~\eqref{TsunamiWave} and~\eqref{TsunamiVolume} suggest that the "entanglement tsunami" may not be sharp. The reason for this is the difference between the two speeds $v_E$ and $v_B$. While the wave front may propagate with $v_B$ it does not necessarily thermally entangle the degrees of freedom it reaches with the outside. 

The bound~\eqref{TsunamiVolume} implies that
\es{tSBound}{
t_S\geq{R_\text{insc}\ov v_B}\,,
}
where  $R_\text{insc}$ was defined after~\eqref{epdetIntro}.
That, as discussed around~\eqref{tsRes}, this bound is saturated in many situations in holography, implies that the tsunami propagating with $v_B$ can become an accurate picture as $t\to t_S$. In~\cite{Mezei:2016wfz} we suggest a microscopic toy model that can produce such behavior. In this paper we prove~\eqref{tSBound} holographically in two very different ways in Secs.~\ref{sec:Review} and~\ref{sec:Shapes}.


\makebox[1.5em][l]{\bf 4.} In Sec.~\ref{sec:Shapes} we discuss entropy growth for arbitrary shapes. By taking the same scaling limit as for spheres (and more trivially strips), we are able to write down a relatively simple PDE governing the time evolution of entropy. The solution of this PDE in the spherical case is~\eqref{AnalyticCurve}, but in general we do not know how to solve it. Nevertheless we have enough control over the system to prove that the rate of growth of entropy is bounded by
\es{BoundIntro}{
{dS\ov dt}\leq s_\text{th}\, v_E\, A_\Sigma\,, \qquad \beta\ll t\,,
}
where the condition on time comes from having taken the scaling limit. It was suggested in~\cite{Liu:2013iza,Liu:2013qca} that this equality may hold in holographic systems. \eqref{BoundIntro} was proved in the quasiparticle model in~\cite{Casini:2015zua}. It plays an important role in the discussions of~\cite{Mezei:2016wfz}. There are rigorous bounds of the form~\eqref{BoundIntro} but with a coefficient that depends on the Hilbert space dimension and operator norms~\cite{2006PhRvL..97e0401B,2014arXiv1411.0680M}. It would be interesting to establish~\eqref{BoundIntro} using field theory techniques.\footnote{See~\cite{Avery:2014dba} for an attempt at a field theory proof.}

The same methods that allowed us to establish~\eqref{BoundIntro} give us an intriguing equation for the rate of growth of entropy. It is easiest to state our result in the boundary state quench model~\cite{Calabrese:2005in}, whose holographic dual is the eternal black hole cut in half by an end of the world brane~\cite{Hartman:2013qma}. The extremal surface determining the entropy at a given time, is a tube connecting the entangling surface $\Sig$ to its image on the brane $\Sig_\text{im}$. The rate of growth of entropy in the scaling limit is
\es{BoundIntro2}{
{dS(t)\ov dt}= s_\text{th}\, v_E\, \text{area}\le[\Sig_\text{im}(t)\ri]\,,
}
where the area is measured in the field theory coordinates $\vec{x}$ defined in~\eqref{BH}. We apply this technology to develop an early time expansion for the entropy, and obtain the first subleading correction to~\eqref{LinGrowth}:
\es{LinGrowth2}{
\hat{S}_\Sig(t)=v_E\, s_\text{th}\, \text{area}(\Sig)\, t\le(1-a\le[\Sig\ri]\,t^2+O(t^5)\ri)\,, \qquad \beta\ll t\ll R\,,  
}
where $a\le[\Sig\ri]$ is given in~\eqref{Agrowth}.\footnote{It was suggested in~\cite{Nahum:2016muy} based on tensor network intuition that the early time expansion would involve these powers of $t$.} \eqref{BoundIntro2} together with the bounds discussed in~\cite{Mezei:2016wfz} is very suggestive of a tensor network interpretation, but we have not been able to make this connection precise.
\\*

The results in this paper provide us with a wealth of information about entropy growth in a quantum quench in holographic systems. The key technical result of our paper is the determination HRT surfaces in the limit~\eqref{Limit}. We find hints that various geometric objects, the largest inscribable ball (appearing in~\eqref{epdetIntro} and~\eqref{tSBound}) and the image on the end of the world brane (appearing in~\eqref{BoundIntro2}) play an important role in the dynamics of entropy. We provide an improved understanding of when the saturation is continuous and when it is not. We study the ``entanglement tsunami" in detail, and find that it propagates with the butterfly effect speed $v_B$. In~\cite{Mezei:2016wfz} we interpret the results of this paper. As a model for the time dependence, we suggest that the entropy is as large as possible given the two constraints~\eqref{TsunamiVolume} and~\eqref{BoundIntro}. We also construct a microscopic toy model based on the chaotic growth of operators that saturates these bounds. It would be very interesting to compute the time evolution of entropy for shapes different from the sphere and the strip, and to see if the bound on saturation time~\eqref{tSBound} is saturated for shapes other than spheres (at least for certain black holes). Another interesting direction is to refine the bounds and the microscopic models presented in~\cite{Mezei:2016wfz}, to bring them closer to reproducing the entropy curve obtained from holographic and other chaotic systems.

\section{Review of entanglement growth from holography}\label{sec:Review}

\subsection{Holographic quench models}

\begin{center}
\begin{figure}[!h]
\includegraphics[scale=0.7]{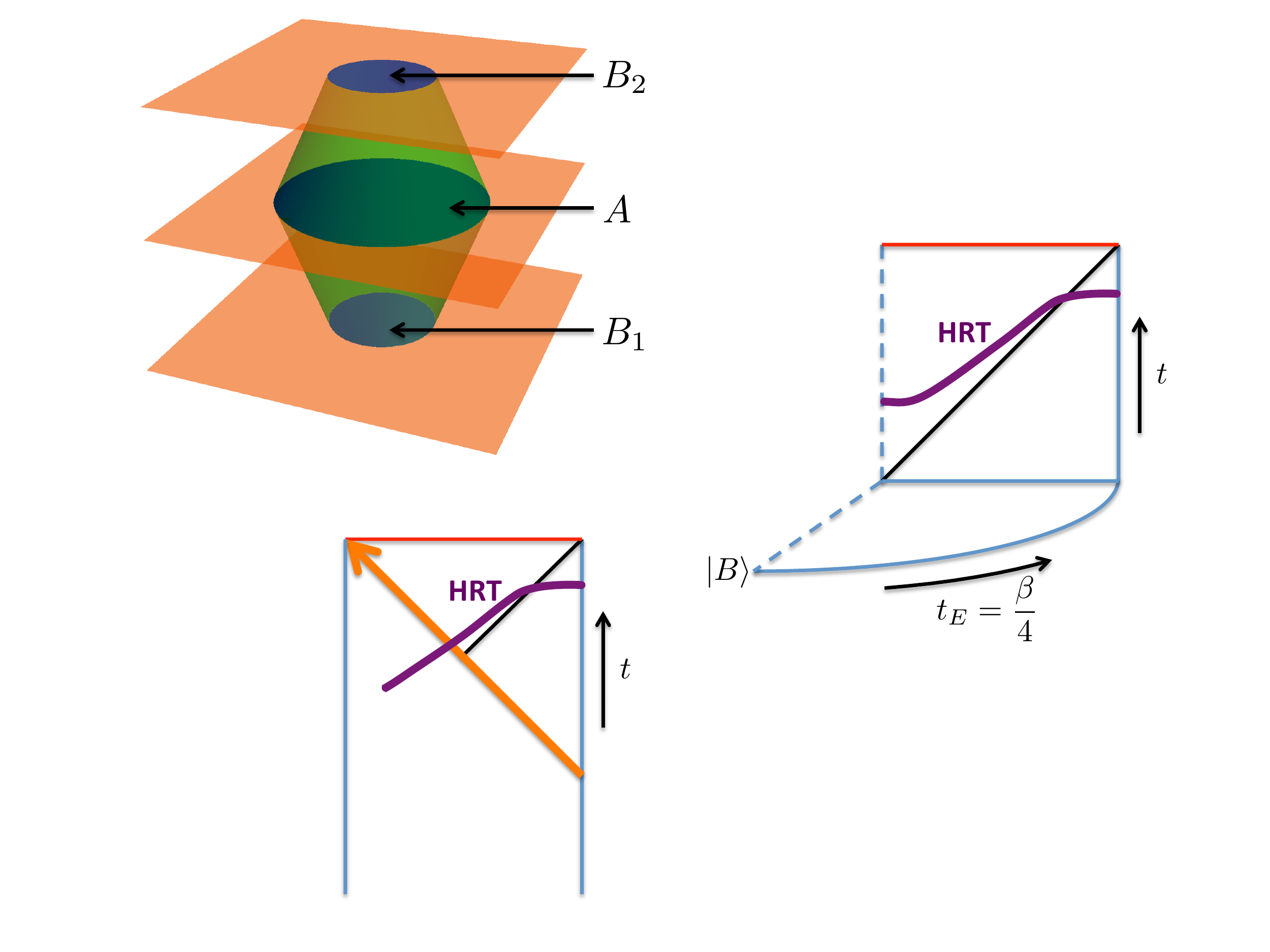} \hspace{2cm}
\includegraphics[scale=0.7]{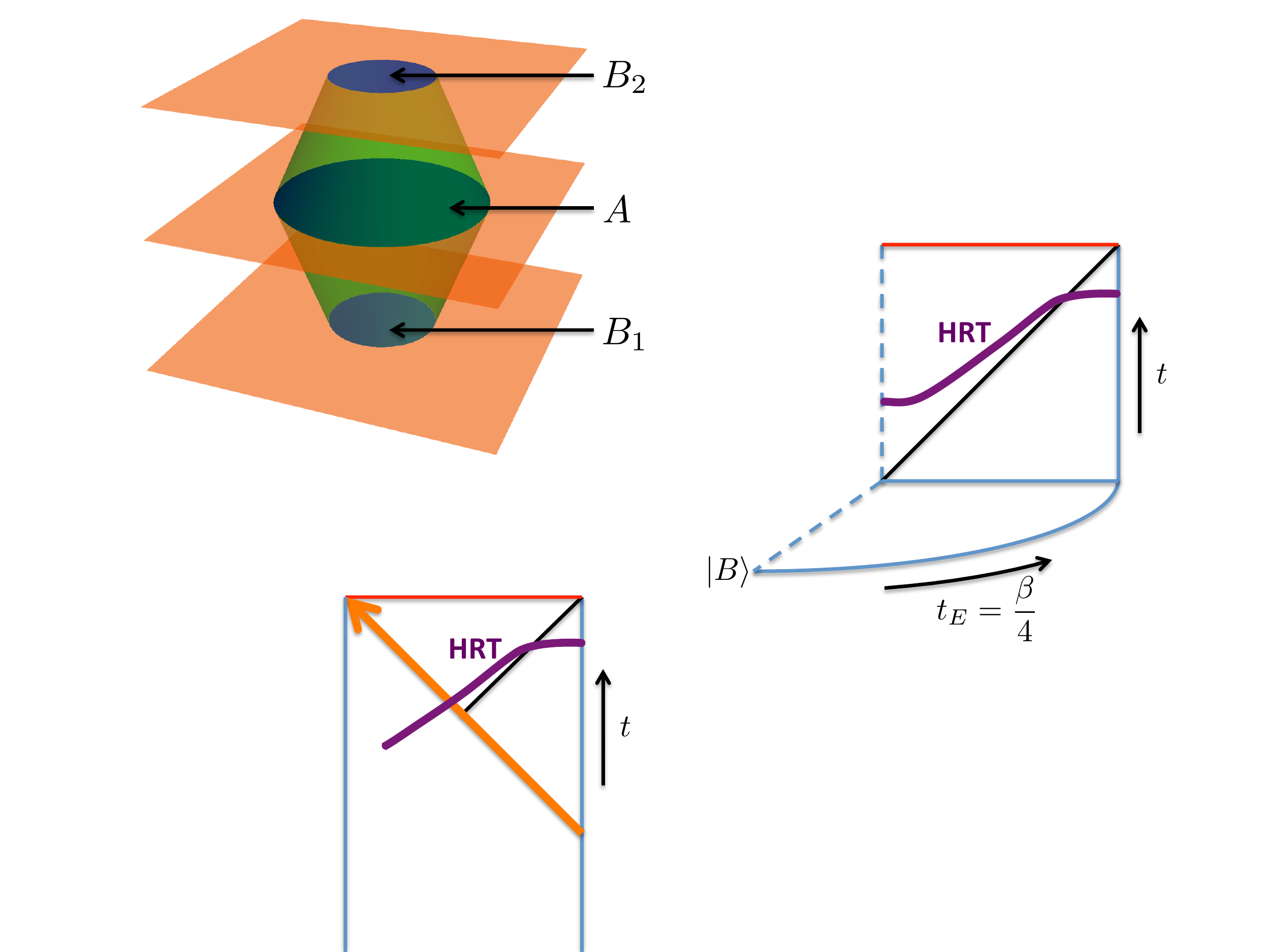}
\caption{{\bf Left:} Vaidya quench model. The infalling null shell is drawn by an orange arrow. Below the shell the geometry is pure AdS, above the shell it is a static black brane. We get time evolution of the entropy in the boundary theory because the HRT surface lives in both parts of the geometry and passes through the null shell. {\bf Right:} The boundary state model of a quench is dual to an end of the world brane cutting the eternal black hole in half. The lower portion of the figure illustrates how the Euclidean path integral prepares a short-range entangled initial state from the boundary state~\cite{Hartman:2013qma}. The time evolution of the entropy comes from the HRT surface entering the black brane horizon and ending on the brane.
\label{fig:twosetups}}
\end{figure}
\end{center}

In a field theory model of a global quench we want to create an initial state that is short-range entangled, translation invariant and has finite energy density. One way to create such a state is to dump in energy in an uncorrelated manner by smearing a local operator over the whole system and acting with it on the vacuum. This is the setup that Liu and Suh considered holographically~\cite{Liu:2013iza,Liu:2013qca}. Another way to model a quench that is more convenient for CFT computations is to consider a conformal boundary state in CFT as the initial state~\cite{Calabrese:2005in}.\footnote{Conformal boundary states are not normalizable, so one first evolves them in Euclidean time for $t_E=\beta/4$. The Euclidean time evolution damps the high energy modes and the resulting state has finite energy density. For holographic field theories, the energy density agrees with that in a thermal ensemble with inverse temperature $\beta$.} The holographic dual of this setup was considered by Hartman and Maldacena~\cite{Hartman:2013qma}. The two setups are shown in Fig.~\ref{fig:twosetups}.

We analyze both setups below for large regions and times $R,t\gg \beta$, and find that the entire time evolution of entanglement entropy is universal: it does not depend on which setup we consider. 
 It has been already been demonstrated in~\cite{Hartman:2013qma,Liu:2013iza,Liu:2013qca} that in the linear regime~\eqref{LinGrowth} the Vaidya and the end of the world brane models  give the same result. Because these setups differ significantly on what conditions the HRT surface has to obey in the bulk, it comes as a surprise that even the saturation behavior agrees between the two models. We discuss the details below.

\subsection{Geometry setup}

 The static geometry of the most general translation and rotation invariant asymptotically AdS black brane is
 \es{Metric}{
ds^2= {1\ov z^2}\le[-{f(z)\ov h(z)} dt^2+ d\vec{x}^2+{dz^2\ov f(z)}\ri]\,,
}
where the boundary is at $z=0$, $\vec{x}$ is the field theory spatial coordinate, and writing $g_{tt}$ in this form is for convenience.
To get an asymptotic AdS spacetime we need to impose the boundary conditions
\es{BdyCond}{
h(0)=f(0)=1\,,
}
and we use the freedom of scaling the coordinates to set the horizon radius $z_h=1$, leading to the other boundary condition:
\es{BdyCond2}{
f(1)=0\,.
}
 A useful example to bear in mind is the one parameter family of Reissner-Nordstrom (RN) black branes with emblackening factor
\es{RNBH}{
h(z)&=1\,,\qquad f(z)=1-M\,z^d+Q^2\, z^{2(d-1)}\,, \qquad M\equiv 1+{d\, q\ov d-2}\,, \qquad Q^2\equiv {d\, q\ov d-2}\,,
} 
where the parameter $0\leq q\leq 1$ characterizes the proximity of the black brane to extremality, $q=(Q/Q_\text{ext})^2$. We get he Schwarzschild geometry by setting $q=0$.

It will be beneficial to work in infalling Eddington--Finkelstein coordinates:
\es{MetricSBH2}{
ds^2&= {1\ov z^2}\le[-{f(z)\ov h(z)}dv^2-{2\ov \sqrt{h(z)}}\,dvdz+ d\vec{x}^2\ri]\\
v&\equiv t-\int_0^z dz'\ {\sqrt{h(z')}\ov f(z')}\,.
}
We can extend the $t$ coordinate inside the black brane using this equation. If we choose an integration contour that goes above the pole at $z'=1$, behind the horizon we have $t=t_I + i{\pi}\, {\sqrt{h(1)}\ov f'(1)}$ with $t_I$ real.

We can obtain a Vaidya geometry by gluing an empty AdS space to the above described black brane along the $v=0$ null plane. The geometry can be written as
\es{Vaidya}{
ds^2&= {1\ov z^2}\le[-{f(v,z)\ov h(v,z)}\,dv^2-{2\ov\sqrt{h(v,z)}}\,dvdz+ d\vec{x}^2\ri]\\
f(v,z)&\equiv1+\theta(v) \le(f(z)-1\ri)\,, \qquad h(v,z)\equiv1+\theta(v) \le(h(z)-1\ri)\,.
}
In the end of the world brane setup the brane is located at the plane of time reflection symmetry, $t_I=0$, hence in the coordinates~\eqref{MetricSBH2} its position is
\es{EndOfWorld}{
v_\text{brane}(z)=-\int_0^z dz'\ {\sqrt{h(z')}\ov f(z')}+ i{\pi}\, {\sqrt{h(1)}\ov f'(1)}\qquad (z>1)\,. 
}

\subsection{Static surfaces}\label{sec:static}

We first want to ask how close RT surfaces anchored on large boundary regions approach the horizon of the static black brane~\eqref{Metric}; the answer will have applications throughout the paper. We take the surface to be parametrized by boundary polar coordinates, $z(\rho,\Omega)$. The area functional in the geometry~\eqref{Metric} is:
\es{AreaFunctStat}{
A&= \int d\rho \, d\Om\ {\rho^{d-2}\ov z^{d-1}}\,\sqrt{Q}\,,\qquad Q\equiv 1+{(\p_\rho z)^2+{1\ov\rho^2}\, (\p_\Om z)^2\ov f(z)}\,,
}
where $\p_\Om$ is the gradient on the unit $S^{d-2}$.
As explained in more detail (for symmetric surfaces) in~\cite{Liu:2013una,Mezei:2016wfz} the minimal surfaces that we are interested in will for mostly lie flat skimming the horizon and then shoot out exponentially to the boundary. Then we can focus on the near horizon region to understand the important physics. It is convenient to define a somewhat peculiar new variable
\es{dezStat}{
 z(\rho,\Omega)\equiv 1-\ep\, s(\rho,\Omega)^2\,, \qquad s(0,\Omega)=1\,,
} 
where $\ep\ll 1$ is the closest approach to the horizon. (For this to be true we have to choose the origin of polar coordinates appropriately.) Expanding the equations of motion corresponding to~\eqref{AreaFunctStat} we get a simple equation governing the evolution of $s(\rho,\Omega)$:
\es{seq}{
\nabla_{\vec{x}}^2 \, s-\mu^2 s&=0\,, \qquad \mu^2\equiv-{d-1\ov2}\, f'(1)\,,
}
where $\nabla_{\vec{x}}^2$ is the Laplacian in boundary coordinates, and $\mu^2>0$ because $f'(1)<0$. 
This equation has solutions blowing up as $s\sim \exp\le(\mu\, x\ri)$ corresponding to the fast shooting out to the boundary explained above. 

One can get a good approximation to the minimal surface by sourcing~\eqref{seq} uniformly around the entangling surface $\Sig$. Using the appropriate Green's function of~\eqref{seq} this approximation amounts to
\es{ssolApp}{
s(\vec{x})\approx C\,\int_\Sig d\vec{y} \ {K_{{d-3\ov 2}}\le(\mu\abs{\vec{x}-\vec{y}}\ri)\ov \le(\mu\abs{\vec{x}-\vec{y}}\ri)^{d-3\ov 2}}\,,
}
where $K_\nu(x)$ is the modified Bessel function of the second kind, and we will fix the constant $C$ below.  Note that~\eqref{ssolApp} defines two disconnected surfaces (for a connected $\Sig$), one that we are looking for and one that runs off to infinity that we should discard.  The solution has the property that it blows up near $\Sig$, which corresponds to the RT surface leaving the near horizon region and arriving to the boundary. It decays exponentially away from the entangling surface, which corresponds to the surface lying flat close to the horizon. Where $s(\vec{x})$ gets big, we should not trust~\eqref{ssolApp}. One can actually solve for the full surface analytically in a double expansion~\cite{Liu:2013una}: it should involve adjusting the source in~\eqref{ssolApp} so that it is smeared an $O(\beta)$ amount around $\Sig$ of characteristic size $R$. The precision of~\eqref{ssolApp} is however enough for our purposes.

The location where $s(\vec{x})$ is the smallest corresponds to the point $\vec{x}$ which is farthest from all points on $\Sig$. For a convex $\Sig$ this is the center of the largest inscribable ball, and this is where we put the origins of the polar coordinate system. To satisfy the boundary condition~\eqref{dezStat} we require that
\es{Cdet}{
C\propto e^{\mu \,R_\text{insc}}\,,
}
where we used the large $x$ asymptotics of $K_\nu(x)$, and we have not written out the prefactor involving powers of $R_\text{insc}$. In order for $\ep\, s(\vec{x})^2\ll 1$ and for the minimal surface to reach the boundary around $\Sig$ we need to take 
\es{epdet}{
\ep\propto e^{-2\mu \,R_\text{insc}}\,.
}
This is the closest approach of the minimal surface anchored on a convex $\Sig$ to the horizon.

Two applications of this result are explored in~\cite{Mezei:2016wfz}, which we briefly describe here. The reader should refer to that paper for a more complete treatment. First, acting with a smeared operator on top of the thermal state is dual to creating a particle near the boundary of AdS, which then falls into the black brane. Let us define $\de z\equiv 1-z$. Its trajectory at late times can be approximated by
\es{ParticleTrajectory}{
\de z(t)\propto \exp\le({ f'(1)\ov \sqrt{h(1)}}\,t\ri) \,,
}
where we have to remember that $f'(1)<0$. In the field theory, the support of the information about the insertion of the operator gets scrambled inside a ball of radius $v_Bt$, and we need to access the density matrix of this entire ball to be able to reconstruct the operator insertion. The dual perspective is that we need to find an entanglement wedge that contains the infalling particle. We found above that two entanglement wedges whose boundary entangling surfaces share the same largest inscribed ball reach equally deep into the bulk, hence the minimal choice of entanglement wedge corresponds to a spherical $\Sig$ concentric with the operator insertion. In order for the entanglement wedge to contain the infalling particle we need
\es{epdez}{
\ep\leq \de z \quad \implies \quad \mu\,R\geq -{ f'(1)\ov \sqrt{h(1)}}\,t\,,
}
where we used~\eqref{epdet} and~\eqref{epdez}. Using the definition of $\mu$~\eqref{seq}, we obtain the expression for the butterfly effect speed
\es{vBres}{
v_B=\sqrt{-{f'(1)\ov 2(d-1)h(1)}}\,.
}
This result agrees with prior determinations of $v_B$ from entirely different computations as discussed in~\cite{Mezei:2016wfz}. For an RN black brane~\eqref{RNBH}
\es{vBRN}{
v_B=\sqrt{(1-q)d\ov 2(d-1)}\,.
}

Second, in the Vaidya setup if the saturation of entropy is continuous, saturation happens when the HRT surface climbs out from behind the shell and its tip is barely touching it~\cite{Liu:2013iza,Liu:2013qca}. If the saturation is discontinuous, then the time when this happens provides a lower bound on saturation time. The shell is following a null line $v=0$, which in Schwarzschild coordinate is at
\es{ShellEq}{
t(z)=\int^z_0 dz' \ {\sqrt{h(z')}\ov f(z')}\approx {\sqrt{h(1)}\ov f'(1)}\, \log\de z\,,
}
where we went to late times or close to the horizon. (Note that before the shell the spacetime is pure AdS, but we do not need to use this fact.)
This formula is identical to~\eqref{ParticleTrajectory}. The (lower bound on) saturation time is obtained by plugging $\de z=\ep$ into~\eqref{ShellEq} giving
\es{tsEq}{
t_S\geq{R_\text{insc}\ov v_B}\,.
}
This bound agrees with~\eqref{tSBound} that we get from field theory in~\cite{Mezei:2016wfz}. For spherical regions in some circumstances we get continuous saturation as explored in Sec.~\ref{sec:cusp}, while for strips the saturation is always discontinuous. One of the most important questions left open in this paper is whether shapes other than the sphere give continuous saturation.

\subsection{Review of the strip}\label{sec:strip}

In the end of the world brane setup the analysis of the strip is particularly simple, the results below are a review of~\cite{Hartman:2013qma} in our set of coordinates. Before saturation the two sides of the strip are disconnected, and we can concentrate on one of them. Because of symmetry, the HRT surface does not move in the boundary spatial directions, and it is just determined by the function $z(v)$.\footnote{In this paper we will often change how we parametrize the extremal surfaces depending on what parametrization sheds more light on the physics and is more convenient for computations. The downside of this choice is that comparing results in different sections involves a change of variables.} The action is
\es{HMStripAction}{
A&=A_\Sig\,\int_{v_\text{brane}(z_c)}^t dv \ {\sqrt{Q }\ov z(v)^{d-1}}\,, \qquad Q\equiv {f(z(v))\ov h(z(v))}+ {2\, z'(v)\ov \sqrt{h(z(v))}}\,,
}
where $z_c$ is the point where the surface ends on the brane.
Because the action does not depend on $v$ we have a conserved quantity:
\es{ConservedStat}{
E={1\ov z^{d-1}\sqrt{Q}}\le({f(z)\ov h(z)}+{z' \ov \sqrt{h(z)}}\ri)\,,
}
which we can solve for $z'$. We can also determine $E$ as a function of $z_c$ using $z' \big\vert_\text{brane}=0$. Plugging back into the area functional and changing integration variables to  $z$ we get:
\es{zsol}{
A(z_c)&=A_\Sig\,\int_0^{z_c} dz\  {1\ov  z^{d-1}\,  \sqrt{f(z)\, \le(1-{f(z_c)/\le(h(z_c)\,z_c^{2(d-1)}\ri)\ov f(z)/\le(h(z)\,z^{2(d-1)}\ri) }\ri)}}\\
t(z_c)&=v_\text{brane}(z_c)-\int_0^{z_c} {dz\ov z'(v)}=i{\pi}\, {\sqrt{h(1)}\ov f'(1)}-\int_0^{z_c} dz\  {\sqrt{ h(z)}\ov f(z)\,\sqrt{1-{ f(z)/\le(h(z)\,z^{2(d-1)}\ri)\ov f(z_c)/\le(h(z_c)\,z_c^{2(d-1)}\ri) }}}\,,
}
where we used~\eqref{EndOfWorld}. One can evaluate the above integrals numerically, and obtain the entropy curve $\le(t(z_c),\,A(z_c)\ri)$. Instead, we proceed analytically.

Let us define $z_\text{HM}$ (after the authors of~\cite{Hartman:2013qma}) as the maximum of $-{ f(z)/\le(h(z)\,z^{2(d-1)}\ri)}$.  We will use frequently in the rest of the paper that at this point
\es{zHMEq}{
f'(z_\text{HM})=f(z_\text{HM})\le({2(d-1)\ov z_\text{HM}}+{h'(z_\text{HM})\ov h(z_\text{HM})}\ri)\,.
}
 An important observation in analyzing the integrals is that as $z_c\to z_\text{HM}$ both $t$ and $A$ diverge~\cite{Hartman:2013qma}. This is exactly the regime that we are interested in, if we want to consider $R,t\gg \beta$ in the field theory. 
 
 We do not even need to obtain~\eqref{zsol} to reach this conclusion. It is easy to see that if~\eqref{ConservedStat} gives $z'(v)=O(1)$ than the HRT surface will reach the boundary in time $t=O(\beta)$. In order to go to long times we need to suppress $z'(v)$. Solving~\eqref{ConservedStat} for small $v$ we obtain:
\es{zExpStrip}{
z(v)=z_c-\le[a\, \le(z_\text{HM}-z_c\ri)+O\le(\le(z_\text{HM}-z_c\ri)^2\ri)\ri]\,v^2+O(v^4)\,,
}
where $a=O(1)$ depends on the details of $f,\, h$ and we do not write down its explicit expression. We see that $z_c\to z_\text{HM}$ indeed suppresses $z'(v)$, and should correspond to late times in the boundary theory. We can actually estimate how small  $\de z_c\equiv z_\text{HM}-z_c$ has to be. Let us define
\es{dez}{
 \de z(v)\equiv  z_c - z(v)\,,
}
and assume that $\de z_c,\, \de z(v)\ll1$. (Note that this $\de z$ is not the same as the one used in Sec.~\ref{sec:static}.) Expanding~\eqref{ConservedStat} in these small quantities and solving the resulting differential equation we get that
\es{dezSol}{
 \de z(v)=2\de z_c\,\sinh^2\le(\sqrt{a\ov 2}\, v\ri)\,.
}
The solution looses its validity once $\de z(v)=O(1)$, hence if we want to go to $v=O(R)$ times, we have to make 
\es{dezc}{
\de z_c\propto e^{-\sqrt{2a}R}\,.
} 
The exponential growth of solutions, and that we have to choose the control parameter $\de z_c$ to be exponentially small is analogous to the exponentially close approach of RT surfaces to the static black brane horizon discussed in Sec.~\ref{sec:static}, and it will appear throughout the paper.

 In summary, we have to choose $\de z_c$ exponentially small in $R$ to make the HRT surface skim $z_\text{HM}$ for a time of $O(R)$. The HRT surface then steeply shoots out to the boundary.
 It is not hard to convince ourselves that this shooting out part gives subleading in $1/R$ contribution to both $t$ and the subtracted extremal surface area
\es{Ahat}{
\hat{A}(t)\equiv A(t)-A_\text{vacuum}\,.
}
We thus conclude that to obtain the entropy as a function of time, we do not need to know the details of how the extremal surface reaches the boundary. We can approximate it with a surface lying on $z_\text{HM}$ for $v\approx t$, and neglect the (subleading) area law contribution to $\hat{A}(t)$, and an $O(1)$ correction to $t$ coming from the shooting out part of the HRT surface. In the rest of the paper we will use a similar philosophy for more complicated geometries. 

\begin{center}
\begin{figure}[!h]
\includegraphics[scale=0.7]{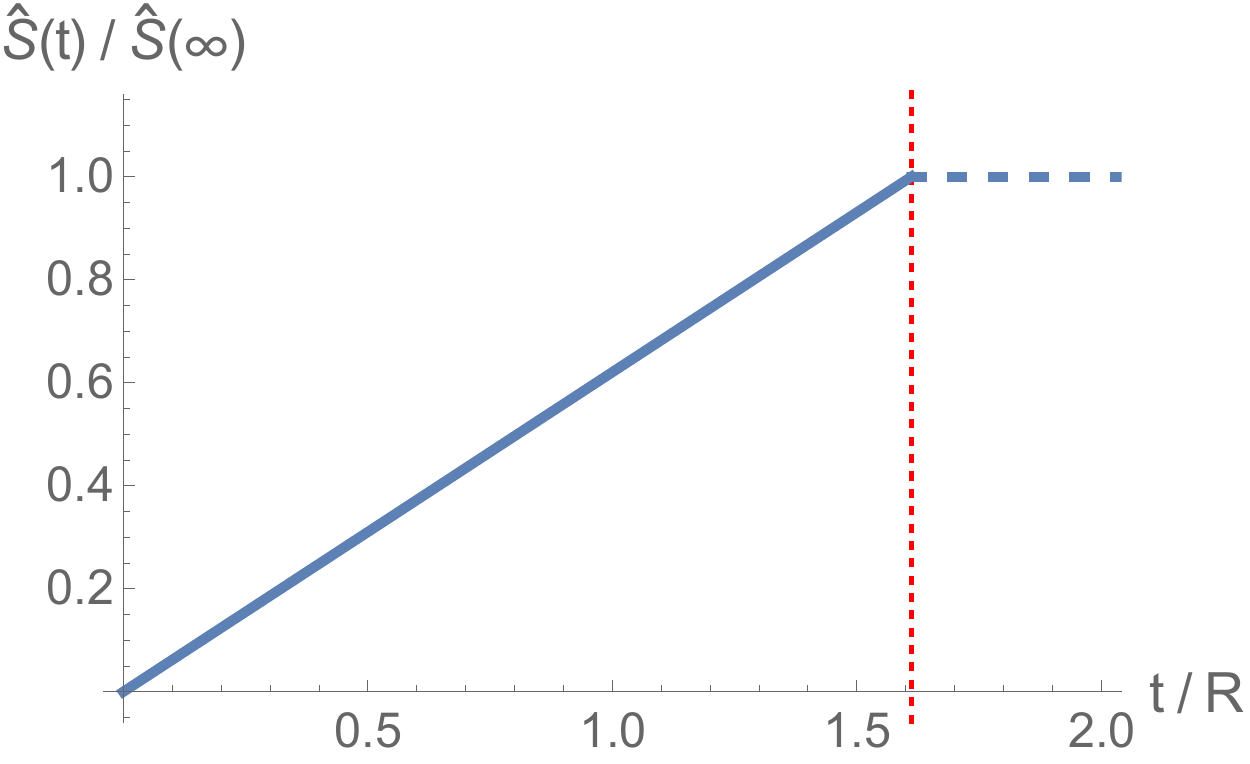} 
\caption{Entropy growth for a strip of width  $2R$ from a $d=4$ Schwarzschild black hole in the limit of large region sizes. The above rescaled curve is exactly linear with slope $v_E$, which gives a saturation time $t_S={R\ov v_E}$. This saturation time satisfies the bound~\eqref{tSBound}, if we use that $v_E\leq v_B$~\eqref{vEvB}.
\label{fig:strip}}
\end{figure}
\end{center}

The entropy is given by the volume element on the surface $z_\text{HM}$~\cite{Hartman:2013qma,Liu:2013iza,Liu:2013qca}:
\es{HRTArea}{
\hat{S}&={\hat{A}\ov 4 G_N}\approx {\sqrt{\ga(z_\text{HM})}\, A_\Sig\, t\ov 4G_N}=\sqrt{-{f(z)\ov h(z) z^{2(d-1)}}}\Bigg\vert_{z=z_\text{HM}}\, {A_\Sigma \,t\ov 4G_N}\\
&\equiv v_E \, s_\text{th} A_\Sigma t\,,
} 
where we used that $s_\text{th}=1/(4G_N)$ with our choice of $z_h=1$. The value of $v_E$ for a Schwarzschild black brane is
\es{vES}{
v_E={\le(d-2\ov d\ri)^{(d-2)/(2d)}\ov \le(2(d-1)\ov d\ri)^{(d-1)/d}}\,.
}
This behavior is valid for any entangling surface at early times~\cite{Hartman:2013qma,Liu:2013iza,Liu:2013qca}, but for the strip it lasts until saturation. We plot the entropy curve in Fig.~\ref{fig:strip}. 

We note that for the $d=2$ BTZ black hole there does not exists a finite value of $z_\text{HM}$, because $-f(z)/z^2$ is monotonically increasing. Hence, the computations above are not valid. Nevertheless, plugging in $d=2$ into our formulas gives the correct result for $d=2$~\cite{Calabrese:2005in,Balasubramanian:2010ce,Balasubramanian:2011ur}.

We have only analyzed the end of the world brane quench model so far. It was shown in~\cite{Liu:2013iza,Liu:2013qca} that the same results apply in the Vaidya setup. They can be derived from integrals very similar to~\eqref{zsol} explicitly, but can also be understood just from a careful analysis of the HRT surfaces. First, the HRT surface passes through the null shell, and caps off in the AdS region of the geometry. This part of the surface is a chunk of an RT surface that we would get in the vacuum, hence only contributes to the area law pieces of the entropy that we discard. In the black hole region of spacetime the equations of motion are the same as in the end of the world brane case, thus the only way that the surface can stay inside the horizon for a long time, is to skim $z_\text{HM}$, as we argued around~\eqref{zExpStrip}. Indeed, we can see from the plots of~\cite{Liu:2013qca} that after crossing the shell the HRT surface reaches the vicinity of  $z_\text{HM}$ exponentially fast, and stays there for a long time, before departing to finally reach the boundary. We can argue in the same way as for the  end of the world brane case that only the part of the surface skimming the $z_\text{HM}$ surface is relevant in the limit  $R,t\gg \beta$. We conclude that the extensive part of the entropy (in the large strip size limit) is the same in the two holographic setups considered above. We will refer to the independence of the quench process as universality, and discuss it in more detail at the end of Sec.~\ref{sec:Universality}.

\section{Holographic bounds on the speed of information spread}\label{sec:Bounds}

The Null Energy Condition (NEC) imposes two conditions on the functions $f(z)$ and $h(z)$ appearing in~\eqref{Metric}:
\es{NECcond}{
h'(z)&\geq 0\\
{d\ov dz}\le({\sqrt{h(z)}\ov z^{d-1}}\, {d\ov dz}\le({f(z)\ov h(z)}\ri)\ri)&\geq 0\,,
}
where $d$ is the number of boundary spacetime dimensions.

The idea that we are going to use is to turn these inequalities into differential equations:
\es{NECeq}{
h'(z)&=a(z)\geq 0\\
{d\ov dz}\le({\sqrt{h(z)}\ov z^{d-1}}\, {d\ov dz}\le({f(z)\ov h(z)}\ri)\ri)&=b(z)\geq 0\,,
}
where $a(z)$ and $b(z)$ are positive functions that we take as given. For example, for the RN black brane~\eqref{RNBH}  $a(z)=0$ and $b(z)=2d(d-1)\,q\, z^{d-3}$.

The differential equations~\eqref{NECeq} can be solved by straightforward integration, and one can fix the three integration constants that arise using the three boundary conditions~\eqref{BdyCond} and~\eqref{BdyCond2}. Naively the solution for $f(z)$ involves a double integral over $b(z)$, but we can get rid of one integral using integration by parts. We can bring the solution to the nice form:
\es{Solution}{
h(z)&=1+\int_0^z dz' \ a(z')\\
f(z)&=h(z)\le[\le(1-{H(z)\ov H(1)}\ri)\le(1-\int_0^1 dz' \ H(z') b(z')\ri)+\int_1^z dz' \ \le(H(z)-H(z')\ri) b(z')\ri]\,,
}
where we introduced the function
\es{HDef}{
H(z)&=\int_0^z dz'\ {z'^{d-1}\ov \sqrt{h(z')}}\,. 
}
It is easy to check that these indeed solve the differential equations~\eqref{NECeq}, and they manifestly satisfy the boundary conditions~\eqref{BdyCond} and~\eqref{BdyCond2}.

Now we are in good shape to prove bounds on the speed of information spread form holography. In Sec.~\ref{sec:Review} we have introduced the quantities $v_E$ and $v_B$ and determined them in terms of the functions $h(z),\, f(z)$.
To bound $v_E$ we will bound $-f(z)/h(z)$ from above. We express
\es{SchMinusf}{
-f^{(S)}(z)+{f(z)\ov h(z)}=\le(z^d-{H(z)\ov H(1)}\ri)+\le({H(z)\ov H(1)}-1\ri)\int_0^1 dz' \ H(z') b(z')+\int_1^z dz' \ \le(H(z)-H(z')\ri) b(z')\,,
}
where $f^{(S)}(z)=1-z^d$ is the Schwarzschild emblackening factor.
From the definition~\eqref{HDef} it is clear that $H(z)$ is a nonnegative monotonically increasing function of $z$, and because $b(z)$ is also nonnegative, the last two terms are readily seen to be positive behind the horizon, $z>1$. We now focus on the first term, and show that it is also nonnegative. Again, from the definition~\eqref{HDef} and using that $h(z)$ is a positive monotonically increasing function of $z$ we see that
\es{HManip}{
{H(z)\ov H(1)}=1+{H(z)-H(1)\ov H(1)}=1+{\int_1^z dz'\ {z'^{d-1}\ov \sqrt{h(z')}}\ov \int_0^1 dz'\ {z'^{d-1}\ov \sqrt{h(z')}}}\leq 1+{(z^d-1)/ (d\sqrt{h(1)})\ov  1/ (d\sqrt{h(1)})}\leq z^d\,,
}
where in the only nontrivial step we increased the numerator by using $1/\sqrt{h(z')}\leq1/\sqrt{h(1)}$ valid for $z'\geq1$, and decreased the denominator by using  $1/\sqrt{h(z')}\geq1/\sqrt{h(1)}$ valid for $0\leq z'\leq 1$. Using the fact that if a function is greater than another one than this is true for their maxima as well, we conclude that
\es{vEFinal}{
-f^{(S)}(z)\geq -{f(z)\ov h(z)} \quad \implies \quad v_E\leq v_E^{(S)}\,,
}
with the Schwarzschild value given in~\eqref{vES}.

Bounding $v_B$ is even more straightforward.
Using~\eqref{Solution} we get the more explicit formula for $v_B$:
\es{cESq}{
v_B^2&=-{1\ov 2(d-1)}\,{f'(1)\ov h(1)}={1\ov 2(d-1)}\,{1- \int_0^1 dz' \ H(z') b(z')\ov \sqrt{h(1)} H(1)}\leq {1\ov 2(d-1)}\, {1\ov \sqrt{h(1)} / (d\sqrt{h(1)})}={d\ov 2(d-1)}\,,
}
which is just the Schwarzschild value~\eqref{vBRN}. In the only nontrivial step above, we increased the numerator and decreased the denominator again. We conclude that
\es{cEBound}{
v_B\leq v_B^{(S)}\,.
}

In~\cite{Mezei:2016wfz} we argue that in any quantum system $v_E\leq v_B$. 
We can test that field theory argument by showing that
\es{RatioBound}{
{v_E\ov v_B}\leq 1\,.
}
To prove this, we first take the explicit formulas from~\eqref{Solution} and~\eqref{cESq}
\es{RatioBound2}{
{v_E^2\ov v_B^2}&={\le[\le({H(z)\ov H(1)}-1\ri)\le(1-\int_0^1 dz' \ H(z') b(z')\ri)-\int_1^z dz' \ \le(H(z)-H(z')\ri) b(z')\ri]/z^{2(d-1)} \bigg\vert_{z=z_\text{HM}} \ov \le(1- \int_0^1 dz' \ H(z') b(z')\ri)/\le( 2(d-1) \sqrt{h(1)} H(1)\ri)}\,.
}
We decrease the numerator by dropping its second term, and we get major cancellations between the numerator and denominator afterwards:
\es{RatioBound3}{
{v_E^2\ov v_B^2}&\leq {2(d-1)\ov z^{2(d-1)}} \bigg\vert_{z=z_\text{HM}}   \sqrt{h(1)} \le({H(z)- H(1)}\ri)\\
&\leq {2(d-1)\ov d }\,\, {z^d-1 \ov z^{2(d-1)}}  \bigg\vert_{z=z_\text{HM}}\,,
}
where in the second line we used the definition of $H(z)$~\eqref{HDef} and that $h(z)$ is a monotonically increasing function. We now plug in the maximum value of the function  ${(z^d-1) / z^{2(d-1)}}$ for $z\geq1$; the value of the function at $z=z_\text{HM}$ is necessarily smaller than this value:
\es{RatioBound4}{
{v_E^2\ov v_B^2}&\leq \le(2(d-1)\ov d-2\ri)^{-(d-2)/d}=\le({v_E^2\ov v_B^2}\ri)^{(S)} \leq 1 \,, 
}
where finally we noticed that the bound that we obtained is equal to the ratio of speeds for the Schwarzschild black brane~\eqref{vBRN} and~\eqref{vES}, which is smaller than 1. Thus, we conclude that this ratio is maximal in an uncharged quench.
We also take this result as a strong confirmation of the field theory argument  for $v_E\leq v_B$ given in~\cite{Mezei:2016wfz}, which perhaps is the most important result that the NEC gives us.

One may wonder whether there is a lower bound in $v_E/v_B$, which would be interesting from the perspective of the bounds analyzed in~\cite{Mezei:2016wfz}. There is no such bound however, as for a near extremal RN black brane we have
\es{NERN}{
\de q\equiv 1-q\,, \quad v_E=\frac12\sqrt{d\ov d-1}\, \de q+O(\de q^2)\,, \quad v_B=\sqrt{d\ov 2(d-1)}\, \sqrt{\de q}\,, \quad {v_E\ov v_B}=\sqrt{\de q\ov 2}+O(\de q^{3/2})\,,
}
hence the ratio can be arbitrary small.

\section{Spherical entangling surfaces}\label{sec:spherical}

\subsection{Equations of motion for the HRT surface}

We will look for the extremal surface parametrized as $(v(\rho),z(\rho))$ in the spherical case. In the Vaidya setup in the AdS region the surface is described by a hemisphere that reaches the shell at $\rho_c$
\es{Hemisphere}{
v_\text{AdS}(\rho)=z_c-z_\text{AdS}(\rho)\,, \qquad z_\text{AdS}(\rho)=\sqrt{z_c^2+\rho_c^2-\rho^2}\,,
}
while in the end of the world picture the extremal surface has the topology of a cylinder and ends on the brane on a sphere of radius $\rho_c$.\footnote{We are using the same symbol $\rho_c$  for the two radii because they play similar roles, but the two setups are different.}

The equation of motion for the HRT surface in black brane background can be obtained by extremizing the area functional
\es{AreaFunct}{
A&= K\, \int^R_{\rho_c} d\rho\ {\rho^{d-2}\ov z_\text{BH}^{d-1}}\sqrt{Q}\,, \qquad Q\equiv 1 -2v'_\text{BH} z'_\text{BH} - f(z_\text{BH}) \le(v'_\text{BH}\ri)^2\,, 
} 
and by satisfying the appropriate boundary conditions, where $K$ is the area of the unit $S^{d-2}$. From here on we will set $h=1$ in~\eqref{Metric} to reduce the complexity of equations. It is straightforward to generalize all our results to that case. Because the area functional does not depend on $v(\rho)$, we have a corresponding conserved quantity
\es{Conserved}{
E={\rho^{d-2}\ov z_\text{BH}^{d-1}\sqrt{Q}}\le(z_\text{BH}'+f(z_\text{BH}) v_\text{BH}'\ri)\,,
}
from which $v'(\rho)$ can be expressed.  The equation of motion of $z(\rho)$ is
\es{EOMs}{
{ z_\text{BH}^{d-1}\sqrt{Q}\ov \rho^{d-2}}\, {d\ov d\rho}\le[{\rho^{d-2}\ov z_\text{BH}^{d-1}\sqrt{Q}}v_\text{BH}'\ri]&={(d-1)Q\ov  z_\text{BH}}+\frac12 {df(z_\text{BH})\ov dz}\, \le(v'_\text{BH}\ri)^2\,,
}
Plugging~\eqref{Conserved} into this equation, we get a single second order ODE for $z_\text{BH}(\rho)$, which can be solved numerically:
\es{FinalEOM}{
0&=\le(f(z_\text{BH})+E^2\,{z_\text{BH}^{2(d-1)}\ov \rho^{2(d-2)}}\ri)\, z_\text{BH}''+ \le(f(z_\text{BH})+\le(z'_\text{BH}\ri)^2\ri)\,\le({d-2\ov \rho}\, z'_\text{BH}+{(d-1)f(z_\text{BH})\ov z_\text{BH}}\ri)\\
&+\le(-\le(z'_\text{BH}\ri)^2+E^2\,{z_\text{BH}^{2(d-1)}\ov \rho^{2(d-2)}}\ri)\,\frac12 {df(z_\text{BH})\ov dz}\,.
}
Below we write down the expression for $E$ explicitly. We note that the RT surface in the static black brane background solves this equation with $E=0$.

For the Vaidya setup the boundary conditions are
\es{VaidyaBC}{
v_\text{BH}(\rho_c)&=0 \qquad\qquad\qquad z_\text{BH}(\rho_c)=z_c  \\
v'_\text{BH}(\rho_c)&=v'_\text{AdS}(\rho_c) \qquad\quad z'_\text{BH}(\rho_c)={1+f\le(z_c\ri)\ov 2}\,z'_\text{AdS}(\rho_c)\,,
}
where the last equation is obtained by integrating the field equations across the null shell~\cite{Liu:2013iza,Liu:2013qca}. For the end of the world brane setup the near brane behavior is dictated by the condition that the HRT surface has to end on the brane perpendicularly.\footnote{In the thermofield double this amounts to requiring a $\Z_2$ symmetry between the two sides.}
\es{EndOfWorldBC}{
v_\text{BH}(\rho)&=v_\text{brane}(z_c)+\sqrt{2\rho_c\ov -(d-2)f(z_c)}\,  \sqrt{\rho-\rho_c}+O(\rho-\rho_c)\\
 z_\text{BH}(\rho)&=z_c-{\le[2(d-1)f(z_c)-z_c f'(z_c)\ri]\rho_c\ov 2(d-2)z_c}  (\rho-\rho_c) +O\le((\rho-\rho_c)^2\ri)\,.
}
Note that $v'_\text{BH}(\rho)$ diverges as $\rho\to\rho_c$. From these boundary conditions we can read off the value of $E$:
\es{EValue}{
E_\text{Vaidya}&=-\frac12\le(\rho_c\ov z_c\ri)^{d-1}\, {1-f(z_c) \ov \sqrt{z_c^2+\rho_c^2}}\\
E_\text{brane}&=-{\rho_c^{d-2}\ov z_c^{d-1}}\, \sqrt{-f(z_c)}
}

\subsection{Numerical solutions}

\begin{center}
\begin{figure}[!h]
\includegraphics[scale=0.6]{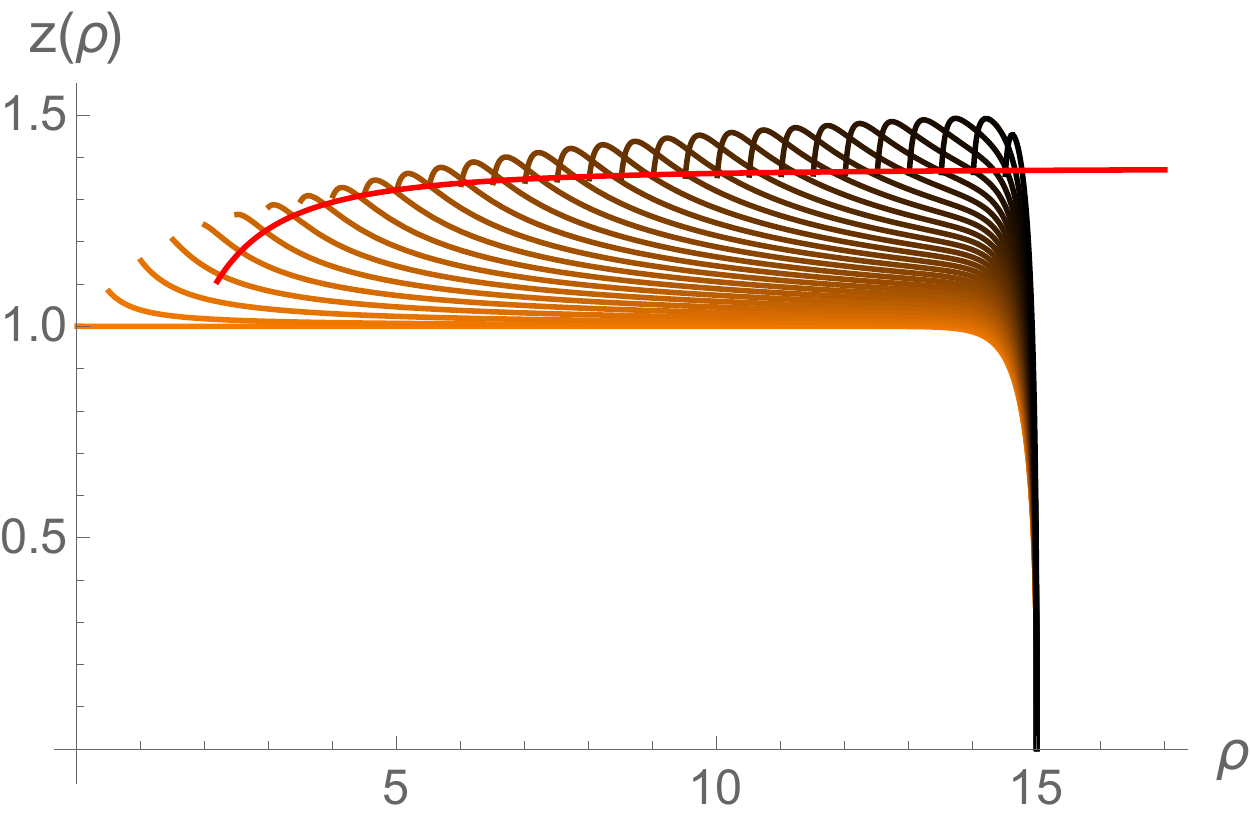} 
\includegraphics[scale=0.6]{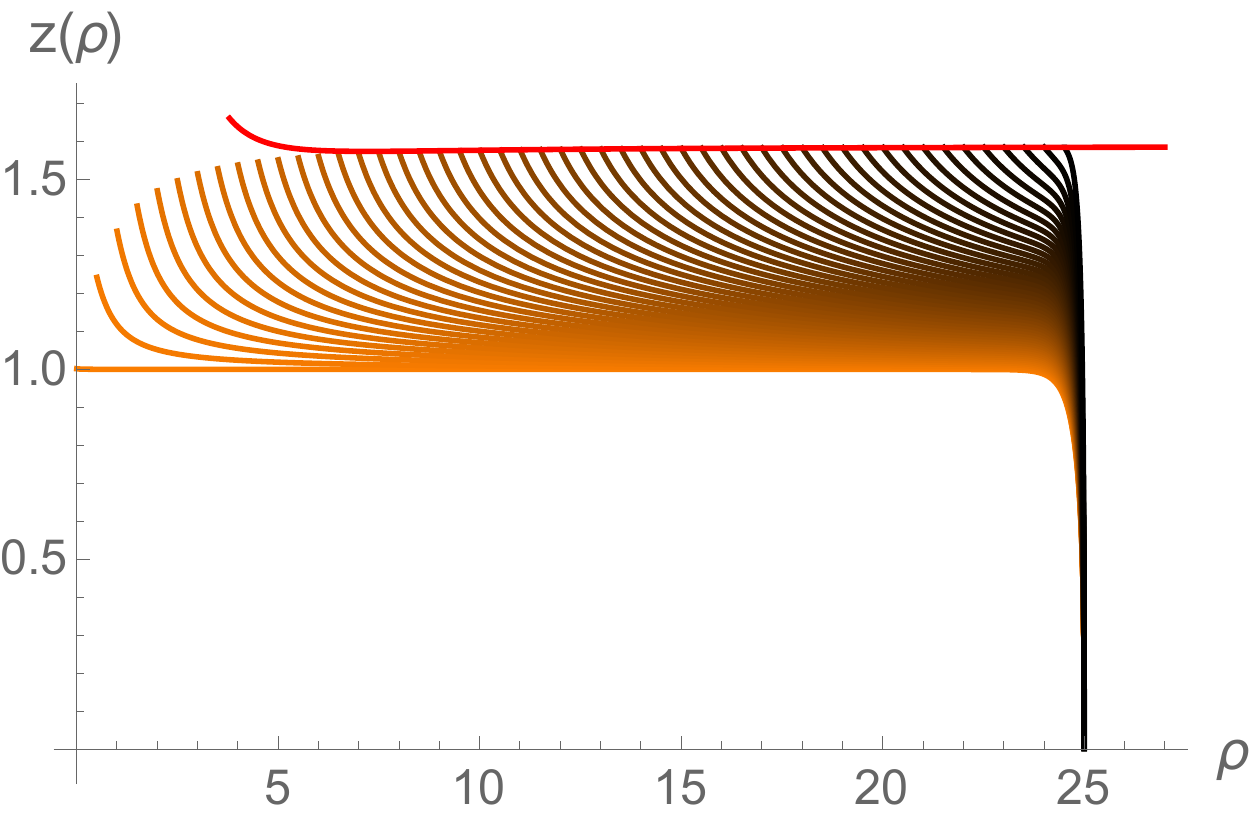}
\caption{Extremal surfaces in the $d=3$ Schwarzschild geometry. {\bf Left:} In the Vaidya setup we show the portion of the extremal surfaces that are in the black brane region for $R=15$. Earlier times are drawn by darker colors. The red line shows~\eqref{ztExpansion2}, which is an expansion for $z_c$ for large $\rho_c$.  {\bf Right:} In the end of the world brane setup we chose $R=25$, and the red line is~\eqref{ztExpansion}.
\label{fig:minsur}}
\end{figure}
\end{center}

\begin{center}
\begin{figure}[!h]
\includegraphics[scale=0.85]{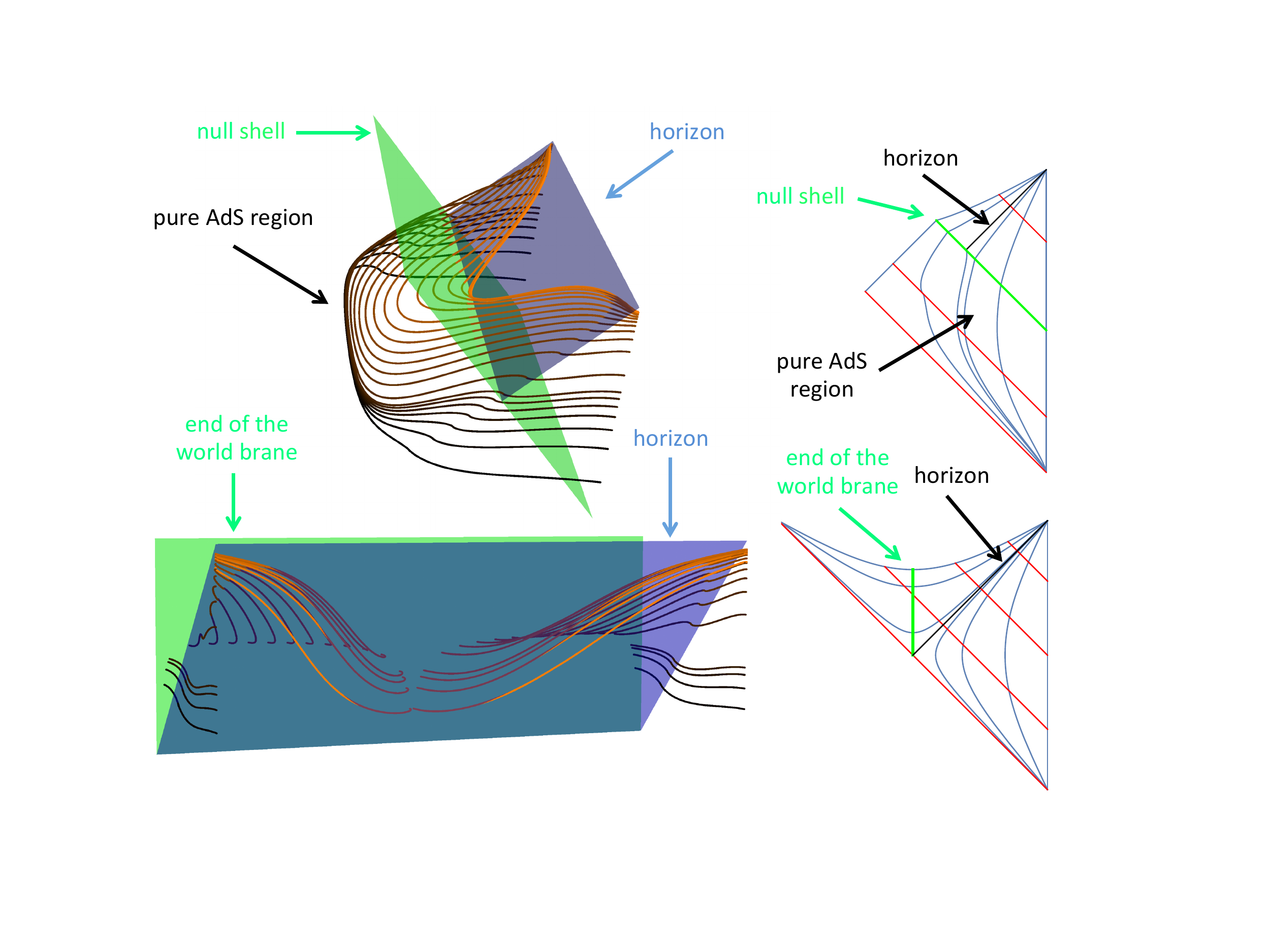} 
\caption{HRT surfaces with $R=4$ in the Schwarzschild geometry corresponding to $d=3$ boundary dimensions. On the left we plotted the HRT surfaces $(v(\rho),z(\rho))$ in coordinates that correspond to the Penrose diagram shown on the right. On the right only the part of the Penrose diagram is shown that is covered by the coordinates $(z,v)$.  {\bf Top:} HRT surfaces in the Vaidya setup, with the null shell shown with green and the horizon with blue.  Darker color correspond to earlier times. At very early times, much of the surface is in the pure AdS region, and is an almost perfect hemisphere in the coordinates~\eqref{Metric}, deformed by the conformal mapping that gives the Penrose diagram on the right. As time evolves the surface goes behind the horizon, most of it lies on $z_\text{HM}$, and near saturation it climbs out from behind the horizon and the entropy saturates, when the surface is only barely touching the shell. {\bf Bottom:}  HRT surfaces in the end of the world brane setup. At early times the HRT surface is a tube connecting the boundary theory entangling surface $\Sig$ to the brane, and the image on the brane is $\Sig_\text{im}\approx\Sig$. The linear regime of entropy growth takes place, when $\Sig_\text{im}$ migrates up to $z_\text{HM}$. We can clearly see that as we go to later times (lighter color) $\Sig_\text{im}$ is shrinking, but staying at $z_\text{HM}$. Finally $\Sig_\text{im}$ migrates towards the bifurcation surface, but this is an $O(\beta/R)$ effect. For similar figures, see~\cite{Hubeny:2013dea}.
\label{fig:3d}}
\end{figure}
\end{center}

For illustration we show extremal surfaces obtained by numerically solving the equations of motions in the two setups in Fig.~\ref{fig:minsur}. We show the HRT surfaces on the Penrose diagram on Fig.~\ref{fig:3d} to provide a better understanding of their fate. We note that saturation in the Vaidya setup  may or may not be continuous depending on the detailed geometry of the black brane. This issue discussed in detail in Sec.~\ref{sec:cusp}. In the end of the world brane setup, the saturation is strictly speaking always discontinuous, as the time dependent HRT surfaces necessarily remain behind the horizon for all times to connect to the brane, while the RT surface that gives the saturation value of the entropy only approaches the horizon (exponentially closely~\eqref{epdet}). The change in slope in the entropy curve from this change of topology is an $O(\beta/R)$ effect, hence in the limit we are working in the saturation can become continuous.

\begin{center}
\begin{figure}[!h]
\includegraphics[scale=1]{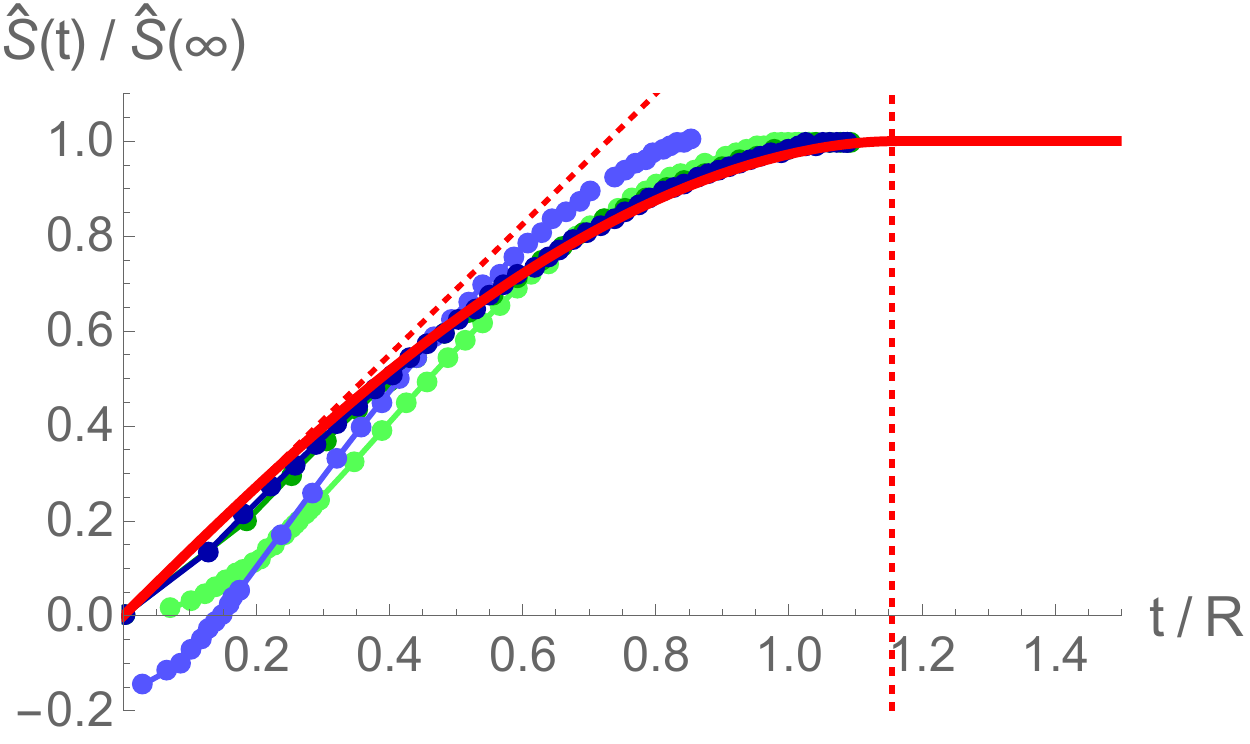} 
\caption{$\hat{S}(t)$ for the Vaidya setup for $R=4$ (light green) and $R=15$ (dark green) and for the end of the world brane setup for $R=4$ (light blue) and $R=25$ (dark blue).   Finally, the red curve corresponds to the analytic prediction for $R\gg\beta$ given in~\eqref{OtherQuant2}. The red dashed lines show how to read off $v_E$ and $v_B$ from the plot. Because in the end of the world brane setup the initial state is different from the CFT vacuum, subtracting the vacuum contribution gives negative $\hat{S}(t)$ for early times. Because this contribution is area law, for larger $R$ this effect goes away.  The data points for $R=4$ come from the HRT surfaces plotted on Fig.~\ref{fig:3d}.
The data points for $R=15,25$ come from the HRT surfaces plotted on Fig.~\ref{fig:minsur}, and are somewhat hard to distinguish due to very accurate overlap. The main message of this plot is that as we increase $R$ the data points collapse onto the red curve~\eqref{OtherQuant2}.
\label{fig:S(t)}}
\end{figure}
\end{center}

The entropy $\hat{S}(t)$ is calculated from the solutions by calculating the area of the HRT surface and subtracting the vacuum entanglement entropy across a circle of the same radius $R$. The results are collected in Fig~\ref{fig:S(t)}.

\subsection{Scaling limit and explanation of universality} \label{sec:Universality}

Based on our experience from the numerical solutions, we make an attempt to capture the limit of large $R$ by making the scaling Ansatz:
\es{ScalingAnsatz}{
z_\text{BH}(\rho)=Z\le({\rho\ov\rho_c} \ri)\,,\qquad (\rho_c\to\infty)\,.
}
In both setups in this limit we expect the following scaling behavior:
\es{ScalingBehavior}{
E&\sim \rho_c^{d-2}\,,\\
z_\text{BH}(\rho)\sim \rho_c^0\,, \quad z'_\text{BH}(\rho)&\sim {1\ov \rho_c}\,, \quad z''_\text{BH}(\rho)\sim {1\ov \rho_c^2}\,,
}
hence to leading order~\eqref{FinalEOM} becomes an algebraic equation:
\es{FinalEOMScaling}{
0&={(d-1)f(Z)^2\ov Z}+{\cal E}^2\,{Z^{2(d-1)}\ov r^{2(d-2)}}\,\frac12 {df(Z)\ov dZ}\,,
}
where for transparency we have introduced the scaled version of $E\equiv \rho_c^{d-2} {\cal E}$ and $\rho\equiv \rho_c\, r$.

\begin{center}
\begin{figure}[!h]
\includegraphics[scale=0.7]{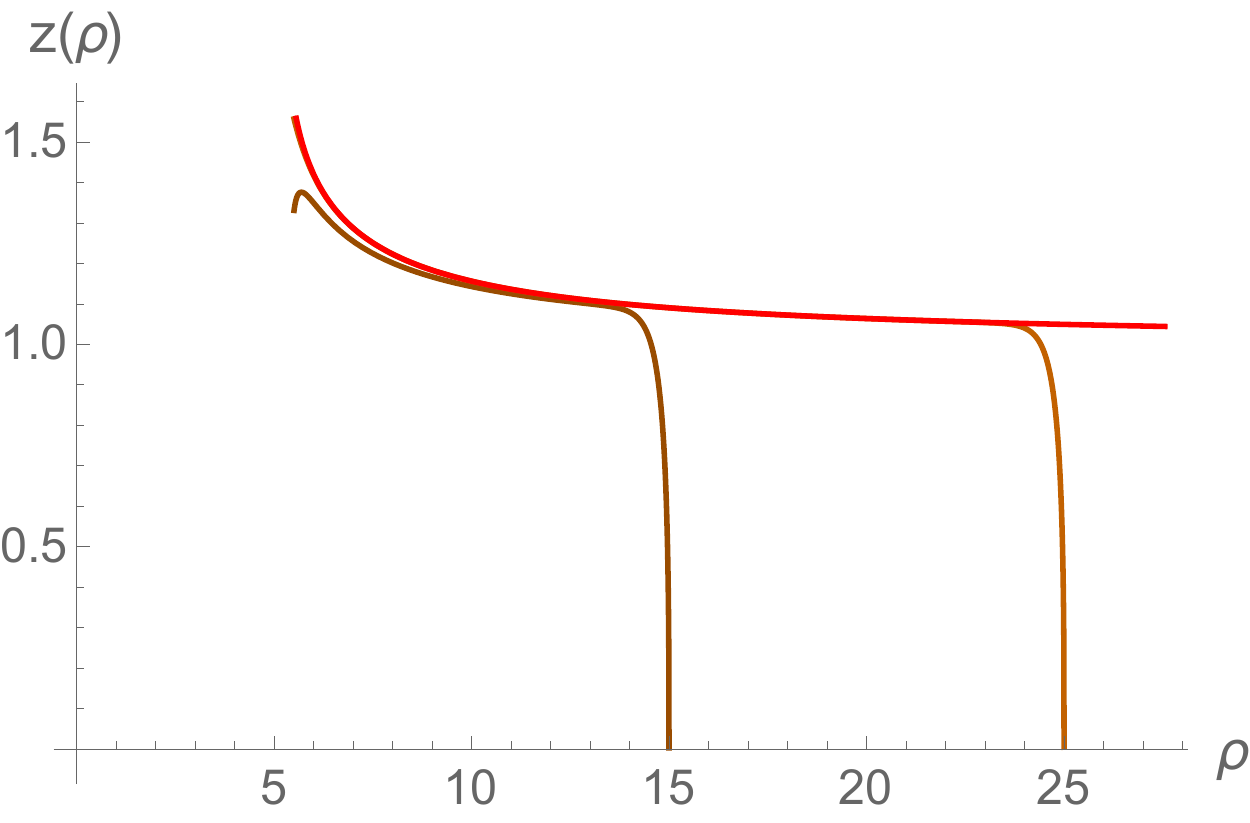}
\caption{Demonstration that the scaling solution~\eqref{FinalEOMScaling2} indeed approximates the real extremal surfaces exponentially well. We took two extremal surfaces from Fig.~\ref{fig:minsur}; for $R=15$ we have a Vaidya, while for $R=25$ an end of the world brane extremal surface, but both corresponding to the same $\rho_c$. Note that in the Vaidya setup the extremal surface shoots up to the scaling solution quickly, as discussed in the main text. To reach the boundary, the extremal surfaces depart the scaling solution exponentially fast, as described by~\eqref{Deviation}.
\label{fig:scalingproof}}
\end{figure}
\end{center}

\begin{center}
\begin{figure}[!h]
\includegraphics[scale=0.5]{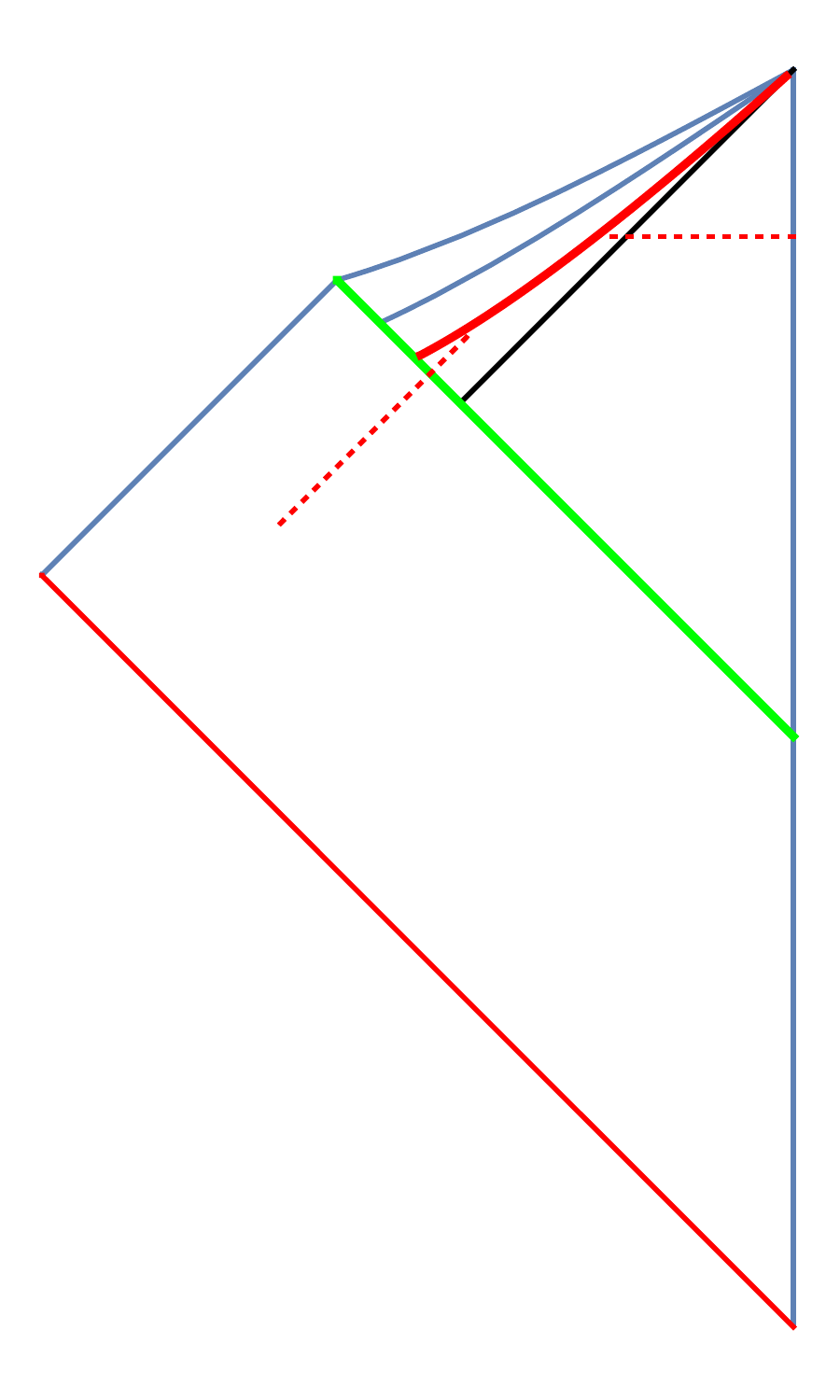} \hspace{1.8cm}
\includegraphics[scale=0.5]{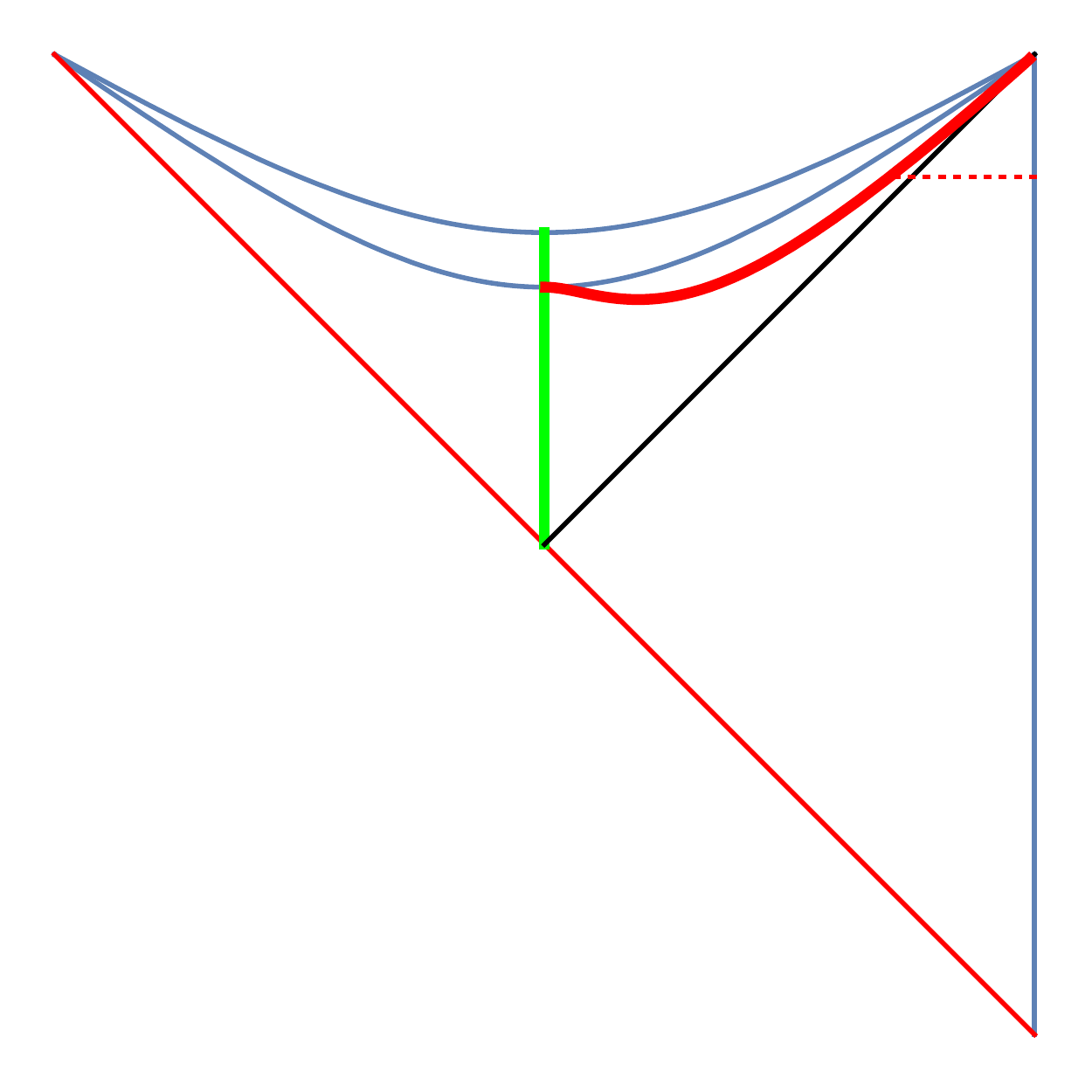} 
\caption{Scaling surface in the $d=3$ Schwarzschild black brane drawn with solid red line on the Penrose diagrams. The first blue line behind the horizon is $z=z_\text{HM}$. (The second is the singularity.)  Otherwise we use the same color coding as on Fig.~\ref{fig:3d}.  For this plot we chose $\rho_c=1$ (for visualization purposes), but for the scaling surface to be a good approximation to the HRT surface, we have to choose a large $\rho_c$.  {\bf Left:} In the Vaidya setup the HRT surface sketched by red dotted line starts in the pure AdS region, crosses the null shell, then skims the scaling surface, and finally reaches the boundary. The extensive part of the entropy is equal to the area of the scaling surface between the points where the red dotted lines reach and depart it.
{\bf Right:} In the end of the world brane setup the HRT surface starts very close to the scaling surface, then shoots out to the boundary sketched with red dotted lines.
\label{fig:1side}}
\end{figure}
\end{center}

Let us take the end of world brane setup first. Plugging in~\eqref{EValue} into~\eqref{FinalEOMScaling} we get
\es{FinalEOMScaling2}{
0&={(d-1)f(Z)^2\ov Z}- {f(z_c) \ov  r^{2(d-2)}}\,\le({Z\ov z_c}\ri)^{2(d-1)} \,\frac12 {df(Z)\ov dZ}\,.
}
This equation in fact determines what $Z(1)=z_c$ is, we simply have to set $r=1$ in~\eqref{FinalEOMScaling2} and use~\eqref{EValue} to get:
\es{ZtDetermine}{
0={df(Z)\ov dZ}\Big\vert_{Z=z_c}-{2(d-1)f(z_c)\ov z_c}\,,
}
which is exactly the equation~\eqref{zHMEq} that $z_\text{HM}$ satisfies.\footnote{When comparing to~\eqref{zHMEq}, recall that we have set $h=1$.} We see that contrary to the strip case, the extremal surface does not like to stay on $z_\text{HM}$ even in the large $R$ limit, instead it moves away from it according to~\eqref{FinalEOMScaling2}, see also Fig.~\ref{fig:scalingproof}. For $r\to\infty$ we get $Z=1$, so the scaled extremal surface does not get out from behind the horizon. The scaling surface is shown on the Penrose diagram on Fig.~\ref{fig:1side}, which summarizes the information on Fig.~\ref{fig:3d} and Fig.~\ref{fig:scalingproof}. The scaling solution we find can be viewed as the precise realization of the ``critical surface" envisioned in~\cite{Liu:2013iza,Liu:2013qca}.

We note that~\eqref{FinalEOMScaling2} can be corrected order by order in $1/\rho_c$. If we wrote $Z=Z_0+{1\ov \rho_c^2}\, Z_1+\dots$, at every order $Z_i$ would be determined by a linear equation only involving the functions determined at previous orders. So without solving differential equations, we get the scaling limit of the HRT surfaces. Without going through the whole procedure, we just write one result for the Schwarzschild black brane corresponding to $d=3$:
\es{ztExpansion}{
z_c(\rho_c)=2^{2/3}-{4\ov3\, \rho_c^2}+{248\times 2^{1/3}\ov9\, \rho_c^4}+\dots\,,
}
where the leading term is $z_\text{HM}$ for the $d=3$ Schwarzschild black brane. The formula~\eqref{ztExpansion} agrees with what we find in numerics as demonstrated on Fig.~\ref{fig:minsur}. The most important observation is that at all orders in $1/\rho_c$ the corrections decay as we go to the horizon, and the scaled surface stays inside the horizon. Then it must be that from this perspective nonperturbative $\exp(-\rho_c)$ effects cause the surface to exit from behind the horizon. This is reminiscent of what we found in the strip case in Sec.~\ref{sec:strip}.  To avoid clutter, we specialize to the Schwarzschild black brane to get:
\es{Deviation}{
z(\rho)&=Z_0\le({\rho\ov\rho_c} \ri)+{1\ov \rho_c^2} \, Z_1\le({\rho\ov\rho_c} \ri)+\dots+\exp\le(-\#\rho_c\ri)\, \tilde z(\rho)\\
0&= \tilde z''(\rho)-d(d-1)Z_0^{d-2}\le[1+{d\ov 2d-2-(d-2)Z_0^d}\ri]\,\tilde z(\rho)+O\le({1\ov\rho_c^2}\ri)\,.
}
For $r\to\infty$ the latter equation simplifies to:
\es{Deviation2}{
0&= \tilde z''(\rho)-2d(d-1)\,\tilde z(\rho)\,,
}
so we get an exponentially growing solution that has to be tamed by the $\exp\le(-\#\rho_c\ri)$ coefficient. We have not analyzed~\eqref{Deviation} in detail, but it is worth noting that the coefficient function of $\tilde z(\rho)$ blows up as $\rho\to\rho_c$. Because this effect amplifies the growth of $z(\rho)$, we leave the coefficient in the exponential suppression undetermined. We have a clear picture of what the surfaces are doing: they follow the scaling solution behind the horizon, and then shoot out to the boundary exponentially fast. This enables us to determine the $S(t)$ curve analytically!

Knowing the surface, it is easy to determine the rest of the quantities of interest. From~\eqref{AreaFunct} and~\eqref{Conserved} we get
\es{OtherQuant}{
v'(r)&={\rho_c\ov \sqrt{f(Z)\le[1- {Z\,{df(Z)/ dZ}\ov 2(d-1)f(Z)}\ri]}}\\
{A'(r)\ov K}&=\rho_c^{d-1}\, { r^{d-2}\ov Z^{d-1}}\, \sqrt{1-{1\ov 1- {Z\,{df(Z)/ dZ}\ov 2(d-1)f(Z)}}}\,,
}
where $Z=Z(r)$ is the solution of the algebraic equation~\eqref{FinalEOMScaling2}.
Because $\tilde z(\rho)$ shoots out exponentially fast to the boundary, we can drop its contribution to both the extensive part of the entropy and the leading order time, and we end up with
\es{OtherQuant2}{
{t\ov R}&=\tau(r_*)\equiv{1\ov r_*}\int_1^{r_*} dr \ {1\ov \sqrt{f(Z)\le[1- {Z\,{df(Z)/ dZ}\ov 2(d-1)f(Z)}\ri]}}\\
{(d-1)\hat{A}\ov K\, R^{d-1}}&=\sA(r_*)\equiv{d-1\ov r_*^{d-1}}\int_1^{r_*} dr \ { r^{d-2}\ov Z^{d-1}}\, \sqrt{1-{1\ov 1- {Z\,{df(Z)/ dZ}\ov 2(d-1)f(Z)}}}\,,
}
where we introduced $r_*\equiv R/\rho_c$, $\hat{A}$ was defined in~\eqref{Ahat}, and we introduced a $(d-1)$ factor for convenience. Below in Sec.~\ref{sec:cusp} we simplify the expression~\eqref{OtherQuant2} somewhat, but for the purposes of this section this formula suffices. By changing $r_*$ from $1$ to $\infty$ we obtain the parametric curve $\hat{S}(t)$. 

\begin{center}
\begin{figure}[!h]
\includegraphics[scale=0.65]{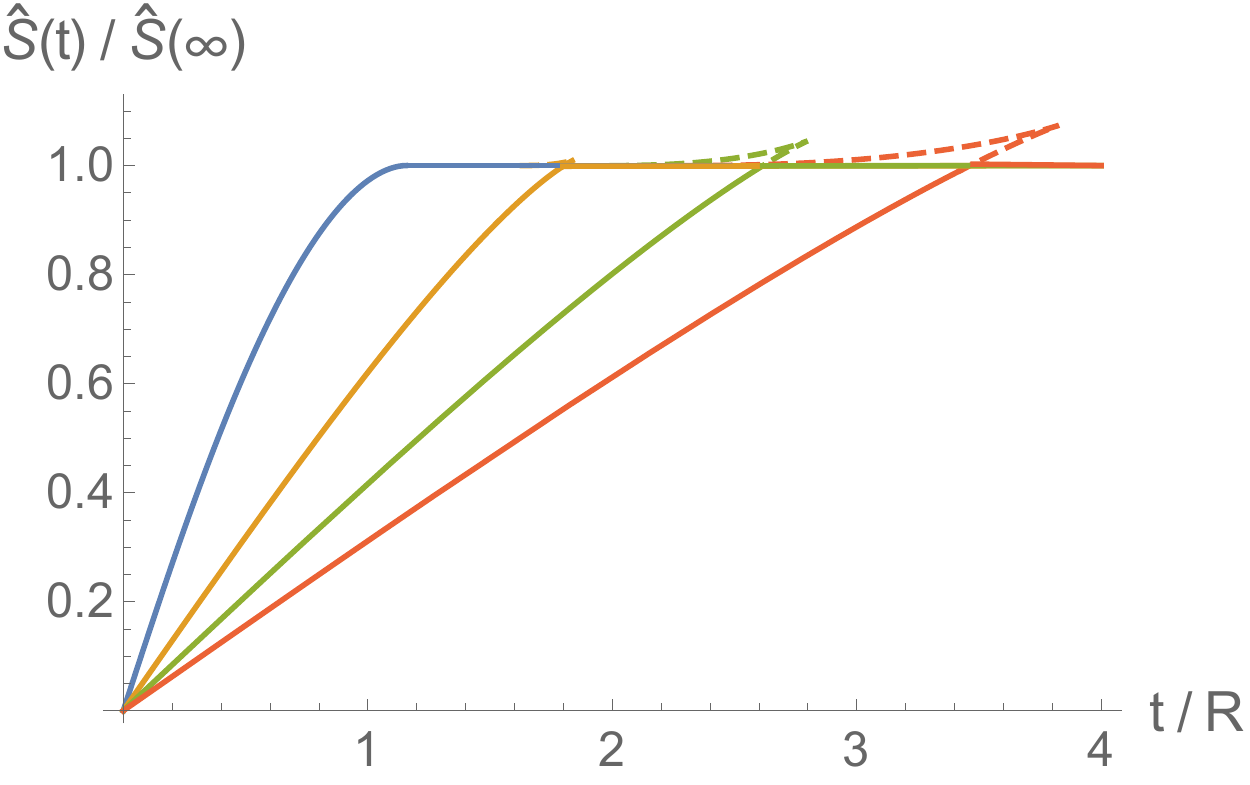}%
\includegraphics[scale=0.65]{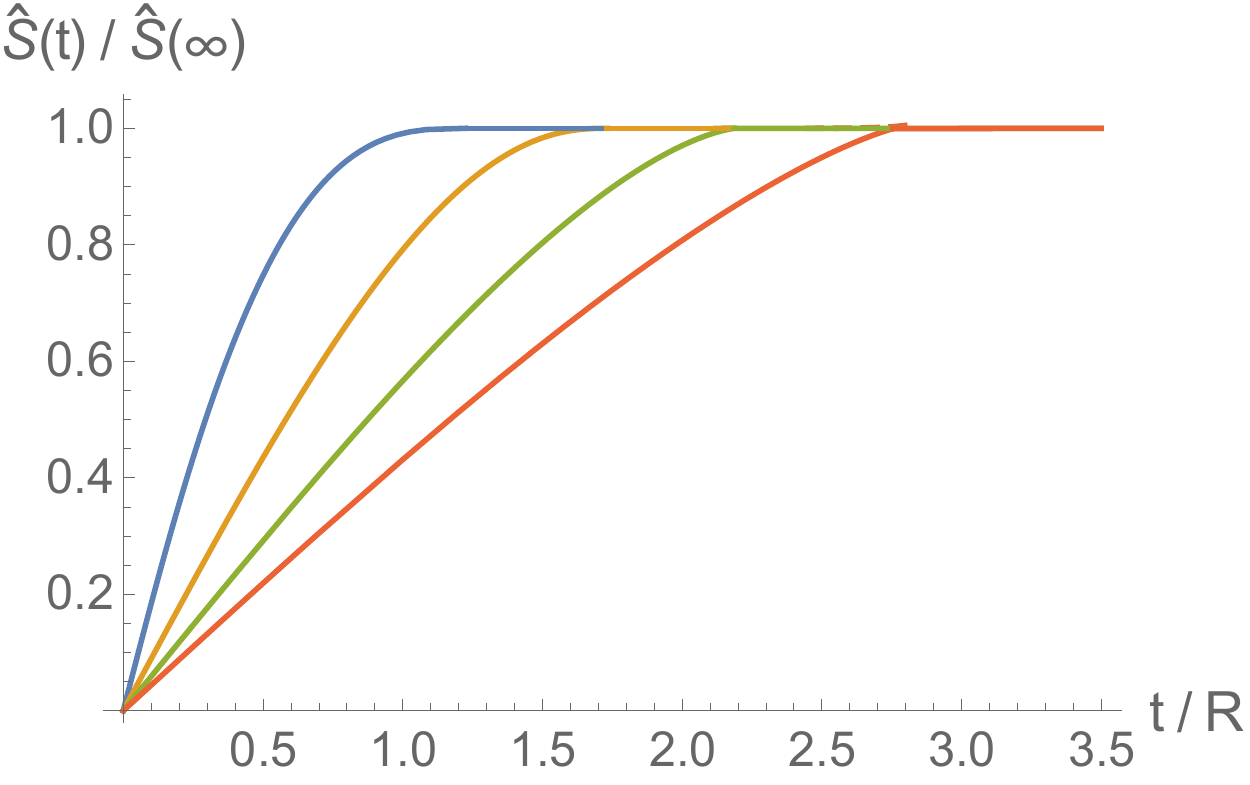}
\caption{Entanglement growth from $d=3,4$ RN black branes with $q=0,1/2,2/3,3/4$ plotted with blue, orange, green, and red respectively. To obtain these curves we numerically integrated~\eqref{OtherQuant2}. The irrelevant parts of the curves $\le(\tau(r_*),\,\sA(r_*)\ri)$ are drawn with dashed lines. {\bf Left:} In $d=3$ saturation is always discontinuous, but in the $q=0$ case to see the cuspy behavior visible for larger $q$ requires extreme numerical precision as discussed in Sec.~\ref{sec:cusp}. {\bf Right:} In  $d=4$ for $q=0,1/2$ saturation is continuous, and discontinuous for  $q=2/3,3/4$, but the dashed parts of the curves are hardly visible.
\label{fig:RN}}
\end{figure}
\end{center}

There is one additional subtlety: it can happen that for a given time $\tau$ there are multiple extremal surfaces, and the holographic entanglement entropy formula requires us to choose the one with the minimal area. In the end of the world brane setup the static surface is available at all times, while for the Vaidya setup it becomes available for $\tau\geq 1/v_B$. It can be shown (see Sec.~\ref{sec:cusp} for details) that 
\es{Limits}{
\tau(\infty)&=\sqrt{-{ 2(d-1)\ov f'(1)}}={1\ov v_B}\,, \qquad \qquad \sA(\infty)=1\,,
}
and a natural minimal assumption is that for $\tau<1/v_B$ we take the extremal surfaces that probe behind the horizon and for $\tau\geq 1/v_B$ we take the static surface.
However, it can happen that the  parametric curve $\le(\tau(r_*),\,\sA(r_*)\ri)$ itself does not give a single valued  $\sA(\tau)$. Then for a given $\tau$, we have to choose the smallest $\sA$.
On Fig.~\ref{fig:S(t)} we show the resulting curve $\le(\tau(r_*),\,\sA(r_*)\ri)$ against the results of numerical calculations  in $d=3$ for the Schwarzschild black brane and find perfect agreement.\footnote{There is an incredibly subtle effect for the $d=3$ for the Schwarzschild black brane. The saturation is not continuous, but this is not visible on Fig.~\ref{fig:S(t)}, as it happens for $(t_S-t)/t_S\sim 10^{-5}$. We zoom in on this neighborhood on Fig.~\ref{fig:nearsat}.} In Fig.~\ref{fig:RN} we plot the resulting curves for RN black branes in $d=3,4$, and we find that the multivalued behavior discussed above does happen in some cases. In Sec.~\ref{sec:cusp} we analyze analytically when such behavior takes place. We note that from the rate of growth bound~\eqref{BoundIntro} (to be proven in Sec.~\ref{sec:Shapes}) and the result~\eqref{Limits} it follows that black branes with ${v_E\ov v_B}<{1\ov d-1}$ will necessarily give a non-single valued entropy curve, as the saturation time in this case has to be $t_S>{R\ov v_B}$. See~\cite{Mezei:2016wfz} for related discussion. 

The Vaidya analysis is a bit more subtle, but the numerics should provide ample guidance. The bottom line is that the scaling solution is the same as in the end of the world brane scenario. At first, this sounds puzzling, as this solution does not satisfy the boundary conditions~\eqref{VaidyaBC}. The resolution is that there is a small, very steep part of the extremal surface that shoots up to $z_\text{HM}$ from the position of the shell crossing $z_c$. This steep part can be seen on Fig~\ref{fig:scalingproof}. This steep part is not captured by the scaling solution, but because it does not give an extensive contribution to $S$ and $t$, we do not need to know its details. To get an identical equation to~\eqref{FinalEOMScaling2} we need the two energies, $E_\text{Vaidya}$ and $E_\text{brane}$ to agree. From this requirement and~\eqref{ztExpansion}, we can derive that in the Schwarzschild case
\es{ztExpansion2}{
z_\text{c,Vaidya}(\rho_c)={\sqrt{3}\ov 2^{1/3}}-{3\sqrt{3}\ov4\, \rho_c^2}+\dots\,.
}
We confirm this prediction with numerical computations on Fig.~\ref{fig:minsur}.

We conclude that the entropy growth in the global quench does not depend on the details of the quench process, the leading order entropy is the same both in the end of the world brane and the Vaidya setup. The analytic argument is backed up by numerical data summarized on Fig.~\ref{fig:S(t)}.
 There are many other quenches that we may consider. We can form a black brane from collapsing massive instead of null matter. We can also smear out the process of a quench in time by a small amount. Because black branes form quickly, which is dual in the field theory to fast local thermalization in time $t_\text{loc}\sim\beta$, the details of the quench should not effect the leading part of the entropy. Indeed, in any quench setup we would find that for times $t\gg \beta$ the extensive part of the entropy is determined by the scaling solution analyzed above. The details of the quench only determine a small part of the surface before it reaches the scaling surface. We demonstrated this phenomenon in the Vaidya case, see Figs.~\ref{fig:minsur} and~\ref{fig:scalingproof}.
   It is a quite satisfying finding that entropy growth for short-range entangled initial states is universal, it does not depend on the state chosen or more generally on the quench process. We found the same universality for strips, and it continues to hold for entangling surfaces of arbitrary shapes as will become clear from the discussion in Sec.~\ref{sec:Shapes}.
   
\subsection{Continuous and discontinuous saturation}\label{sec:cusp}

Let us examine~\eqref{OtherQuant2} for large $r_*$. For this analysis it is more convenient to regard $r$ as the dependent variable. From~\eqref{FinalEOMScaling2} we get
\es{FinalEOMScaling3}{
r(z)&=\le[{f(z_\text{HM}) \ov f(z) }\,\le({z\ov z_\text{HM}}\ri)^{2(d-1)}  {z\, f'(z)\ov 2(d-1)f(z) }\ri]^{1/2(d-2)}\,.
}
We rewrite the integrals as 
\es{OtherQuant3}{
\tau(z_f)&=-{1\ov r(z_f)}\int_{z_f}^{z_\text{HM}} dz \ r'(z)\, {1\ov \sqrt{f(z)\le[1- {z\, {f'(z)}\ov 2(d-1)f(z) }\ri]}}\equiv -{1\ov r(z_f)}\int_{z_f}^{z_\text{HM}} dz \ \sI_\tau(z)\\
\sA(z_f)&=-{d-1\ov r(z_f)^{d-1}}\int_{z_f}^{z_\text{HM}} dz \ r'(z)\,  { r(z)^{d-2}\ov z^{d-1}}\, \sqrt{1-{1\ov 1- {z\, {f'(z)}\ov 2(d-1)f(z) }}}\equiv-{1\ov r(z_f)^{d-1}}\int_{z_f}^{z_\text{HM}} dz\ \sI_\sA(z)\,,
}
where we defined $z_f$ such that $r(z_f)=r_*$, and from now on we will parametrize the $\sA(\tau)$ curve as $\le(\tau(z_f),\,\sA(z_f)\ri)$. Remarkably, we can reduce the number of integrals we need to perform, as $\sA(z_f)$ can be rewritten as follows:\footnote{For Schwarzschild black branes the integrals are computable in terms of Appel functions. Below we perform the integrals in some limits explicitly, where the results are simpler.} 
\es{sArewrite}{
\sA(z_f)&={v_E \ov r(z_f)^{d-2}}\le[\sqrt{{1-{z_f f'(z_f)\ov 2(d-1)f(z_f) } \ov f(z_f)}}+\tau(z_f)\ri]\,,\qquad 
v_E=\sqrt{-{f(z_\text{HM})\ov z_\text{HM}^{2(d-1)}} }\,.
}
From this equation an interesting relation follows. To derive it we note that from~\eqref{OtherQuant} it follows that
\es{derRelation}{
A'(z_f)= (\rho_c\, r)^{d-2}\,{\sqrt{-z_f f'(z_f)\ov2(d-1)}\ov z_f^{d-1}}\, t'(z_f)\,.
}
Combing this relation with~\eqref{sArewrite} and the definition of $\tau,\, \sA$ in terms of $t,\, A$~\eqref{OtherQuant2}, we obtain the derivative relation
\es{derRelation2}{
\sA'(z_f)&={(d-1)v_E\ov r^{d-2}}\, \tau'(z_f)\,.
}
A consequence of this relation is the possibility of cuspy behavior of the entropy curve: $\tau'(z_\text{cusp})=0$ implies  $\sA'(z_\text{cusp})=0$, leading to a cusp in the parametric curve $\le(\tau(z_f),\,\sA(z_f)\ri)$. The appearance of the formula~\eqref{derRelation2} from complicated manipulations of integrals is somewhat magical. In Sec.~\ref{sec:GrowthRate} we give a conceptually clean derivation of~\eqref{derRelation2} generalized to arbitrary shapes.

From the experience with RN black branes, discontinuous saturation is correlated with the near saturation behavior of $\tau(z_f)$ (and hence $\sA(z_f)$ using~\eqref{sArewrite}), thus we investigate the behavior near saturation, $z_f\to 1$. We emphasize that  it is logically possible, to have more complicated behavior than we analyze below, and it would be interesting to understand whether the criterion that we give for discontinuous saturation is necessary or only sufficient. In the limit $z_f\to 1$ both the integrals and  $r(z_f)$ diverge, but $\tau,\, \sA$ have finite limits~\eqref{Limits}.

We first examine $r(z_f)$:
\es{rzfExp}{
z_f&\equiv 1+\de z_f\\
r(z_f)&={a_1\ov \de z_f^{1/(d-2)}}\le(1+a_2\,\de z_f +\dots\ri)\,,
}
where $a_i$ are determined by the emblackening factor $f(z)$ and $d$, and we will avoid writing the explicit expressions down to avoid clutter. 
 We now focus on the integrand $\sI_\tau$ near $z=1$, where it diverges, and write down the potentially divergent terms:
\es{zIExp}{
z&\equiv 1+\de z\\
\sI_{\tau,\infty}(z)&={a_3\ov \de z^{(d-1)/(d-2)}}\le(1+a_4\,\de z \ri)\,.
}
Subtracting these terms from the integrand and adding them back, we obtain the integral in a form that is easier to treat:
\es{IntegralsNewForm}{
\int_{z_f}^{z_\text{HM}} dz \ \sI_\tau(z)&= \int_{z_f}^{z_\text{HM}} dz \ \sI_{\tau,\infty}(z)+\int_{z_f}^{z_\text{HM}} dz \ \le(\sI_\tau(z)-\sI_{\tau,\infty}(z)\ri)\\
&={\tilde{a}_3\ov \de z_f^{1/(d-2)}}+{\tilde{a}_4\, \de z_f^{(d-3)/(d-2)}}+ a_5+a_6\, \de z_f+\dots\\
a_5&\equiv -{\tilde{a}_3\ov  (z_\text{HM}-1)^{1/(d-2)}}-{\tilde{a}_4\, (z_\text{HM}-1)^{(d-3)/(d-2)}}+\int_{1}^{z_\text{HM}} dz \ \le(\sI_\tau(z)-\sI_{\tau,\infty}(z)\ri)\\
a_6&\equiv-\le(\sI_\tau(z)-\sI_{\tau,\infty}(z)\ri)\Big\vert_{z=1}\,,\qquad \dots\,,
}
where in the second line we evaluated the remaining integral $\int_{z_f}^{z_\text{HM}} dz \ \sI_{\tau,\infty}(z)$. In $d=3$, the $\de z_f^{(d-3)/(d-2)}\ $ term becomes a logarithm.

\begin{center}
\begin{figure}[!h]
\includegraphics[scale=0.4]{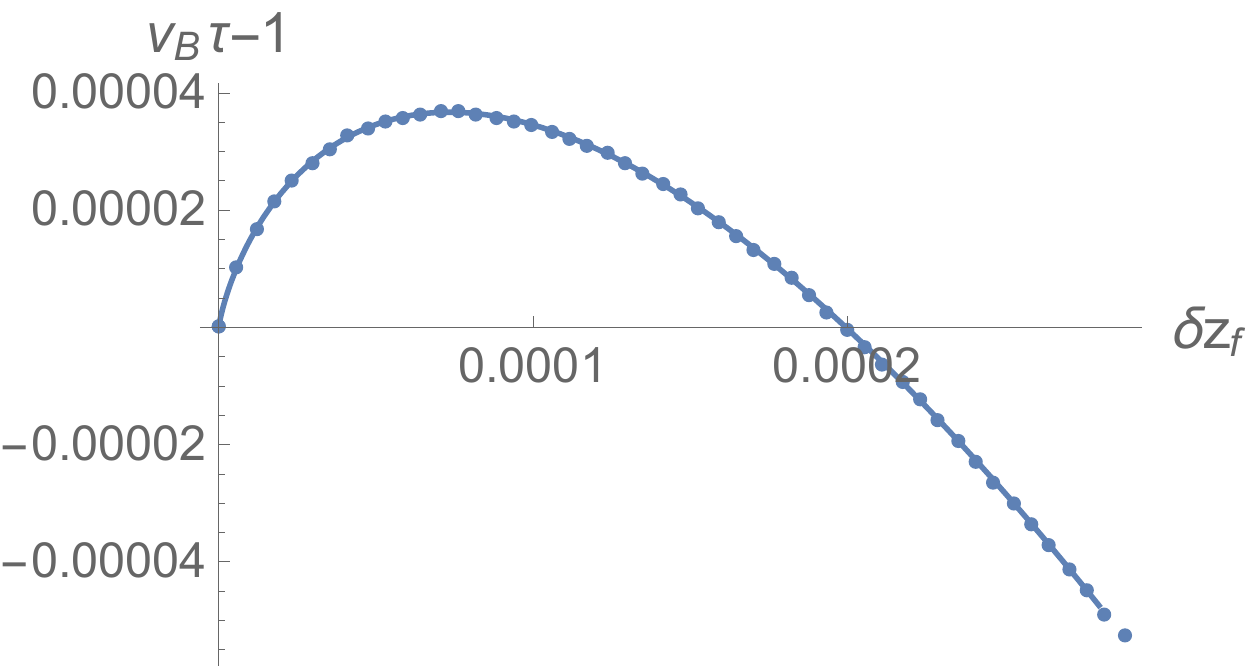}\hspace{0.3cm}
\includegraphics[scale=0.4]{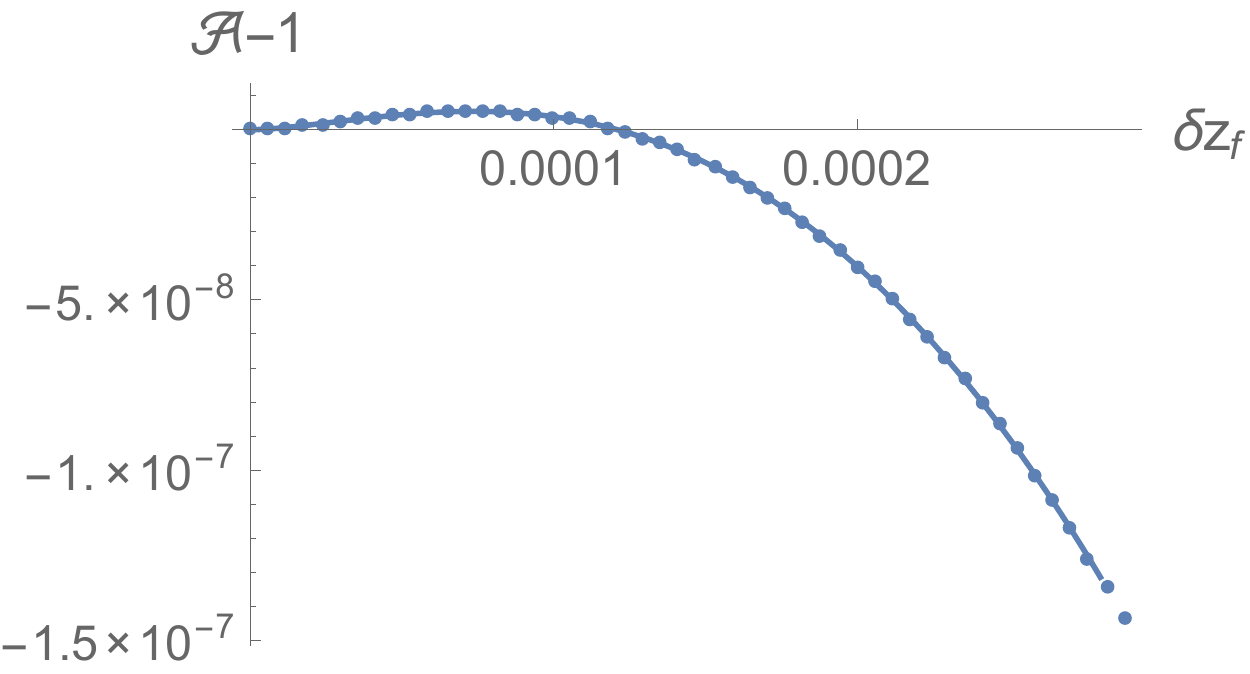}\hspace{0.3cm}
\includegraphics[scale=0.4]{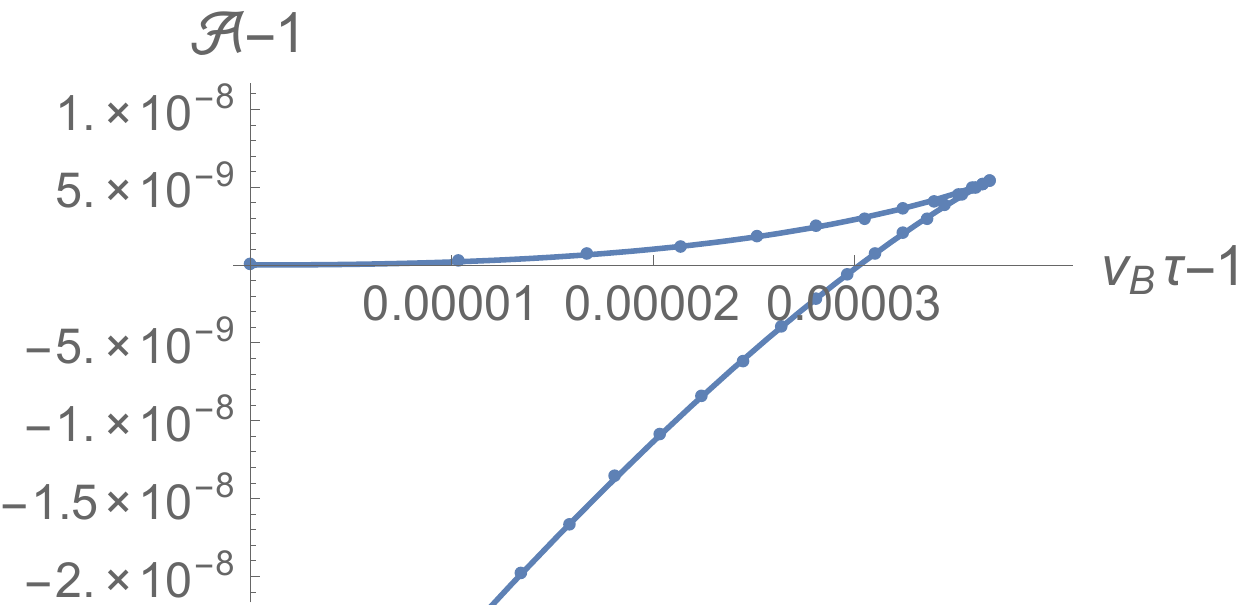}\vspace{0.3cm}\\
\includegraphics[scale=0.4]{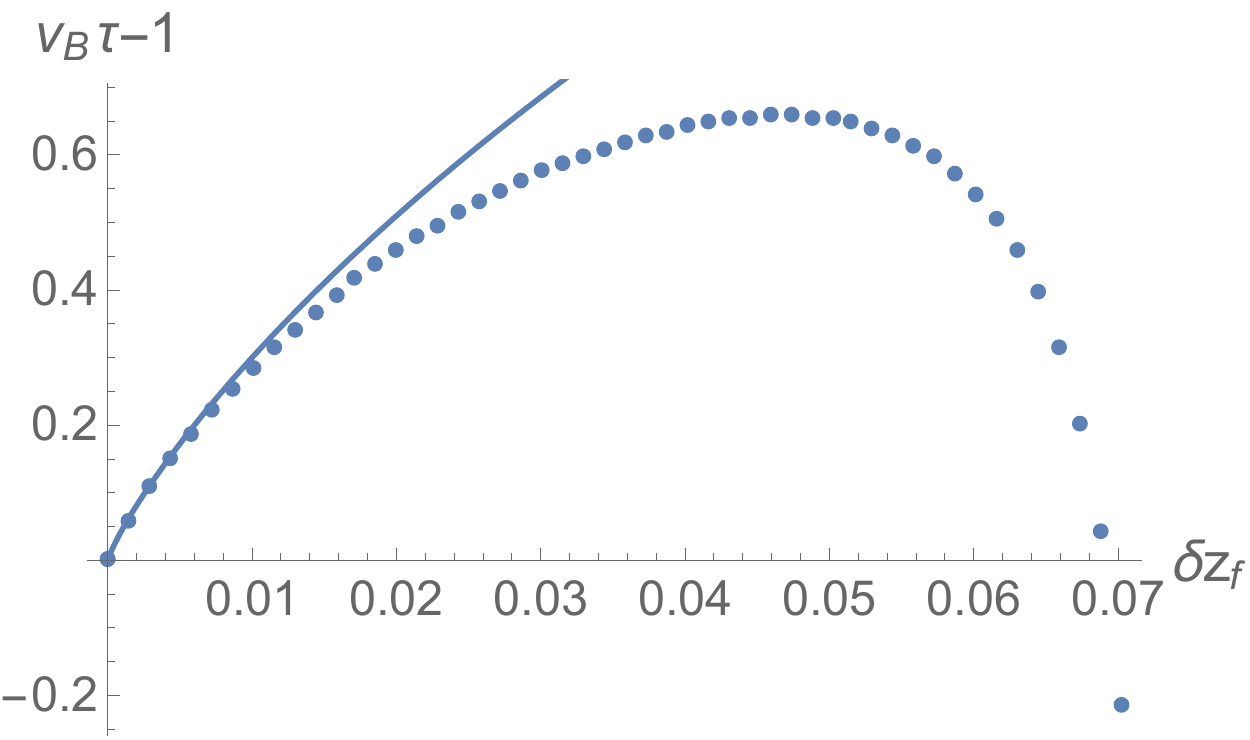}\hspace{0.3cm}
\includegraphics[scale=0.4]{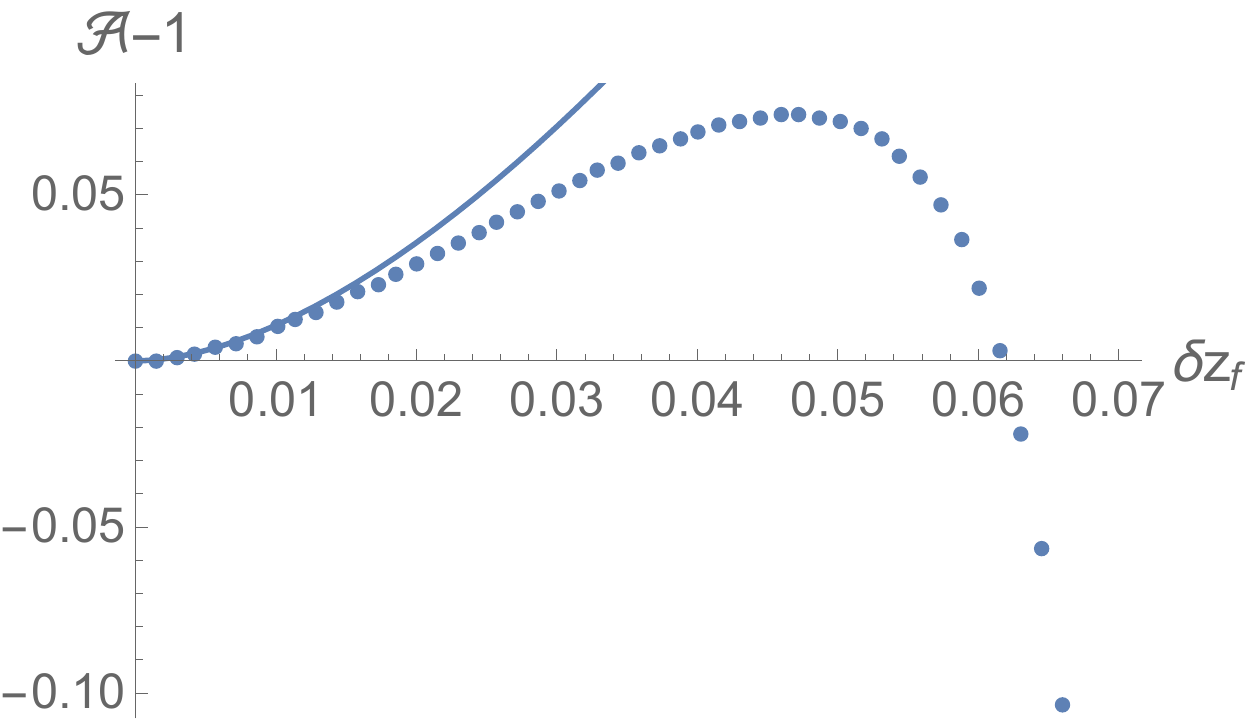}\hspace{0.3cm}
\includegraphics[scale=0.4]{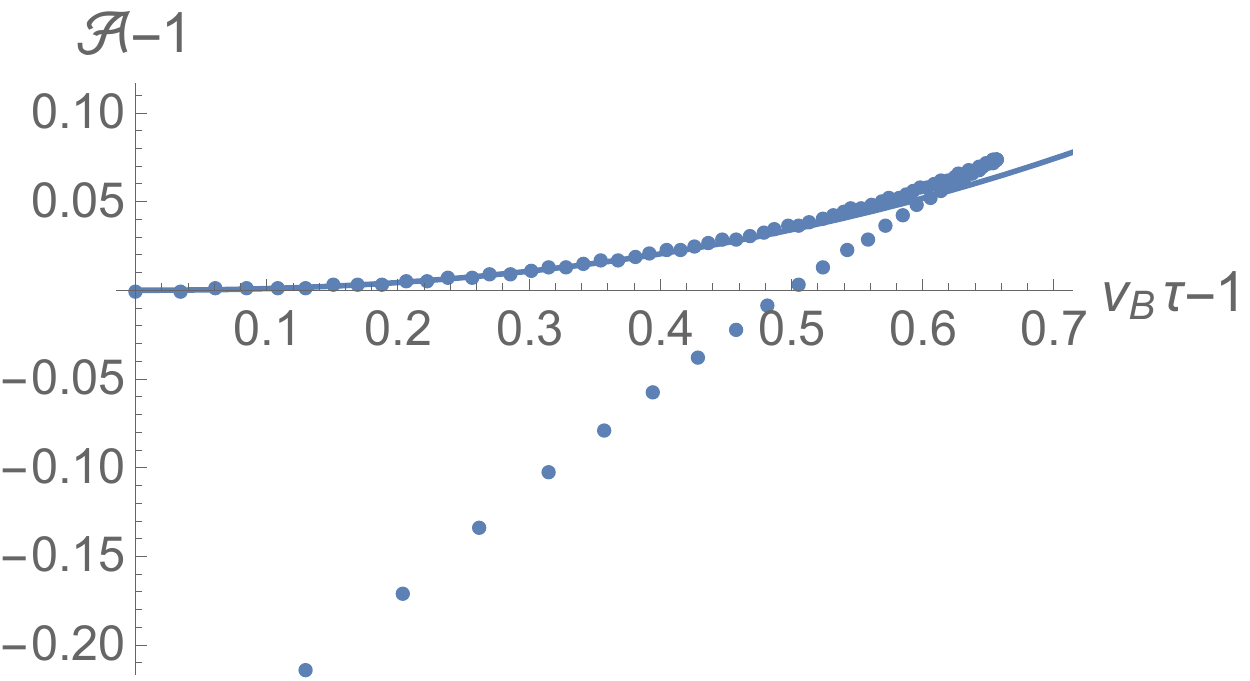}\vspace{0.3cm}\\
\includegraphics[scale=0.4]{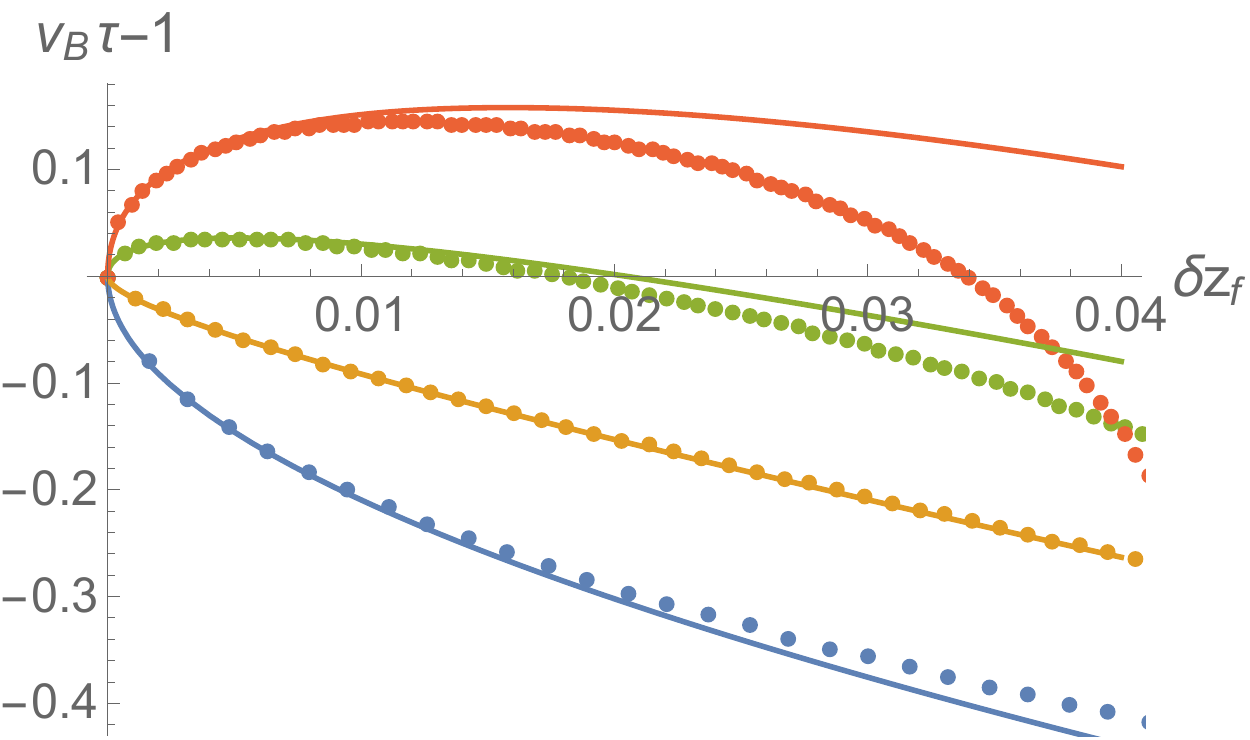}\hspace{0.3cm}
\includegraphics[scale=0.4]{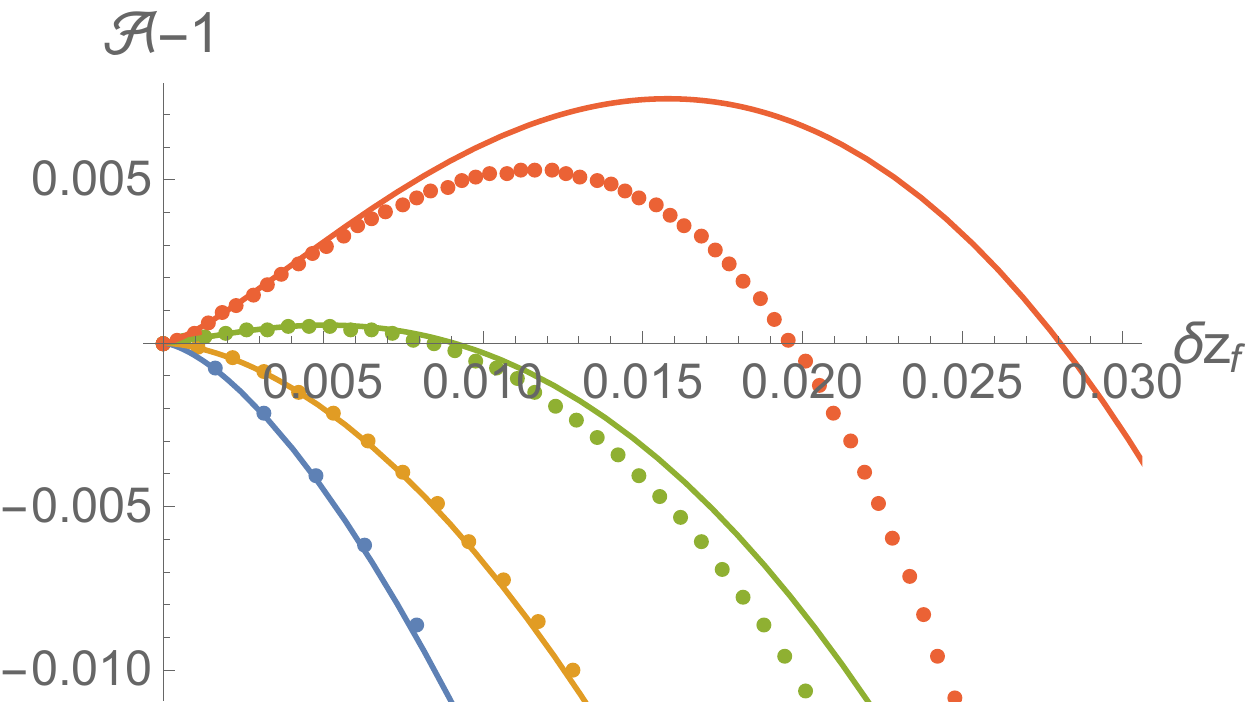}\hspace{0.3cm}
\includegraphics[scale=0.4]{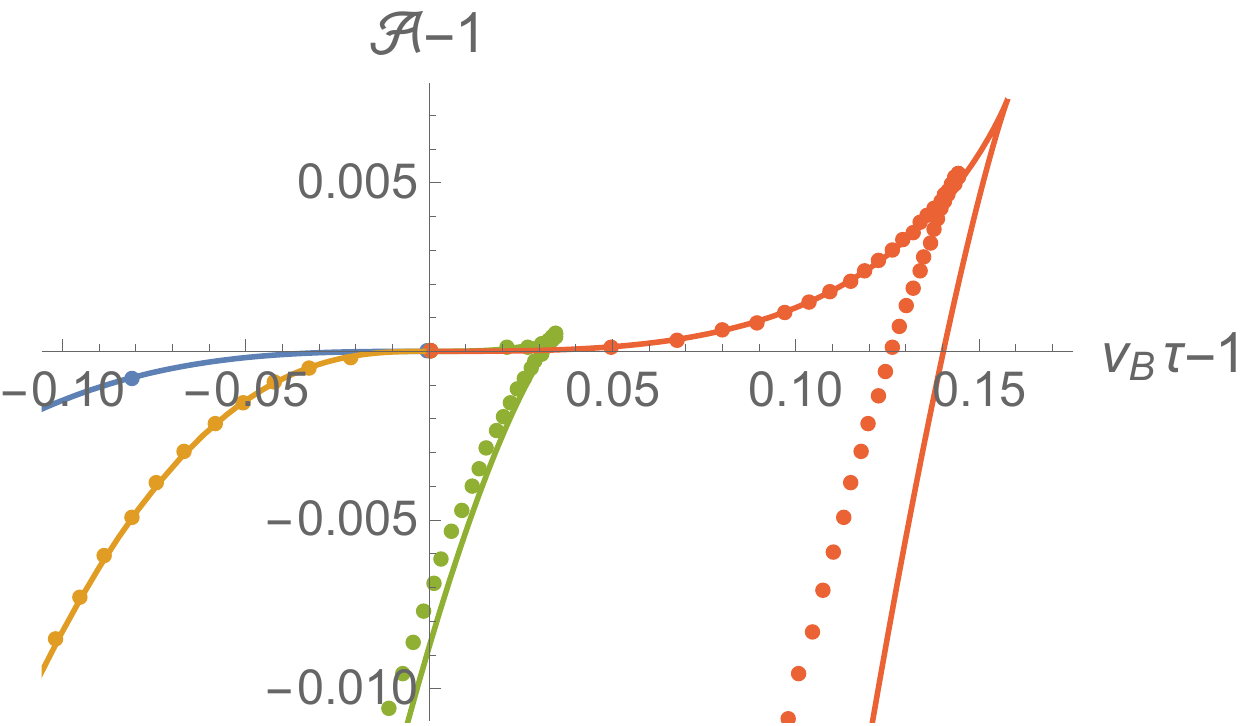}
\caption{Comparison of the $\de z_f$ expansion (solid curve)~\eqref{tauzfExp} and~\eqref{tauzfExp2} (combined with the series expansion of~\eqref{sArewrite})  with the numerical evaluation of~\eqref{OtherQuant3} (data points). For different black branes from left to write we plot $v_B\tau-1$ and $\sA-1$ as a function of $\de z_f$, and finally the parametric curve $(v_B\tau-1,\sA-1)$ showing the cuspy behavior seen on Fig.~\ref{fig:RN}, but magnified.
{\bf Top:} $d=3$ Schwarzschild black brane, which was the primary example discussed in this section. Note that the extremely small values on the axes is the reason why we were not able to see discontinuous saturation on Figs.~\ref{fig:S(t)} and~\ref{fig:RN}. The curve plotted on the first plot is $v_B\tau-1=-\frac12\de z_f\log \de z_f - 4.2598 \de z_f$. The zero of this curve is hence approximately at $\de z_f=e^{-8.52}$ explaining the small scale of the graph. The rest of the curves are easy to obtain from this formula. There is a perfect agreement between the expansion and the numerical evaluation of the integrals~\eqref{OtherQuant3}.
{\bf Middle:} $d=3$ RN black brane with $q=3/4$, for which $v_B\tau-1=-{13\ov 2}\de z_f\log \de z_f + 0.03822 \de z_f+\dots$. We see that the scale of the curves is a lot bigger than what we got for the Schwarzschild case, and the series expansion only matches the numerical data for small values of $\de z_f$.
{\bf Bottom:} $d=4$ RN black branes with $q=0,1/2,2/3,3/4$ plotted with blue, orange, green, and red respectively. The values of $c_4$ defined in~\eqref{tauzfExp2} are 
$c_4=-1.9955,-0.5164,1.0005,2.5126$ respectively, and the sign of $c_4$ determines whether saturation is continuous or discontinuous: for $q=0,1/2$ we get continuous saturation, while the other two cases lead to discontinuity. The solid curves include the subleading term from~\eqref{tauzfExp2} as well.
\label{fig:nearsat}}
\end{figure}
\end{center}

We put all this back together to obtain an expansion for $\tau(z_f)$. In $d=3$ we get:
\es{tauzfExp}{
\tau(z_f)&=\tau(1)\le(1-{3f'(1)-f''(1)\ov 2 f'(1)}\, \de z_f\,\log\le({\de z_f}\ov z_\text{HM}-1\ri)+c_3\, \de z_f+ O\le(\de z_f^2\, \log{\de z_f}\ri)\ri)\,,
}
where $c_3$ involves $a_5$ defined in~\eqref{IntegralsNewForm}, and hence requires numerical integration. The leading behavior is however determined by the coefficient of the $\de z_f\,\log{\de z_f}$ term: combining the Null Energy Condition~\eqref{NECeq} evaluated at $z=1$ with $f'(1)<0$ implies that
\es{c3cond}{
{3f'(1)-f''(1)\ov 2 f'(1)}>{1\ov2}\,.
}
Near saturation $\tau(z_f)$ grows with $\de z_f$, thus saturation is discontinuous for any black brane. In the caption of Fig.~\ref{fig:nearsat} we give $c_3$ for the black branes analyzed there, and find perfect agreement with the numerical evaluation of~\eqref{OtherQuant3}.

In higher dimensions we get:
\es{tauzfExp2}{
\tau(z_f)&=\tau(1)\le(1-{(2d-3)f'(1)-f''(1)\ov 2(d-3) f'(1)}\, \de z_f+O\le(\de z_f^2\ri)\ri)+
c_d\,  \de z_f^{1/(d-2)}\le[1+O(\de z_f)\ri] \qquad (d>3)\,,
}
where $c_d$ has a complicated expression that involves $a_5$ defined in~\eqref{IntegralsNewForm}, and hence requires numerical integration. Because the nonanalytic term with coefficient $c_d$ gives the leading correction, we are unable to determine analytically whether the saturation is continuous or not. If it is continuous, we near saturation we get a power law behavior
\es{PowerLaw}{
1-\sA(\tau)=\# \le({1\ov v_B}-\tau\ri)^{d-1}+\dots\,.
}
This behavior is valid for ${1\ov v_B}-\tau\ll1$. This is the same power law that the upper bound~\eqref{TsunamiVolume} or the combined bounds discussed in~\cite{Mezei:2016wfz} give, but the overall coefficients do not agree. In~\cite{Liu:2013iza,Liu:2013qca} instead of this regime, they zoomed in on times ${1\ov v_B}-\tau\ll \beta/R$, and found a different power law in the Vaidya setup. For such time differences the details of the quench matter, as can also be seen from their analysis. Here instead we are focusing on universal features.

\subsection{Early time growth and some worked out examples}

The early time growth of entropy is significantly easier to work out that the late time behavior: we have to expand $r(z)$ and the integrand $\sI_\tau(z)$ in~\eqref{OtherQuant3} for $z=z_\text{HM}-\de z$, integrate, and combine to get an expression for $\tau(z_f)$. We plug into~\eqref{sArewrite}, obtain $\sA(z_f)$ as a power series, finally invert the relation  $\tau(z_f)$ perturbatively, which gives
\es{sAtau}{
\sA(\tau)=(d-1)v_E\,\tau\le(1-a\, \tau^2+O(\tau^4)\ri)\,, \qquad a\equiv -{(d-2)^2\ov 6}\, f(z_\text{HM})>0\,.
}
It was suggested in~\cite{Nahum:2016muy} based on tensor network intuition that the early time expansion would take this form. This expansion could provide clues to how to improve the field theory bounds discussed in~\cite{Mezei:2016wfz}, which give exact linear growth at early times.

\begin{center}
\begin{figure}[!h]
\includegraphics[scale=0.6]{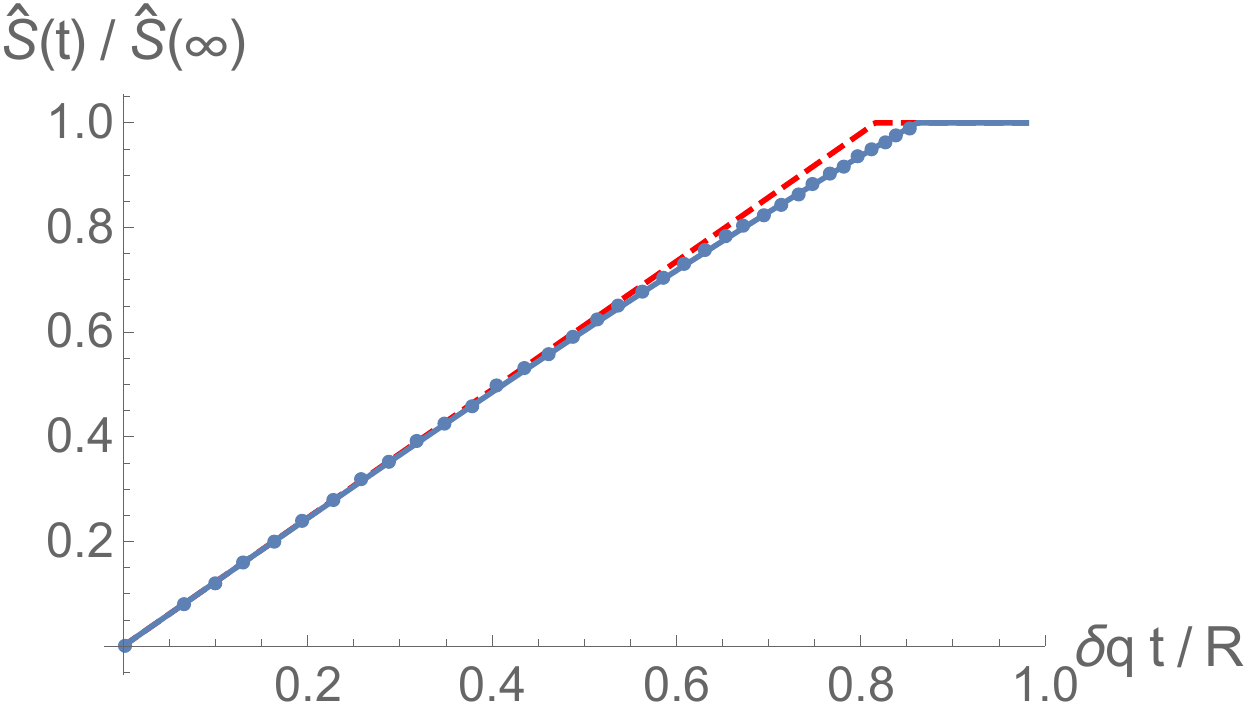}\hspace{0.3cm}
\includegraphics[scale=0.6]{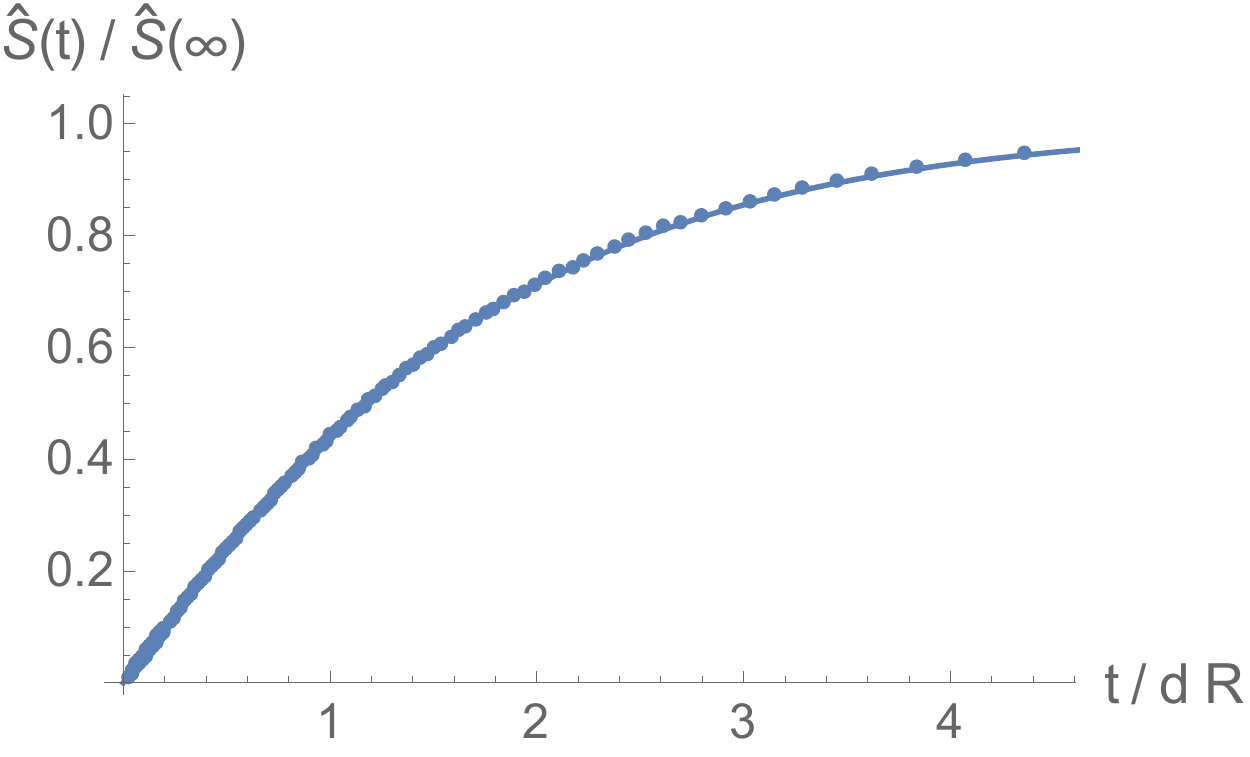}
\caption{{\bf Left:} Entropy growth for $d=3$ near extremal RN black holes. The data points are for $\de q={1\ov 10}$, while the solid curve is~\eqref{deqCurve}. The red dashed line has slope $2v_E$, where $v_E=\sqrt{3\ov 8}\, \de q$ as given in~\eqref{NERN}. Note that time has been rescaled by $\de q$. The curve is extremely straight and saturation happens parametrically later than the lower bound given in~\eqref{tSBound}, as $v_B=\sqrt{3\ov4}\, \sqrt{q}\gg v_E$. {\bf Right:} Entropy growth for a large $d$ Schwarzschild black hole. The data points are for $d=20$ and the solid curve is~\eqref{deqCurveLargeD}.  Note that time has been rescaled by $d$, hence the curve has finite initial slope, but saturation takes infinitely long (in rescaled time). In unscaled time $t_S=\sqrt{2}\,R+O(1/d)$.
\label{fig:deqRN}}
\end{figure}
\end{center}

Next we investigate two limits that give some simplification. In Fig.~\ref{fig:RN} we observe that as $q\to 1$ the $\sA(\tau)$ becomes more and more linear. Motivated by this observation, we investigate this limit in some detail in an expansion in $\de q\equiv1-q$. We obtain the simplest formulas in $d=3$, and hence we will restrict to this case. Because in the extremal limit $z_\text{HM}\to 1$, it is easier to work in terms of a new bulk radial coordinate, $z\equiv 1+(z_\text{HM}-1)(1-\xi)$. In the limit $\de q\to 0$  we find that the integrand  $\sI_\tau(\xi)$ is dominated by the small $\xi$ region: 
\es{ItauScaling}{
\xi&\equiv \de q\, \zeta\\
\tau(\zeta_f)&={2\ov \sqrt{3}\, \de q}\int_0^{\zeta_f} d\zeta \ {1\ov \sqrt{\zeta(1+2\zeta)}}\,.
}
Using~\eqref{sArewrite} we finally arrive at the parametric form of the entropy curve:
\es{deqCurve}{
(\de q \, \tau(\zeta_f),\, \sA(\zeta_f))=\le(\sqrt{8\ov 3}\, {{\rm arcsinh}(\sqrt{2\zeta})\ov \sqrt{1+2\zeta}},\,  \sqrt{2\zeta\ov 1+2\zeta}+{{\rm arcsinh}(\sqrt{2\zeta})\ov 1+2\zeta}\ri)\,.
} 
This curve looks extremely straight, as seen in Fig.~\ref{fig:deqRN}, and saturates at the value $\zeta_f=0.2337$.

It has been recently observed that general relativity simplifies in the large $d$ limit~\cite{Emparan:2015rva,Bhattacharyya:2015dva}. Motivated by this observation, we investigate the large $d$ limit of the entropy curve for the Schwarzschild black brane. In this limit, the black brane becomes a membrane and $z_\text{HM}\to 1$, as in the $q\to 1$ limit discussed above. Hence, it is convenient to change the bulk radial coordinate to $z\equiv 1+(z_\text{HM}-1)(1-\xi)$, and expand in $1/d$. A straightforward computation gives:
\es{deqCurveLargeD}{
(d \, \tau(\xi_f),\, \sA(\xi_f))&=\le(\sqrt{8}\,{\rm arccosh}\le({1\ov \sqrt{{2\ov u^2}-1}}\ri)-3\, {\rm arccosh}( u) ,\, \frac12\, u^2\sqrt{u^2-1} \ri)\,, \qquad u\equiv 2^{\xi_f/2}\,.
}

\section{Two-sided mutual information }\label{sec:MutInf}

As explained in the Introduction, we would like to determine the two-sided mutual information to probe the tsunami picture of~\cite{Liu:2013iza,Liu:2013qca}. We are looking for two-sided surfaces with $R_L,\, t_L=0$ and $R_R,\, t_R$. We will calculate the mutual information between these two regions. As in the one-sided setup, the area and time of the extremal surface is dominated by the (doubled) scaling surface. Here, because we do not have the constraint of $\Z_2$ symmetry, we are allowed to boost this (doubled) scaling surface, which amounts to a shift in its $v$ coordinate. 
 In this section we will only discuss in detail the case, when the black brane under consideration gives continuous saturation in the one-sided setup. We will comment briefly on the discontinuous case, which is more complicated to analyze.

 \begin{center}
\begin{figure}[!h]
\subfloat[][The point where the scaling surface touches $z_\text{HM}$ is to the left of the connection points. \label{figa}]{\includegraphics[scale=0.44]{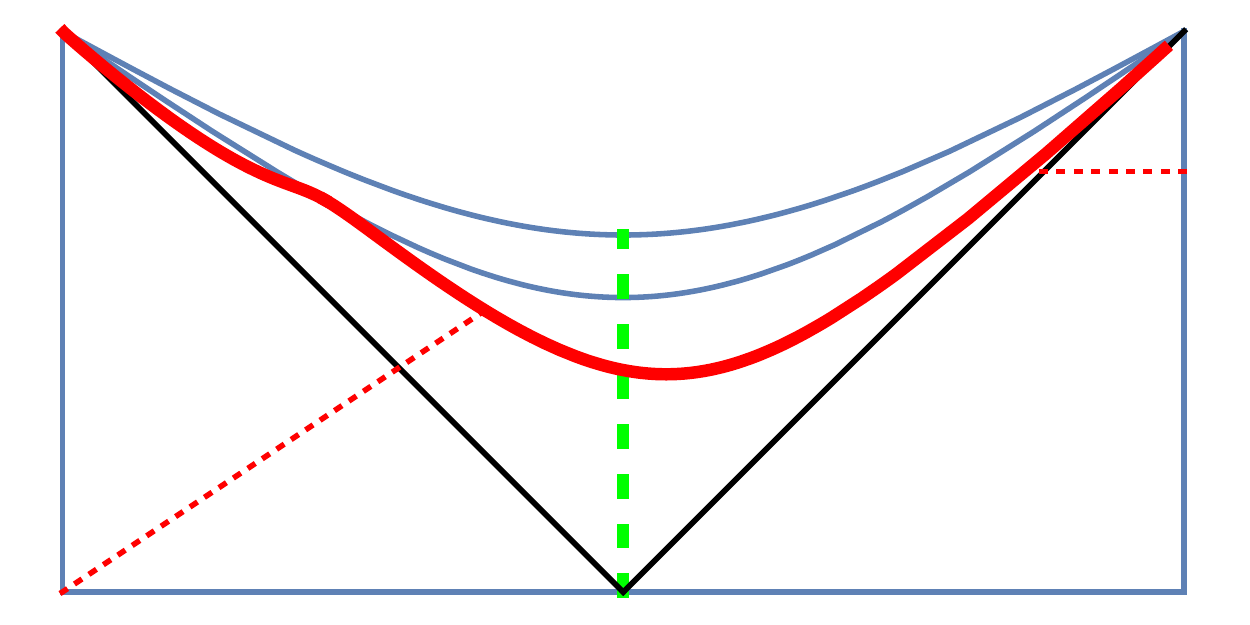}} 
\subfloat[][The touching point is between the connection points.\label{figb}]{\includegraphics[scale=0.44]{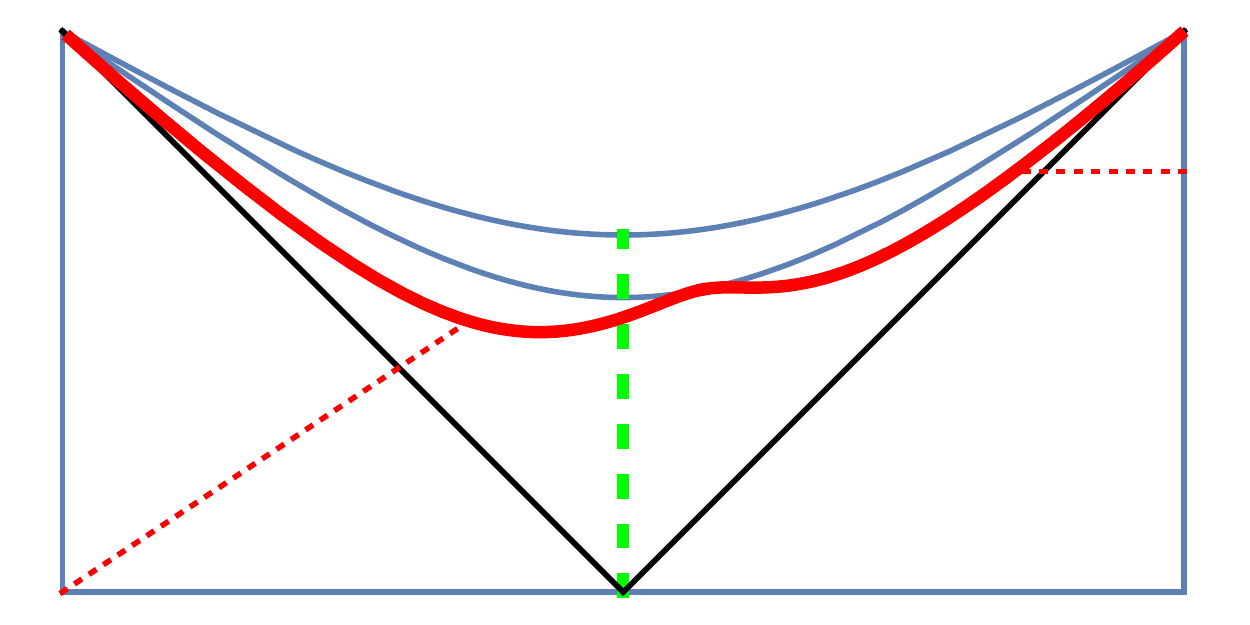}}
\subfloat[][The touching point is to the right of the connection points. \label{figc}]{\includegraphics[scale=0.44]{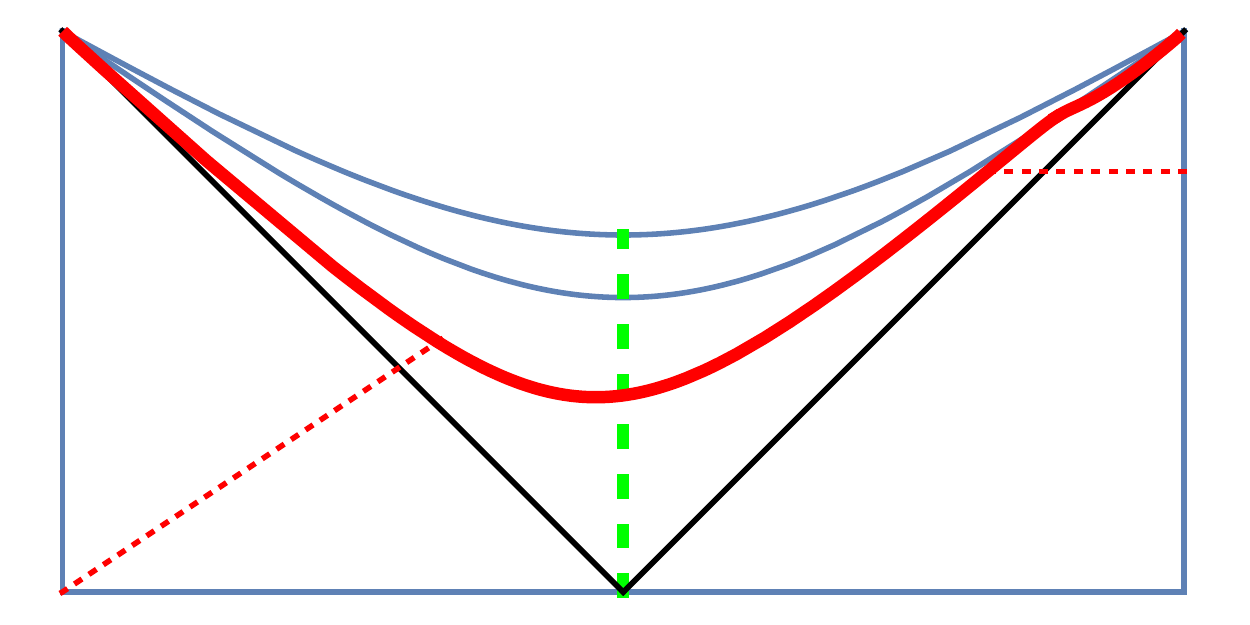}}
\caption{ Doubled scaling surfaces in the $d=3$ Schwarzschild black brane drawn with red on the Penrose diagram. The first blue line behind the horizon is $z=z_\text{HM}$. (The second is the singularity.)  The scaling surface touches $z_\text{HM}$ at one point, where $v=v_\text{shift}$. The connection of the scaling surface to the boundary does not contribute to the extensive part of the entropy, and is sketched with red dotted lines. The scaling surface area between the connection points is what computes the two-sided entropy. For this plot we chose $\rho_c=1$ (for visualization purposes), but for the scaling surface to be a good approximation to the HRT surface, we have to choose a large $\rho_c$.
\label{fig:2side}}
\end{figure}
\end{center}
 
 The scaling surface has a size on $z_\text{HM}$ that we will keep denoting by $\rho_c$, and we will use the rescaled field theory radial coordinate $r$ defined below~\eqref{FinalEOMScaling}. The left and right boundaries are connected to the scaling surface at some $r_{*,L}={R_L/ \rho_c}$ and $r_{*,R}={R_R/\rho_c}$ respectively. Depending on where these connections are compared to where the scaling surface touches $z_\text{HM}$, we have three distinct possibilities. We explain these possibilities below, and illustrate them on Fig.~\ref{fig:2side}.

\makebox[1.5em][l]{\bf (a)} The midpoint is to the left of both shooting out points, see Fig.~\ref{figa}.
Using the quantities introduced in~\eqref{OtherQuant2}  the equations governing this situation are:
\es{MidpointL}{
0&={t_L}=-R_L\,\tau(r_{*,L})+v_\text{shift}\\
{t_R }&=R_R\,\tau(r_{*,R})-v_\text{shift}\\
{(d-1)A_\text{conn}\ov K}&=-R_L^{d-1}\sA(r_{*,L})+R_R^{d-1}\sA(r_{*,R})\,,
}
where $v_\text{shift}$ quantifies the boost we made. 
We found the easiest to think about the left side by mirroring the Penrose diagram. On the mirrored spacetime the shift in $\tilde v_\text{vshift}=2v_\text{brane}(z_\text{HM})- v_\text{vshift}\approx-v_\text{vshift}$, as we know from  the scaling relations~\eqref{ScalingBehavior} that $v_\text{vshift}=O(\rho_c)$.
This then explains where the sign in the first equation of~\eqref{MidpointL} comes from. 

We introduce the natural rescaled variables
\es{NaturalResc}{
{\cal R}&\equiv {R_R\ov R_L}\,, \qquad {\cal T}\equiv{t\ov R_L}\,, \qquad \sI\equiv (d-1)\,{A_\text{disconn}-A_\text{conn}\ov K R_L^{d-1}}\,,
}
where $0\leq\sI \leq2$ is the two-sided mutual information in the focus of our interest.
Using the relation
\es{rsRelation}{
r_{*,R}={R_R\ov \rho_c}={\cal R}\, r_{*,L}\,,
}
\eqref{MidpointL} can be rewritten in terms of the rescaled variables as:
\es{MidpointL2}{
{\cal T}&={\cal R}\,\tau({\cal R}r_{*})-\tau(r_{*})\\
{\sI }&=1+\sA(r_{*})+{\cal R}^{d-1}\le[1-\sA(r_{*})\ri]\,,
}
where we suppressed the $L$ subscript on $r_*$. Henceforth, we will always use~\eqref{rsRelation} to eliminate $r_{*,R}$.

\begin{center}
\begin{figure}[!h]
\includegraphics[scale=0.6]{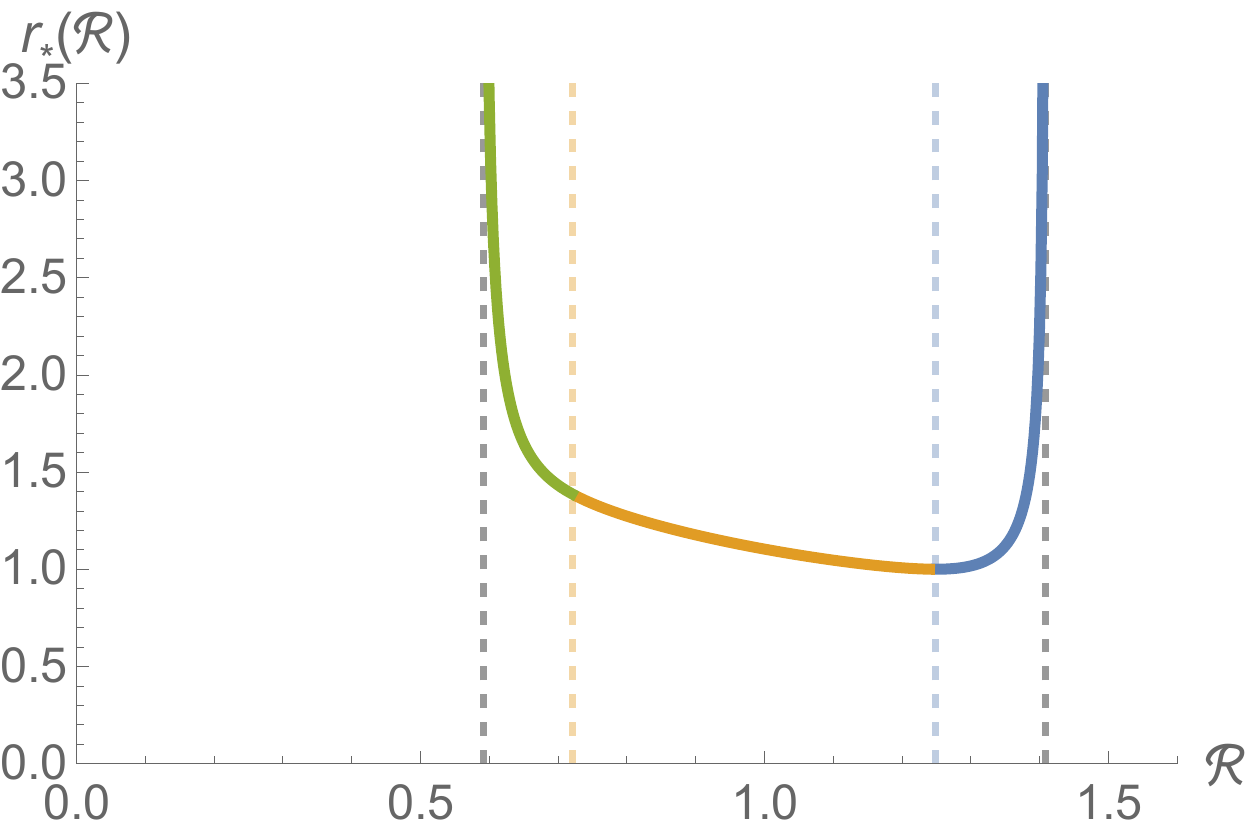}\hspace{0.3cm}
\includegraphics[scale=0.6]{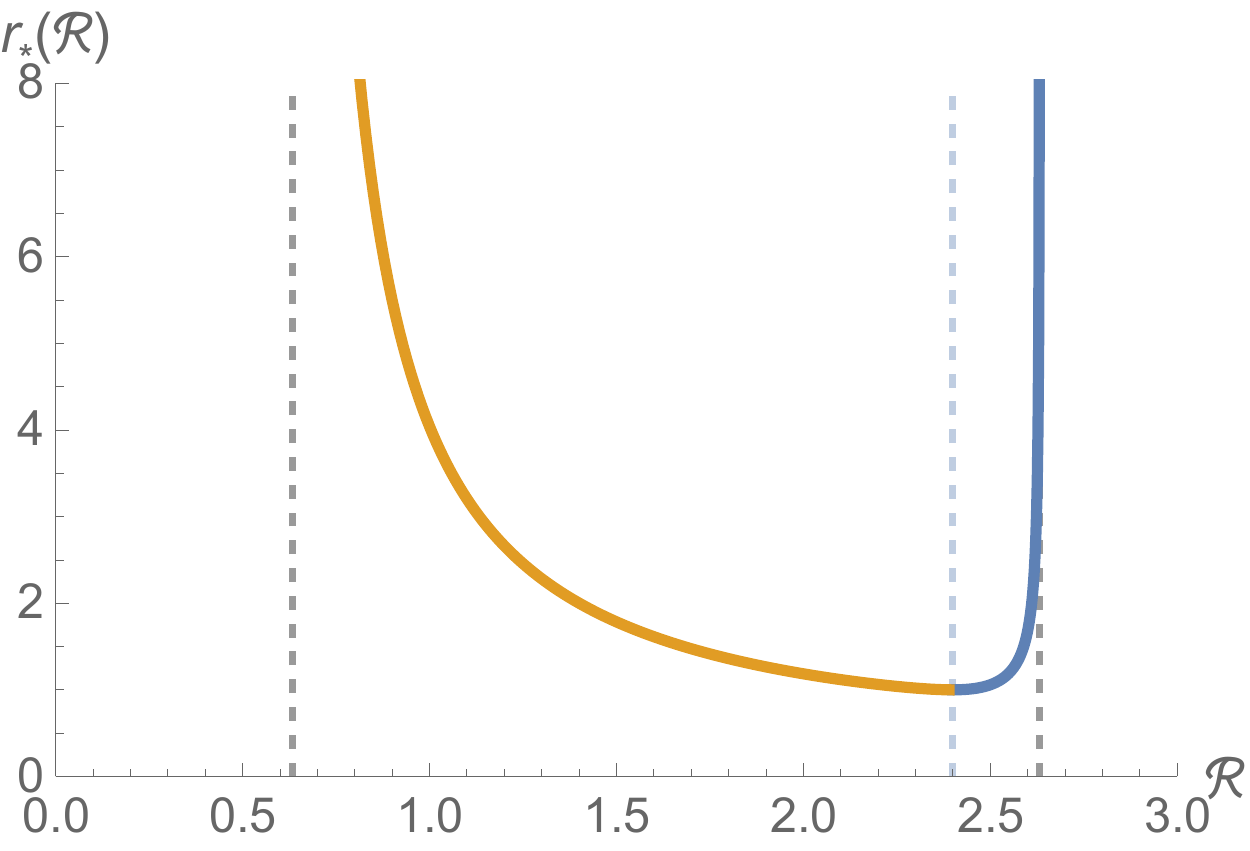}\vspace{0.3cm}
\includegraphics[scale=0.6]{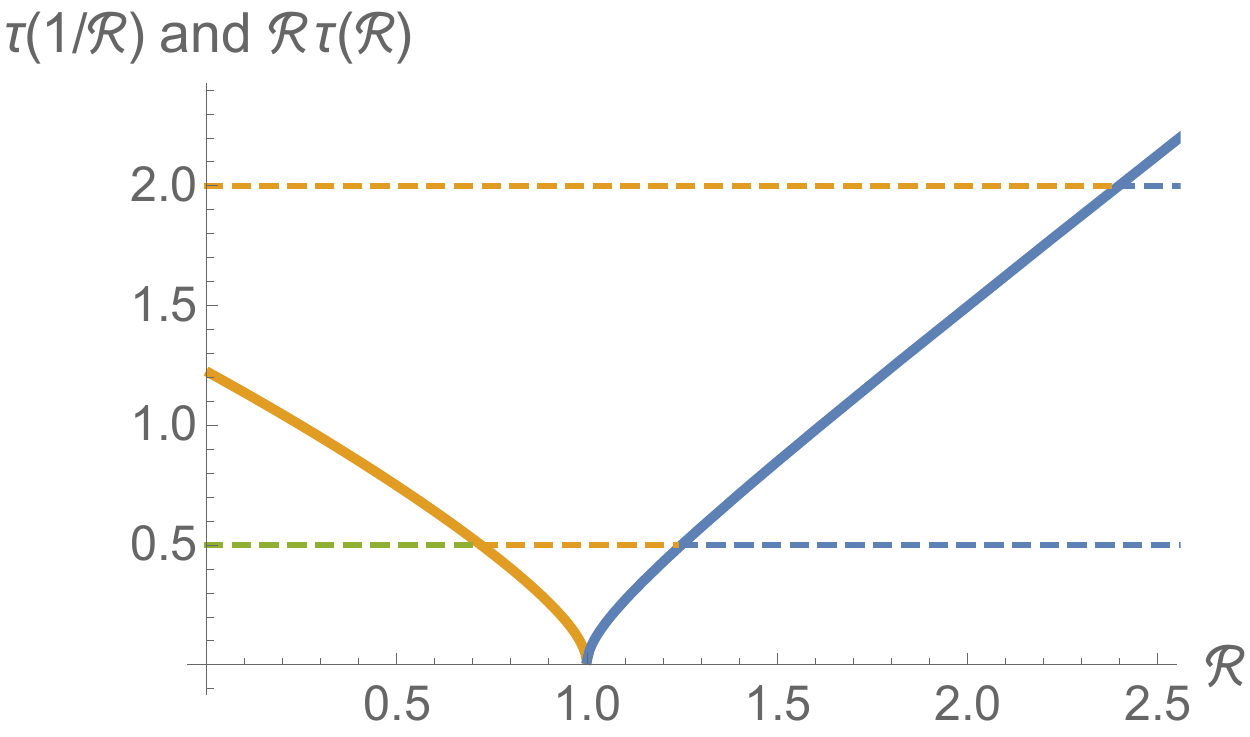}
\caption{On the first two figures we plot $r_*({\cal R})$ for ${\cal T}=1/2$ (left) and ${\cal T}=2$ (right)  for the $d=4$ Schwarzschild black brane. The cases (a),(b), and (c) are drawn by blue, orange, and green respectively. The vertical dashed lines indicate the boundaries between the different cases, and can be determined by drawing a horizontal line at ${\cal T}$ on the third graph.  The horizontal dashed lines at ${\cal T}=1/2,2$ are colored according to the color of the curves in the top row. In more detail, the left black dashed vertical line is at ${\cal R}=\abs{v_B{\cal T}-1}$, the orange one (if it exists) at ${\cal R}_{2,{\cal T}}$, the blue one at ${\cal R}_{1,{\cal T}}$, finally the right black one at $v_B{\cal T}+1$.
\label{fig:twoside}}
\end{figure}
\end{center}

In the following we want to fix ${\cal T}$ and determine the function $\sI ({\cal R})$. To do this we have to determine the relation $r_*({\cal R})$ for fixed ${\cal T}$. This is easily done numerically. 
There is one subtlety, the domain $\sD$ of the function $r_*({\cal R})$ is bounded. The function $\tau(r_{*})$ is only defined for $ r_*\in [1,\infty)$\footnote{We note that $r_*\geq1$ from its definition $r_*=R/\rho_c$. }
 and takes values in $[0,1/v_B)$. First, let us plug in $ r_*=1$ into the first equation of~\eqref{MidpointL2}, and define  ${\cal R}_{1,{\cal T}}$  through it:
\es{Rmin}{
{\cal T}={\cal R}_{1,{\cal T}}\,\tau({\cal R}_{1,{\cal T}})\,.
}
Second, taking $ r_*\to\infty$ in the first equation of~\eqref{MidpointL2} gives
\es{Rmin2}{
{\cal T}={{\cal R}-1\ov v_B}\,.
}
From~\eqref{Rmin},~\eqref{Rmin2} and the blue curves on Fig.~\ref{fig:twoside}, we conclude that  the domain of the function $r_*({\cal R})$
is 
\es{Da}{
\sD^{(a)}=\le({\cal R}_{1,{\cal T}},\, 1+v_B{\cal T} \ri)\,.
}
At the endpoints of this domain $r_*=1$ and $r_*=\infty$ respectively. What we said above remains true even in the case of discontinuous saturation.

\makebox[1.5em][l]{\bf (b)}  The next case we consider is when the midpoint of the scaling surface is between the left and right shooting out points,  see Fig.~\ref{figb}. Only some signs change compared to~\eqref{MidpointL2}:
\es{MidpointM}{
{\cal T}&={\cal R}\,\tau({\cal R}r_{*})+\tau(r_{*})\\
{\sI }&=1-\sA(r_{*})+{\cal R}^{d-1}\le[1-\sA(r_{*})\ri]\,.
}
Again, the domain of $r_*({\cal R})$ is a bit subtle. Let us define ${\cal R}_{2,{\cal T}}$ as the root of the equation
\es{Rmin3}{
{\cal T}=\tau\le(1\ov {\cal R}_{2,{\cal T}}\ri)\,.
}
The equation has a solution only  if $ v_B{\cal T}<1$. By a similar logic as above, and analyzing the orange curves on Fig.~\ref{fig:twoside}, we conclude that the  domain of the function $r_*({\cal R})$ is 
\es{Db}{
\sD^{(b)}=\begin{cases}\le({\cal R}_{2,{\cal T}},\, {\cal R}_{1,{\cal T}} \ri) \qquad &(v_B{\cal T}<1)\,,\\
  \le(v_B\, {\cal T}-1,\, {\cal R}_{1,{\cal T}} \ri)  \qquad &(v_B{\cal T}>1)\,.
  \end{cases}
}  
The last equation is not true in the case of discontinuous saturation, and it has important consequences.

\makebox[1.5em][l]{\bf (c)}  Finally, the midpoint of the scaling surface can be to the right of both shooting out point, see Fig.~\ref{figc}. This immediately implies that ${\cal R}<1$ the same way as the setup in (a) implied that ${\cal R}>1$. Again there are only some sign changes:
\es{MidpointR}{
{\cal T}&=-{\cal R}\,\tau({\cal R}r_{*})+\tau(r_{*})\\
{\sI }&=1-\sA(r_{*})+{\cal R}^{d-1}\le[1+\sA(r_{*})\ri]\,.
}
The domain of $r_*({\cal R})$ can be read off from the green curve on Fig.~\ref{fig:twoside}, and we get
\es{Dc}{
\sD^{(c)}=\begin{cases}
\le(1-v_B\, {\cal T},\, {\cal R}_{2,{\cal T}}\ri) \qquad &(v_B{\cal T}<1)\,,\\
\varnothing  \qquad &(v_B{\cal T}>1)\,.
  \end{cases}
}

There are two more cases to understand. When $r_*=\infty$ the important part of the surface lies flat on the horizon. Because the horizon is flat, we can glue to it a tube that connects to the boundary at any time. Formally, we can take the cases (a) and (c), and drop the equation relating ${\cal T}$ to $r_*$, and plug $r_*=\infty$ into the expression for the mutual information. From there we get
\es{BlackLine}{
&\text{Case (a):} \quad {\sI }=2 \qquad\qquad\hspace{0.1cm} \le({\cal R}>1\ri)\,,\\
&\text{Case (c):} \quad {\sI }=2{\cal R}^{d-1} \qquad \le({\cal R}<1\ri)\,.
}
Whenever none of the previous cases apply, we do not have a connected surface bridging the two sides. Hence the mutual information vanishes.

\begin{center}
\begin{figure}[!h]
\includegraphics[scale=1]{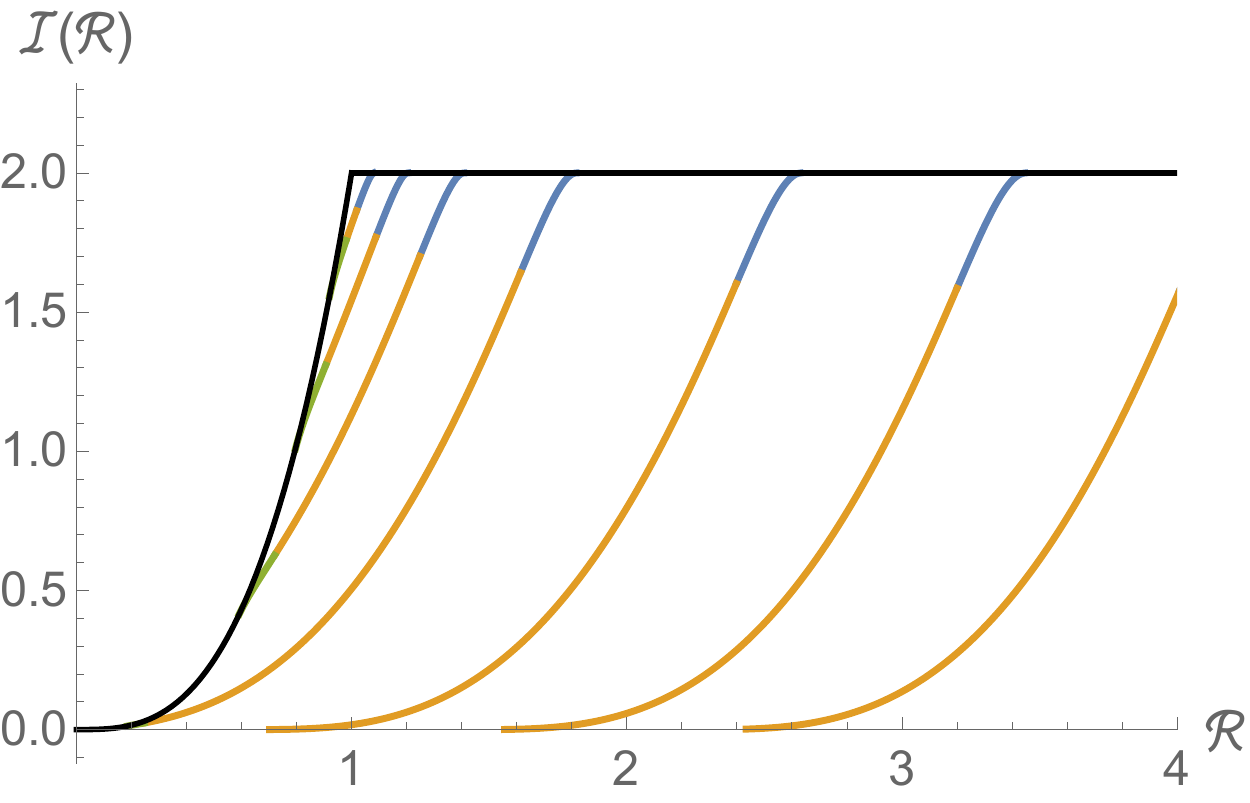}
\caption{Mutual information from the two-sided extremal scaling surface  for the $d=4$ Schwarzschild black brane. The black line shows the contribution of the surfaces determined by~\eqref{BlackLine}, and it gives the $t/R_L={\cal T}=0$ value of ${\cal I}({\cal R})$. The other times plotted are ${\cal T}=1/10,1/4,1/2,1,2,3,4$, different colors denote the cases (a),(b), and (c), as in Fig.~\ref{fig:twoside}. E.g. the ${\cal T}=1/10$ curve is obtained by first going along the black line, then cutting the corner with the colored curve, and then continuing with the black line, while the ${\cal T}=4$ curve starts as 0, and gets up to the (black) saturation value on an orange and blue curve.
\label{fig:twoside2}}
\end{figure}
\end{center}

\begin{center}
\begin{figure}[!h]
\includegraphics[scale=0.7]{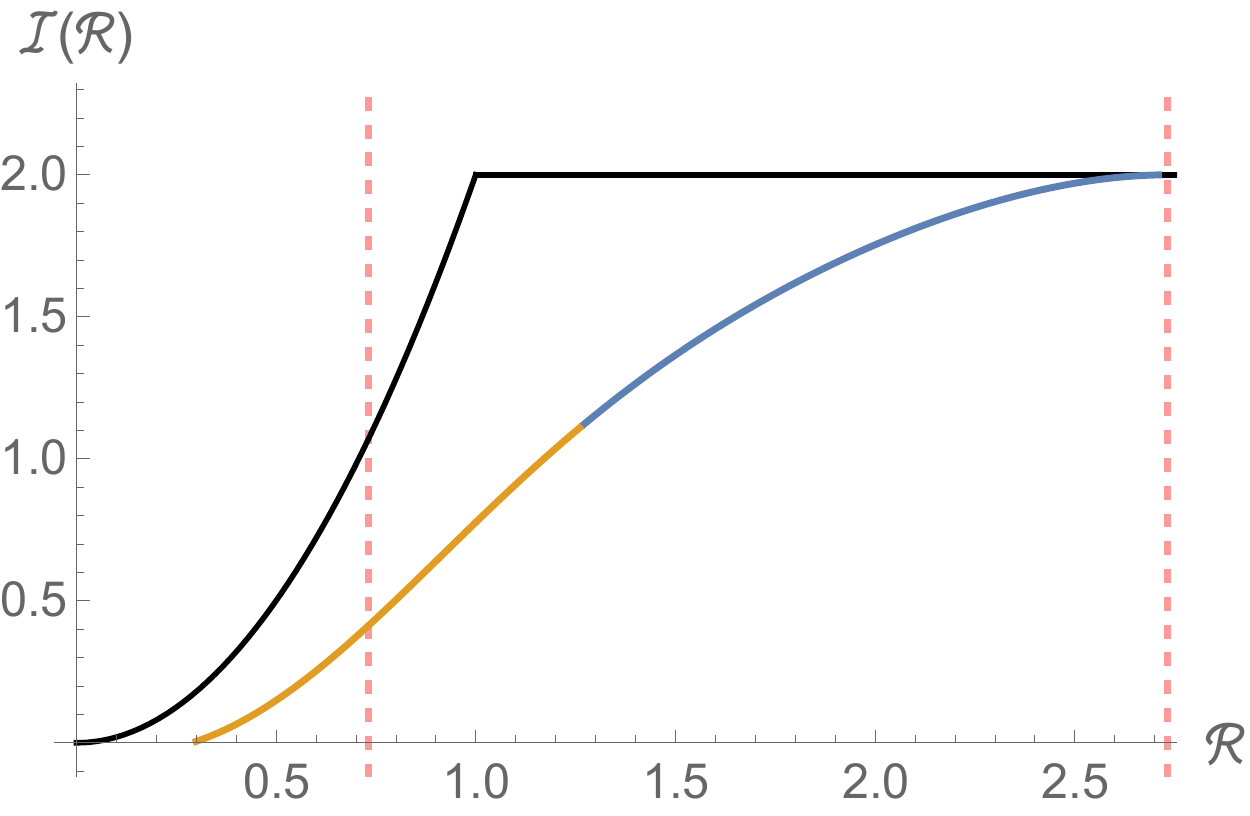}
\caption{Two-sided mutual information for the $d=3$ RN black brane with $q=3/4$, which does not give continuous saturation in the one-sided setup. On Fig.~\ref{fig:RN} it corresponds to the red curve with the largest cusp. Besides~\eqref{BlackLine}, we have plotted ${\cal I}$ for ${\cal T}=4$, and marked the points~\eqref{NotablePoints} with dashed red lines. While ${\cal I}$ saturates at the point $ {\cal R}^\text{(saturates)}=v_B{\cal T}+1$, it starts is nonzero earlier than what we get for black branes with continuous saturation. Also, while the ${\cal I}$ curve starts with zero slope on Fig.~\ref{fig:twoside2}, here the slope is nonzero.
\label{fig:discont}}
\end{figure}
\end{center}

Finally, we plot ${\sI }({\cal R})$ for the $d=4$ Schwarzschild black brane on Fig.~\ref{fig:twoside2}. Different parts of the curves come from the different cases above, and they are plotted with the corresponding colors. There are two notable points: the values of ${\cal R}$  where the mutual information starts to be nonzero, and where it saturates
\es{NotablePoints}{
{\cal R}^\text{(first nonzero)}=\max(0,v_B{\cal T}-1)\,, \qquad {\cal R}^\text{(saturates)}=v_B{\cal T}+1\,.
}
Multiplying the second equation by $R_L$ and recalling the definitions of ${\cal R}, \, {\cal T}$ given in~\eqref{NaturalResc}, we get
\es{NotablePoints2}{
 R_R^\text{(saturates)}=R_L+v_B\, t\,.
 }
This result was quoted in the Introduction~\eqref{TsunamiWave}.
In the discontinuous case the mutual information becomes nonzero at an earlier time, but saturates, where~\eqref{NotablePoints2} says so, see Fig.~\ref{fig:discont}.\footnote{Even thought we have not explained the computation for this case in detail above, this is an important point deserving a figure.}  This is quite interesting, because in the  discontinuous case we do not know how to extract the value of $v_B$ from the entropy growth in a quench.\footnote{Recall that in the continuous case $v_B$ is encoded in the saturation time, $v_B=R/t_S$} The two-sided mutual information can reveal the value of $v_B$. Similar comments apply to spin chains~\cite{Mezei:2016wfz}.

We now comment on how these results refine the  ``entanglement tsunami" picture proposed in~\cite{Liu:2013iza,Liu:2013qca}, see also~\cite{Mezei:2016wfz} for further comments. It was suggested that at early times only degrees of freedom behind the tsunami wavefront spreading with the entanglement velocity $v_E$ would become entangled. This picture would imply that for fixed $R_L,\, t$ the radius $R_R$ at which the mutual information  saturates is $R_L+v_E\, t$, which is smaller than the result~\eqref{NotablePoints2}.\footnote{Recall that we proved in Sec.~\ref{sec:Bounds} that $v_E\leq v_B$ in holographic theories satisfying the NEC. This inequality is valid in any unitary quantum system~\cite{Mezei:2016wfz}.} However, we cannot simply modify the propagation speed of a sharp tsunami to $v_B$, as it would imply an early time growth of entropy exceeding the result~\eqref{LinGrowth} established in~\cite{Hartman:2013qma,Liu:2013iza,Liu:2013qca}. We suggest that there does not exist a sharp tsunami wave: the wave front has to propagate with $v_B$ to match~\eqref{NotablePoints2}, but degrees of freedom do not get fully entangled behind it as can be deduced from~\eqref{LinGrowth}. Thus $v_E$ is an average speed; it is obtained by weighting the newly entangled degrees of freedom by their degree of entanglement.    It would be interesting to provide an interpretation of $R_R^\text{(first nonzero)}$ given in~\eqref{NotablePoints}, which is only valid in the discontinuous case.

\begin{center}
\begin{figure}[!h]
\includegraphics[scale=0.7]{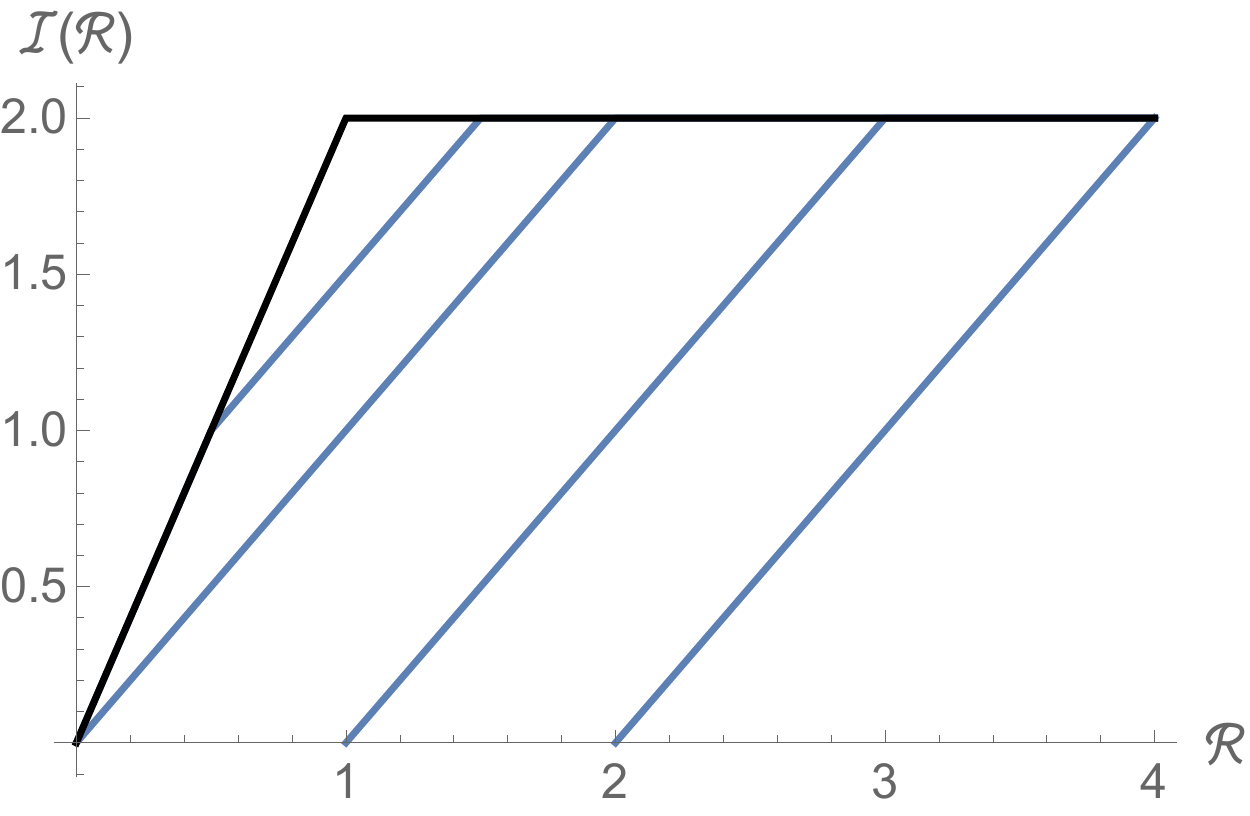}
\caption{Mutual information from the two-sided extremal surfaces in $d=2$. The black line is the result for time ${\cal T}=0$, while the blue lines correspond to ${\cal T}=0,1/2,1,2,3$. As explained in the caption of Fig.~\ref{fig:twoside2}, the blue curves may start and end as black.
\label{fig:twoside3}}
\end{figure}
\end{center}

It is interesting to compare the result on Fig.~\ref{fig:twoside2} to what we get in $d=2$, see Fig.~\ref{fig:twoside3}. Recall that in $d=2$ there is no $z_\text{HM}$ and the analysis is different (and well-understood) from the higher dimensional case. We briefly describe the computation.

We consider AdS$_3$ as a two-sided Rindler space (BTZ black brane): 
\es{BTZ}{
ds^2 = {1\ov z^2}\le[-f(z) dt^2 + dx^2+\frac{dz^2}{f(z)}\ri]\,,\qquad f(z) = 1-z^2\,.
}
We will use embedding coordinates
\es{Embedding}{
T_1 = \sqrt{1-z^2\ov z^2}\sinh t\,, \qquad T_2 ={1\ov z}\cosh x\,, \qquad X_1 = \sqrt{1-z^2\ov z^2}\cosh t\,, \qquad X_2 = r \sinh x\,.
}
These give the coordinates for points on the $R$ side. To get the points on the $L$ side, we add $i\pi$ to the time coordinate. This switches the sign of $T_1$ and $X_1$. The length of a geodesic connecting two points is given by
\es{GeodLength}{
\cosh d = T_1T_1' + T_2 T_2' - X_1 X_1' - X_2 X_2'\,.
}
We are interested in the mutual information between intervals on the $L$ and $R$ side. We want to connect the endpoints of intervals separated in time and centered at the origin on the two sides, thus we are looking for a geodesic connecting the point $(t_L=0,R_L)$ on the $L$ side, with $(t_R,R_R)$ on the $R$ side. Note that the length of the intervals is $2R_{L,R}$. In the computation below we assume that $R_R>R_L$, the $R_L<R_R$ is easily obtained from the result.

Subtracting a universal $\log(z_\text{cutoff})$ term from the length, we get
\es{dLR}{
\hat{d}_{LR} = \log\left[2\cosh(R_R - R_L) + 2 \cosh t\right]\approx \text{max}(R_R-R_L,t).
}
To compute the mutual information, we also need to know the length connecting the ends of the intervals to themselves. This is e.g.  $d_{LL} = \log\left[2\cosh (2R_L) - 2\right]$. The mutual information is then
\es{Mut2d}{
I_{LR}  =& \frac{1}{4 G_N} \Big\{\log\left[2\cosh (2R_L) - 2\right] + \log\left[2\cosh (2R_R) - 2\right] - 2\log\left[2\cosh(R_R - R_L) + 2 \cosh t\right]\Big\}\\
\approx& \frac{1}{4G_N}\left[2R_L + 2 R_R - 2\cdot\text{max}(R_R-R_L,t)\right].
}
This formula is valid as long as the answer is positive. If it is negative, a disconnected surface dominates and the correct answer is zero. The rescaled mutual information~\eqref{NaturalResc} is plotted in Fig.~\ref{fig:twoside3}.

\section{Arbitrary shapes}\label{sec:Shapes}

\subsection{Scaling limit for arbitrary shapes}

In this section we will restrict to $d=3$ in order to alleviate the notation; all our results generalize straightforwardly to higher dimensions. It will be advantageous to consider the infalling time $v$ and the angular coordinate $\theta$ as independent variables. The extremal surface is then given by two functions $\rho(v,\theta),\, z(v,\theta)$, and we can write down a complicated action in terms of these variables. Instead of doing this, using our experience with spherical surfaces, we do a rescaling:
\es{Scaling}{
v&\equiv R\,V\\
 \rho(v,\theta) & \to R \, r(V,\theta) \\
 z(v,\theta) & \to Z(V,\theta)\,,
}
where $R$ is an arbitrary scale associated to the entangling surface $\Sig$.
We will consider the end of the world brane setup only, but the arguments at the end of Sec.~\ref{sec:Universality} imply that we get the same extensive part of the entropy for any quench for arbitrary shapes. The scaling solution is determined by an image on the brane $r_\text{im}(\theta)$ and the boundary shape $r_\text{bdy}(\theta)$. The boundary conditions on $ r(V,\theta) $ hence are
\es{ScalingBC}{
 r(V_i=0,\theta) &=r_\text{im}(\theta)\\
 r\le(V_f={t\ov R},\theta\ri) &=r_\text{bdy}(\theta)\,.
}
 Recall that the actual HRT surface consists of the scaling solution connected to the AdS boundary with a straight tube that does not change the (leading order) value of the boundary time and does not contribute to the extensive part of the entropy. One consequence of this is that $Z(V_f,\theta)$ does not have to satisfy any boundary condition, it can be always glued to a straight tube. Henceforth we concentrate on the scaling solution.
 
 It may be instructive to compare to our treatment to the discussion in Sec.~\ref{sec:Universality} of the spherical case. The approach used in this section could also be used to derive the results for the spherical case  in a streamlined fashion. In Sec.~\ref{sec:Universality} we scaled with  $\rho_c$ instead of $R$, which meant that we fixed $r_\text{im}(\theta)=1$, obtained one scaling solution that we connected to the boundary at some $r_*$, finally we had to scale the answer with the ratio $r_*=R/\rho_c$. Here we find it more convenient to scale with $R$, which has the consequence of  $r_\text{bdy}(\theta)$ staying fixed, but the image on the brane $r_\text{im}(\theta)$ changing as a function of time. 

In terms of the scaled variables~\eqref{Scaling} the action is:
\es{AreaFunctScaled}{
A&=R^2 \int dVd\theta\ {r\ov Z^2}\sqrt{Q}+O(\rho_c)\,, \qquad Q\equiv (\p_V r)^2- f(Z)\le(1+{(\p_\theta r)^2\ov r^2}\ri)\,.
} 
Note that the large $R$ limit brings major simplification: no derivatives of $Z$ appear in the action. This implies that the equation of motion for $Z$ is algebraic. We write it in a suggestive form:
\es{ZEOM}{
\p_V r&=\sqrt{f(Z)-{Z f'(Z)\ov 4}}\, \sqrt{1+{(\p_\theta r)^2\ov r^2}}\,.
}
This equation is describing a ``tsunami wave" in polar coordinates, where $r(V,\theta)$ describes the tsunami wavefront at time $V$, and $v_\text{ts}(V,\theta)\equiv\sqrt{f(Z)-{Z f'(Z)\ov 4}}$ is a time and space dependent local propagation velocity of the wave front. If $v_\text{ts}$ was a constant this equation would describe a real tsunami: the wave front would move in the normal direction with speed $v_\text{ts}$. An important property of this equation is that it smoothes out the wavefront.  For a fixed boundary shape $r_\text{bdy}(\theta)$ we want to find the corresponding HRT surface, hence we want to evolve~\eqref{ZEOM}  backward in time. This leads to singularities (at least for constant $v_\text{ts}$), as illustrated on Fig.~\ref{fig:ellipse}.  We will make the tsunami analogy sharper in Sec.~\ref{sec:simpcases} below.

\begin{center}
\begin{figure}[!h]
\includegraphics[scale=0.7]{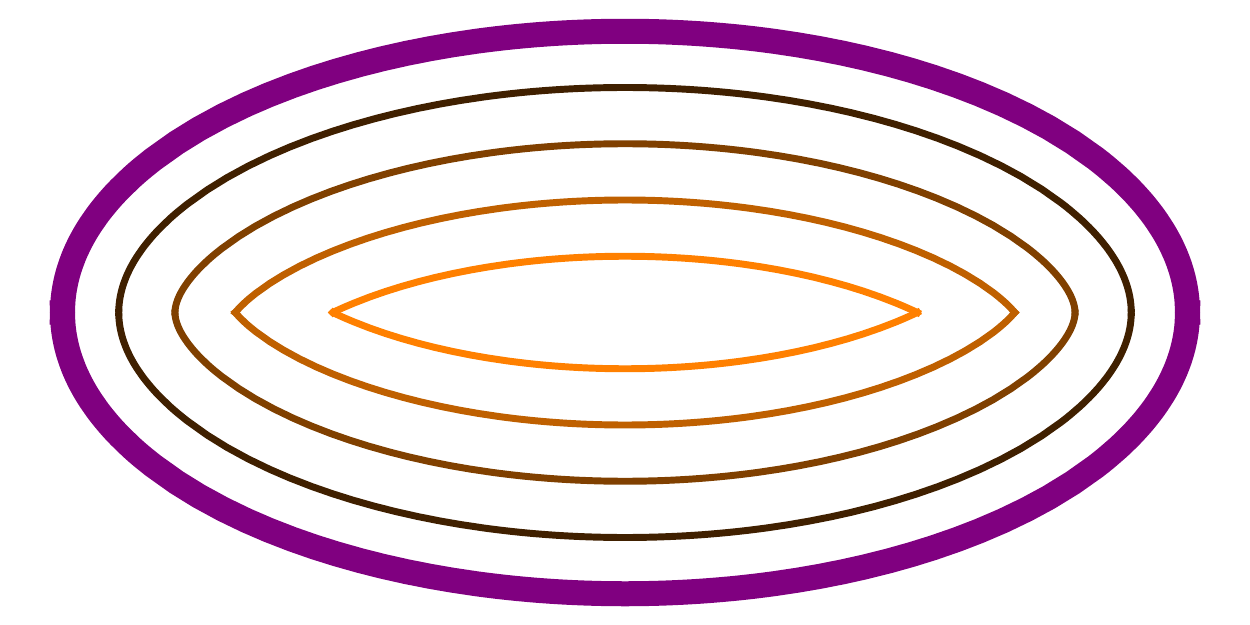}
\caption{Tsunami evolution (with constant $v_\text{ts}$) for an ellipse drawn with purple. The wavefront is colored lighter as we go to later time. The wavefront develops a swallowtail singularity at a finite time, which complicates the numerical solution of the differential equation~\eqref{ZEOM}. 
\label{fig:ellipse}}
\end{figure}
\end{center}

The other equation of motion is more complicated:
\es{rEOM}{
{(\p_V r)^2-f(Z)\ov Z^2\sqrt{Q} }&=\p_V\le(r\, \p_V r\ov Z^2\sqrt{Q}\ri)-\p_\theta \le(f(Z)\,\p_\theta r/r\ov Z^2\sqrt{Q}\ri)\,.
}
Using~\eqref{ZEOM} we see that some terms in this equation simplify:
\es{Simp}{
\sqrt{Q}&=\sqrt{-{Zf'(Z)\ov 4}}\, \sqrt{1+{(\p_\theta r)^2\ov r^2}}\,, \qquad \p_V\le(r\, \p_V r\ov Z^2\sqrt{Q}\ri)=\p_V\le({\sqrt{f(Z)-{Z f'(Z)\ov 4}}\ov Z^2\,\sqrt{-{Zf'(Z)\ov 4}}}\, r\ri)\,,
}
and we arrive at the following final form of the equation:
\es{FinalForm}{
{-f(Z)\ov Z^2\, \sqrt{-{Zf'(Z)\ov 4}}}\, {1\ov \sqrt{1+{(\p_\theta r)^2\ov r^2}}}&=r\,\p_V\le({\sqrt{f(Z)-{Z f'(Z)\ov 4}}\ov Z^2\,\sqrt{-{Zf'(Z)\ov 4}}}\ri)-\p_\theta \le({f(Z)\ov Z^2\sqrt{-{Zf'(Z)\ov 4}}}\, {\p_\theta r/r\ov \sqrt{1+{(\p_\theta r)^2\ov r^2}}}\ri)\,.
}
Note that we have eliminated $\p_V r$. One can reduce the number of equations from two to one, by expressing $Z$ as a function of $r$ and its derivatives from~\eqref{ZEOM}, but the resulting equation is very complicated even for the Schwarzschild black brane. Before making some comments about the behavior of this complicated equation, we analyze some simple cases.

\subsection{Simple cases}\label{sec:simpcases}

Let us first take the example of a spherical surface. \eqref{ZEOM} and~\eqref{FinalForm} simplify drastically to
\es{SphereEOM}{
{dr\ov dV} &=\sqrt{f(Z)-{Z f'(Z)\ov 4}}\\
{-f(Z)\ov Z^2\, \sqrt{-{Zf'(Z)\ov 4}}}&=r\,{d\ov dV} \le({\sqrt{f(Z)-{Z f'(Z)\ov 4}}\ov Z^2\,\sqrt{-{Zf'(Z)\ov 4}}}\ri)\,.
}
Let us see how these agree with the result of our analysis in Sec.~\ref{sec:spherical}. There we were using $r$ as the independent coordinate, so we have to account for the change of independent coordinate.
 The first equation above is the same as the first equation of~\eqref{OtherQuant}, if we use that ${dr\ov dV} =1/ {dV\ov dr}$. The second equation above is the derivative of the first, if we use the relation between $r$ and $Z$~\eqref{FinalEOMScaling3}.
 
Second,  we can find a special solution to the equations. Setting $Z=1$ solves~\eqref{FinalForm}, and~\eqref{ZEOM2} becomes
\es{ZEOM2}{
\p_V r&=v_B\, \sqrt{1+{(\p_\theta r)^2\ov r^2}}\,,
}
a true tsunami equation, with $v_\text{ts}=v_B$~\eqref{vBres}. From our experience with the spherical case in Sec.~\ref{sec:spherical} we know that this HRT surface is only relevant, if saturation is continuous. We do not know, if saturation can be continuous for any shape other than the sphere. If it can be, then the saturation time is immediately seen to be $t_S={R_\text{insc}/ v_B}$, the time it takes the tsunami (propagating from $\Sig$ inwards) to cover the interior of the region. This is the same time it takes  for the static surface to become available in the Vaidya setup, as discussed around~\eqref{tsEq}. If the saturation is discontinuous, this solution still provides a lower bound on the saturation time:
\es{tSBound2}{
t_S\geq{R_\text{insc}\ov v_B}\,.
}
The above argument is a second  holographic proof (the first being~\eqref{tsEq}) of the same inequality proven in field theory in~\cite{Mezei:2016wfz}.

\subsection{Comments on the general case}

We can solve~\eqref{ZEOM} and~\eqref{rEOM} for small $V$ in a series form:
\es{rSeries}{
r(V,\theta)&=r_\text{im}(\theta)+{{-f(z_\text{HM}) \ov 2}\, {  {1\ov r_\text{im}}-\le(r_\text{im}'\ov r_\text{im}^2\ri)' \ov 1+\le(r_\text{im}'\ov r_\text{im}\ri)^2}}\, V^2+O(V^4)\\
Z(V,\theta)&=z_\text{HM}+{{4z_\text{HM}f(z_\text{HM})^2 \ov 12f(z_\text{HM})-z_\text{HM}^2f''(z_\text{HM})}\, { \le( {1\ov r_\text{im}}-\le(r_\text{im}'\ov r_\text{im}^2\ri)' \ri)^2\ov \le(1+\le(r_\text{im}'\ov r_\text{im}\ri)^2\ri)^3}}\, V^2+O(V^4)\,.
}
Note that the expansion enforces the scaling surface to start from $z_\text{HM}$. We have seen this phenomenon already for the strip in~\eqref{zExpStrip} and for the sphere in~\eqref{ZtDetermine}.
As a check, in the spherical case $r_\text{im}(\theta)=1$ and we get back the large $\rho_c$ limit of~\eqref{EndOfWorldBC}. 
Note that for an image on  the brane approaching a strip 
\es{StripLim}{
r_\text{im}(\theta)\to {1\ov\cos\theta} \quad \implies \quad {1\ov r_\text{im}}-\le(r_\text{im}'\ov r_\text{im}^2\ri)' \to 0\,.
}
which agrees with the result in Sec.~\ref{sec:strip} that for the strip the scaling surface stays on $z_\text{HM}$ forever. Let us fix  $r_\text{im}(\theta)$ for the following argument.  From~\eqref{StripLim} it follows that for angles $\theta$ for which $r_\text{im}(\theta)$ is more curved will depart form $z_\text{HM}$ faster, while flatter parts hang around longer. At least initially, the surface becomes more wiggly. Eventually, we expect that every bit of the surface reaches the vicinity of the horizon, where we can perform another expansion.

For large $V$ we have only found one admissible behavior, which implies that the surface approaches the horizon.\footnote{The scaling solution for the strip is an exception to the story sketched here, as it lies on $z_\text{HM}$ for all $V$.} The form of the near horizon expansion is:
\es{NHExp}{
r(V,\theta)&={\sqrt{-f'(1)}\ov2}\, V+ a(\theta)\, \log V+ b(\theta)+  {r_1(\theta)\, \log^2 V+r_2(\theta)\, \log V+ r_3(\theta)\ov V}+\dots\\
Z(V,\theta)&=1+{4\sqrt{-f'(1)}\ov3f'(1)-f''(1)}\, {a(\theta)\ov V}+  {z_1(\theta)\log V+z_2(\theta)\ov V}+\dots\,,
}
where $a(\theta),\, b(\theta)$ are arbitrary functions, while $r_i(\theta),\, z_i(\theta)$ are functions of $a(\theta),\, b(\theta)$ and their derivatives that we have determined, but do not write down here. As usual in an asymptotic analysis, the freedom of choosing $a(\theta),\, b(\theta)$ can be used to obtain a regular solution near the brane at  $z_\text{HM}$ with an arbitrary $r_\text{im}(\theta)$. Note that for fixed $r_\text{im}(\theta)$ the surface becomes spherical for large $V$ according to~\eqref{NHExp}. It would be a very interesting to understand the details of how the HRT surface interpolates between the two regimes~\eqref{rSeries} and~\eqref{NHExp}. A straightforward numerical solution of~\eqref{ZEOM} and~\eqref{rEOM} is prevented by the formation of singularities for intermediate times. We understand the reason for the formation of cusps from two perspectives:~\eqref{rSeries} leads to a wigglier wave front as $V$ grows, while evolving backwards in $V$ in the near the horizon where~\eqref{NHExp} holds, should also lead to singularities according to what was explained around~\eqref{ZEOM} and in Fig.~\ref{fig:ellipse}. 

For the above discussion we have fixed  $r_\text{im}(\theta)$, but we are actually interested in keeping  $r_\text{bdy}(\theta)$ fixed, which requires adjusting  $r_\text{im}(\theta)$ as time evolves. For early times, the expansion~\eqref{rSeries} relates the two, and this is explored further in Sec.~\ref{sec:ShapeEarly}. We speculate that for intermediate times we have to start with an almost spherical $r_\text{im}(\theta)$ to reach a more deformed $r_\text{bdy}(\theta)$. For late times, presumably we start with some (possibly cuspy) $r_\text{im}(\theta)$, which under evolution in $V$ becomes more wiggly and can develop cusps, finally it gets smoothed out to end as $r_\text{bdy}(\theta)$ at $V=V_f$. It would be important to understand, if this is indeed what happens. While our understanding is incomplete, we have enough control to prove an important result below.

\subsection{Bound on the rate of growth}\label{sec:GrowthRate}

Let us state explicitly the variational problem that we are solving: we regard $V$ as time, and in the end of the world brane setup we are fixing $V_i=0,\, \p_V r(V,\theta)\vert_{V=0}=0$ and
$V_f=t/R,\, r(V_f,\theta)=r_\text{bdy}(\theta)$.\footnote{We expect that different types of quenches give identical results for the entropy curve, as in the strip and sphere cases. Thus, the bound that we derive below should apply to any quench setup.} 
In the following we use the Hamilton-Jacobi formalism to analyze the dependence of the area on $V_f$. Let us first write the area as:
\es{Lagrangian}{
{A\ov R^2}=\int_0^{V_f} dVd\theta\ \sL(r,Z)\,, \qquad \sL(r,Z)\equiv{r\ov Z^2}\sqrt{Q}\,.
}
The canonical momenta and the Hamiltonian are:
\es{Momenta}{
\Pi_r&={r\, \p_V r\ov Z^2\sqrt{Q}}\,,\qquad \Pi_Z=0\,,\\
H&=\int d\theta\ { r\, f(Z)\ov Z^2\, \sqrt{Q}}\,\le(1+{(\p_\theta r)^2\ov r^2}\ri)\,. 
}
$ \Pi_Z=0$ because the Lagrangian does not depend on the derivative of $Z$. The Hamilton-Jacobi equation govern the change of on-shell action:
\es{HJ}{
{1\ov R^2}\, {dA\ov dV_f}=-H(V_f)+\int d\theta\ \le[\Pi_r(V_f,\theta)\, {d r(V_f,\theta)\ov dV_f}-\Pi_r(V_i,\theta) \, {d r(V_i,\theta)\ov dV_f}\ri]\,,
}
where we also accounted for the possible change in boundary conditions. Only the Hamiltonian term remains, as $ r(V_f,\theta)=r_\text{bdy}(\theta)$ by assumption (hence it does not change as we change $V_f$), and $\Pi_r(V_i,\theta)=0$ by plugging $\p_V r(V,\theta)\vert_{V=0}=0$ into~\eqref{Momenta}.
Thus,~\eqref{HJ} reduces to
\es{HJ2}{
{1\ov R^2}\, {dA\ov dV_f}=-H(V_f)\,.
}

 A series of simple inequalities now proves a bound on the rate of growth. First,
\es{Bound1}{
&\sqrt{Q}=\sqrt{(\p_V r)^2- f(Z)\le(1+{(\p_\theta r)^2\ov r^2}\ri)}\geq\sqrt{- f(Z)}\,\sqrt{1+{(\p_\theta r)^2\ov r^2}}\\
&\implies \qquad -H(V_f)=-\int d\theta\ { r\, f(Z)\ov Z^2\, \sqrt{Q}}\,\le(1+{(\p_\theta r)^2\ov r^2}\ri)\leq \int d\theta\ \sqrt{- f(Z)\ov Z^4}\,\sqrt{r^2+{(\p_\theta r)^2}}\,.
}
Using that $v_E=\sqrt{- f(Z)/ Z^4}\vert_\text{max}$, we can further bound this from above as:
\es{Bound2}{
-H(V_f)&\leq v_E\,\int d\theta\ \sqrt{r^2+{(\p_\theta r)^2}}=v_E\, \text{area}[r_\text{bdy}(\theta)]\,.
}
Undoing the rescalings~\eqref{Scaling}, we get a bound on the rate of growth
\es{Bound3}{
{dS\ov dt}\leq s_\text{th}\, v_E A_\Sigma\,,
}
where $A_\Sigma$ is the area of the entangling surface. Recall that $s_\text{th}={1/ 4 G_N}$ is what converts area to entropy. Note that at late times, when $Z(V_f)\to 1$, this is a huge overestimate, as the prefactor in~\eqref{Bound1}, $\sqrt{- f(Z)/ Z^4}\to 0$.

There is an interesting result that we can derive along the same lines. Unfortunately, we have not found a clear field theory interpretation of it.\footnote{Conversely, we do have not found a holographic proof of the inequality~\eqref{TsunamiVolume} proven in~\cite{Mezei:2016wfz}. This situation may be analogous to the case of the monotonicity of renormalized entanglement entropy~\cite{Liu:2012eea}, where there exists a field theory proof~\cite{Casini:2012ei}, but no holographic argument.} Note that in~\eqref{Simp} we have obtained a simple expression for $Q$ valid on-shell that we have not used above. Using that expression  we get:
\es{HV}{
 -H(V_f)&=-\int d\theta\ { r\, f(Z)\ov Z^2\, \sqrt{Q}}\,\le(1+{(\p_\theta r)^2\ov r^2}\ri)\Big\vert_{V_f}=\int d\theta\ {- f(Z)\ov Z^2\sqrt{-{Z f'(Z)/ 4}}}\,\sqrt{r^2+{(\p_\theta r)^2}}\Big\vert_{V_f}\,.
}
Now using that energy is conserved as the Lagrangian does not depend on $V$ explicitly, we can write
\es{HV2}{
 -H(V_f)&=-H(V_i=0)=\int d\theta\ {- f(Z)\ov Z^2\sqrt{-{Z f'(Z)/ 4}}}\,\sqrt{r^2+{(\p_\theta r)^2}}\Big\vert_{V_i}\\
&={- f(z_\text{HM})\ov z_\text{HM}^2\sqrt{-{z_\text{HM} f'(z_\text{HM})/ 4}}}\, \text{area}[r_\text{im}(\theta)]\\
&=v_E\,  \text{area}[r_\text{im}(\theta)]\,, 
}
where we used that $Z(V_i=0,\theta)=z_\text{HM}$ according to~\eqref{rSeries}, and that~\eqref{zHMEq} relates $f(z_\text{HM})$ to $f'(z_\text{HM})$. Note that the area of the image on the brane is measured in the field theory coordinates $\vec{x}$ defined in~\eqref{Metric}. This is also a somewhat different proof of the rate of growth bound~\eqref{Bound3}, as $\text{area}[r_\text{im}(\theta)]<\text{area}[r_\text{bdy}(\theta)]$. The higher dimensional generalization is straightforward, and gives the area of the image on the brane on the right-hand side of~\eqref{HV2}.

In the spherical case, by manipulating integrals, we found a mysterious relation~\eqref{derRelation2}. Now we see that it is a simple consequence of~\eqref{HV2}. For the spherical case it reproduces~\eqref{derRelation2} 
\es{HV3}{
{d\sA\ov d\tau}=v_E\, {2\ov r_\text{im}}\,,
} 
where we used that $\sA=A/(2\pi R^2)$. 

\subsection{Early time growth}\label{sec:ShapeEarly}

Combining~\eqref{rSeries} and~\eqref{HV2} with minimal effort we can obtain the early time expansion of the entropy, where by early times we mean $\beta\ll t\ll R$. Requiring that $r(V,\theta)=r_\text{bdy}(\theta)$, we can obtain how the image on the brane changes its shape as a function of time by inverting~\eqref{rSeries} perturbatively
\es{rimV}{
r_\text{im}(V,\theta)&=r_\text{bdy}(\theta)-{{-f(z_\text{HM}) \ov 2}\, {  {1\ov r_\text{bdy}}-\le(r_\text{bdy}'\ov r_\text{bdy}^2\ri)' \ov 1+\le(r_\text{bdy}'\ov r_\text{bdy}\ri)^2}}\, V^2+O(V^4)\,.
} 
Using~\eqref{HV2} this gives the early time growth of entropy is given by:
\es{Agrowth}{
{A(\tau)\ov R^2}=v_E\, \text{area}[r_\text{bdy}(\theta)]\,\tau\, \le[1-a\,\tau^2+O(\tau^4)\ri]\,, \qquad a\equiv{{-f(z_\text{HM}) \ov 6}\, {1\ov \text{area}[r_\text{bdy}(\theta)]}\int d\theta\ {  {1\ov r_\text{bdy}}-\le(r_\text{bdy}'\ov r_\text{bdy}^2\ri)' \ov 1+\le(r_\text{bdy}'\ov r_\text{bdy}\ri)^2}}\,.
}
For the spherical case $a={-f(z_\text{HM}) / 6}$, which is in complete agreement with~\eqref{sAtau}, if we take into account the difference in normalization between $A/R^2$ and $\sA$ used in Sec.~\ref{sec:spherical}. Perhaps matching this expansion from a tensor network perspective could help us understand the relation between tensor networks and the geometry of the bulk gravitational spacetime.

\section*{Acknowledgments}

I am grateful to Douglas Stanford for extremely helpful discussions, suggestions, and an enjoyable collaboration on the companion paper~\cite{Mezei:2016wfz}. I also thank Hong Liu, Juan Maldacena, Dan Roberts, and Wilke van der Schee for useful discussions.  My research was supported in part by the U.S. Department of Energy under grant No. DE-SC0016244.


\bibliographystyle{ssg}
\bibliography{speeds}

\begingroup\raggedright\begin{thebibliography}{10}

\bibitem{Mezei:2016wfz}
M.~Mezei and D.~Stanford, ``{On entanglement spreading in chaotic systems},''
  \href{http://xxx.lanl.gov/abs/1608.05101}{{\tt 1608.05101}}.

\bibitem{Maldacena:1997re}
J.~M. Maldacena, ``{The Large N limit of superconformal field theories and
  supergravity},'' {\em Adv.Theor.Math.Phys.} {\bf 2} (1998) 231--252,
  \href{http://xxx.lanl.gov/abs/hep-th/9711200}{{\tt hep-th/9711200}}.

\bibitem{Gubser:1998bc}
S.~Gubser, I.~R. Klebanov, and A.~M. Polyakov, ``{Gauge theory correlators from
  noncritical string theory},'' {\em Phys.Lett.} {\bf B428} (1998) 105--114,
  \href{http://xxx.lanl.gov/abs/hep-th/9802109}{{\tt hep-th/9802109}}.

\bibitem{Witten:1998qj}
E.~Witten, ``{Anti-de Sitter space and holography},'' {\em
  Adv.Theor.Math.Phys.} {\bf 2} (1998) 253--291,
  \href{http://xxx.lanl.gov/abs/hep-th/9802150}{{\tt hep-th/9802150}}.

\bibitem{Ryu:2006bv}
S.~Ryu and T.~Takayanagi, ``{Holographic derivation of entanglement entropy
  from AdS/CFT},'' {\em Phys.Rev.Lett.} {\bf 96} (2006) 181602,
  \href{http://xxx.lanl.gov/abs/hep-th/0603001}{{\tt hep-th/0603001}}.

\bibitem{Ryu:2006ef}
S.~Ryu and T.~Takayanagi, ``{Aspects of Holographic Entanglement Entropy},''
  {\em JHEP} {\bf 0608} (2006) 045,
  \href{http://xxx.lanl.gov/abs/hep-th/0605073}{{\tt hep-th/0605073}}.

\bibitem{Hubeny:2007xt}
V.~E. Hubeny, M.~Rangamani, and T.~Takayanagi, ``{A Covariant holographic
  entanglement entropy proposal},'' {\em JHEP} {\bf 07} (2007) 062,
  \href{http://xxx.lanl.gov/abs/0705.0016}{{\tt 0705.0016}}.

\bibitem{Calabrese:2005in}
P.~Calabrese and J.~L. Cardy, ``{Evolution of entanglement entropy in
  one-dimensional systems},'' {\em J. Stat. Mech.} {\bf 0504} (2005) P04010,
  \href{http://xxx.lanl.gov/abs/cond-mat/0503393}{{\tt cond-mat/0503393}}.

\bibitem{Asplund:2015eha}
C.~T. Asplund, A.~Bernamonti, F.~Galli, and T.~Hartman, ``{Entanglement
  Scrambling in 2d Conformal Field Theory},'' {\em JHEP} {\bf 09} (2015) 110,
  \href{http://xxx.lanl.gov/abs/1506.03772}{{\tt 1506.03772}}.

\bibitem{Cardy:2015xaa}
J.~Cardy, ``{Quantum Quenches to a Critical Point in One Dimension: some
  further results},'' {\em J. Stat. Mech.} {\bf 1602} (2016), no.~2 023103,
  \href{http://xxx.lanl.gov/abs/1507.07266}{{\tt 1507.07266}}.

\bibitem{DeChiara:2005wb}
G.~De~Chiara, S.~Montangero, P.~Calabrese, and R.~Fazio, ``{Entanglement
  entropy dynamics in Heisenberg chains},'' {\em J. Stat. Mech.} {\bf 0603}
  (2006) P03001, \href{http://xxx.lanl.gov/abs/cond-mat/0512586}{{\tt
  cond-mat/0512586}}.

\bibitem{Calabrese:2007rg}
P.~Calabrese and J.~Cardy, ``{Quantum Quenches in Extended Systems},'' {\em J.
  Stat. Mech.} {\bf 0706} (2007) P06008,
  \href{http://xxx.lanl.gov/abs/0704.1880}{{\tt 0704.1880}}.

\bibitem{2008PhRvA..78a0306F}
M.~{Fagotti} and P.~{Calabrese}, ``{Evolution of entanglement entropy following
  a quantum quench: Analytic results for the XY chain in a transverse magnetic
  field},'' {\em \pra} {\bf 78} (July, 2008) 010306,
  \href{http://xxx.lanl.gov/abs/0804.3559}{{\tt 0804.3559}}.

\bibitem{2008JSMTE..05..018L}
A.~M. {L{\"a}uchli} and C.~{Kollath}, ``{Spreading of correlations and
  entanglement after a quench in the one-dimensional Bose Hubbard model},''
  {\em Journal of Statistical Mechanics: Theory and Experiment} {\bf 5} (May,
  2008) 05018, \href{http://xxx.lanl.gov/abs/0803.2947}{{\tt 0803.2947}}.

\bibitem{Kim:2013bc}
H.~Kim and D.~A. Huse, ``{Ballistic spreading of entanglement in a diffusive
  nonintegrable system},'' {\em Phys. Rev. Lett.} {\bf 111} (2013) 127205,
  \href{http://xxx.lanl.gov/abs/1306.4306}{{\tt 1306.4306}}.

\bibitem{2016PhRvB..94b4306L}
Y.~{Lemonik} and A.~{Mitra}, ``{Entanglement properties of the critical quench
  of O (N ) bosons},'' {\em \prb} {\bf 94} (July, 2016) 024306,
  \href{http://xxx.lanl.gov/abs/1512.02749}{{\tt 1512.02749}}.

\bibitem{Cotler:2016acd}
J.~S. Cotler, M.~P. Hertzberg, M.~Mezei, and M.~T. Mueller, ``{Entanglement
  Growth after a Global Quench in Free Scalar Field Theory},''
  \href{http://xxx.lanl.gov/abs/1609.00872}{{\tt 1609.00872}}.

\bibitem{Hartman:2013qma}
T.~Hartman and J.~Maldacena, ``{Time Evolution of Entanglement Entropy from
  Black Hole Interiors},'' {\em JHEP} {\bf 05} (2013) 014,
  \href{http://xxx.lanl.gov/abs/1303.1080}{{\tt 1303.1080}}.

\bibitem{Liu:2013iza}
H.~Liu and S.~J. Suh, ``{Entanglement Tsunami: Universal Scaling in Holographic
  Thermalization},'' {\em Phys. Rev. Lett.} {\bf 112} (2014) 011601,
  \href{http://xxx.lanl.gov/abs/1305.7244}{{\tt 1305.7244}}.

\bibitem{Liu:2013qca}
H.~Liu and S.~J. Suh, ``{Entanglement growth during thermalization in
  holographic systems},'' {\em Phys. Rev.} {\bf D89} (2014), no.~6 066012,
  \href{http://xxx.lanl.gov/abs/1311.1200}{{\tt 1311.1200}}.

\bibitem{AbajoArrastia:2010yt}
J.~Abajo-Arrastia, J.~Aparicio, and E.~Lopez, ``{Holographic Evolution of
  Entanglement Entropy},'' {\em JHEP} {\bf 11} (2010) 149,
  \href{http://xxx.lanl.gov/abs/1006.4090}{{\tt 1006.4090}}.

\bibitem{Albash:2010mv}
T.~Albash and C.~V. Johnson, ``{Evolution of Holographic Entanglement Entropy
  after Thermal and Electromagnetic Quenches},'' {\em New J. Phys.} {\bf 13}
  (2011) 045017, \href{http://xxx.lanl.gov/abs/1008.3027}{{\tt 1008.3027}}.

\bibitem{Balasubramanian:2010ce}
V.~Balasubramanian, A.~Bernamonti, J.~de~Boer, N.~Copland, B.~Craps,
  E.~Keski-Vakkuri, B.~Muller, A.~Schafer, M.~Shigemori, and W.~Staessens,
  ``{Thermalization of Strongly Coupled Field Theories},'' {\em Phys. Rev.
  Lett.} {\bf 106} (2011) 191601, \href{http://xxx.lanl.gov/abs/1012.4753}{{\tt
  1012.4753}}.

\bibitem{Balasubramanian:2011ur}
V.~Balasubramanian, A.~Bernamonti, J.~de~Boer, N.~Copland, B.~Craps,
  E.~Keski-Vakkuri, B.~Muller, A.~Schafer, M.~Shigemori, and W.~Staessens,
  ``{Holographic Thermalization},'' {\em Phys. Rev.} {\bf D84} (2011) 026010,
  \href{http://xxx.lanl.gov/abs/1103.2683}{{\tt 1103.2683}}.

\bibitem{Shenker:2013pqa}
S.~H. Shenker and D.~Stanford, ``{Black holes and the butterfly effect},'' {\em
  JHEP} {\bf 03} (2014) 067, \href{http://xxx.lanl.gov/abs/1306.0622}{{\tt
  1306.0622}}.

\bibitem{Roberts:2014isa}
D.~A. Roberts, D.~Stanford, and L.~Susskind, ``{Localized shocks},'' {\em JHEP}
  {\bf 03} (2015) 051, \href{http://xxx.lanl.gov/abs/1409.8180}{{\tt
  1409.8180}}.

\bibitem{Maldacena:2015waa}
J.~Maldacena, S.~H. Shenker, and D.~Stanford, ``{A bound on chaos},'' {\em
  JHEP} {\bf 08} (2016) 106, \href{http://xxx.lanl.gov/abs/1503.01409}{{\tt
  1503.01409}}.

\bibitem{Gu:2016oyy}
Y.~Gu, X.-L. Qi, and D.~Stanford, ``{Local criticality, diffusion and chaos in
  generalized Sachdev-Ye-Kitaev models},''
  \href{http://xxx.lanl.gov/abs/1609.07832}{{\tt 1609.07832}}.

\bibitem{Hartman:2015apr}
T.~Hartman and N.~Afkhami-Jeddi, ``{Speed Limits for Entanglement},''
  \href{http://xxx.lanl.gov/abs/1512.02695}{{\tt 1512.02695}}.

\bibitem{Kundu:2016cgh}
S.~Kundu and J.~F. Pedraza, ``{Spread of entanglement for small subsystems in
  holographic CFTs},'' \href{http://xxx.lanl.gov/abs/1602.05934}{{\tt
  1602.05934}}.

\bibitem{Casini:2015zua}
H.~Casini, H.~Liu, and M.~Mezei, ``{Spread of entanglement and causality},''
  {\em JHEP} {\bf 07} (2016) 077,
  \href{http://xxx.lanl.gov/abs/1509.05044}{{\tt 1509.05044}}.

\bibitem{Nahum:2016muy}
A.~Nahum, J.~Ruhman, S.~Vijay, and J.~Haah, ``{Quantum Entanglement Growth
  Under Random Unitary Dynamics},''
  \href{http://xxx.lanl.gov/abs/1608.06950}{{\tt 1608.06950}}.

\bibitem{Maldacena:2001kr}
J.~M. Maldacena, ``{Eternal black holes in anti-de Sitter},'' {\em JHEP} {\bf
  04} (2003) 021, \href{http://xxx.lanl.gov/abs/hep-th/0106112}{{\tt
  hep-th/0106112}}.

\bibitem{Morrison:2012iz}
I.~A. Morrison and M.~M. Roberts, ``{Mutual information between thermo-field
  doubles and disconnected holographic boundaries},'' {\em JHEP} {\bf 07}
  (2013) 081, \href{http://xxx.lanl.gov/abs/1211.2887}{{\tt 1211.2887}}.

\bibitem{Hosur:2015ylk}
P.~Hosur, X.-L. Qi, D.~A. Roberts, and B.~Yoshida, ``{Chaos in quantum
  channels},'' {\em JHEP} {\bf 02} (2016) 004,
  \href{http://xxx.lanl.gov/abs/1511.04021}{{\tt 1511.04021}}.

\bibitem{2006PhRvL..97e0401B}
S.~{Bravyi}, M.~B. {Hastings}, and F.~{Verstraete}, ``{Lieb-Robinson Bounds and
  the Generation of Correlations and Topological Quantum Order},'' {\em
  Physical Review Letters} {\bf 97} (Aug., 2006) 050401,
  \href{http://xxx.lanl.gov/abs/quant-ph/0603121}{{\tt quant-ph/0603121}}.

\bibitem{2014arXiv1411.0680M}
M.~{Mari{\"e}n}, K.~M.~R. {Audenaert}, K.~{Van Acoleyen}, and F.~{Verstraete},
  ``{Entanglement Rates and the Stability of the Area Law for the Entanglement
  Entropy},'' {\em ArXiv e-prints} (Nov., 2014)
  \href{http://xxx.lanl.gov/abs/1411.0680}{{\tt 1411.0680}}.

\bibitem{Avery:2014dba}
S.~G. Avery and M.~F. Paulos, ``{Universal Bounds on the Time Evolution of
  Entanglement Entropy},'' {\em Phys. Rev. Lett.} {\bf 113} (2014), no.~23
  231604, \href{http://xxx.lanl.gov/abs/1407.0705}{{\tt 1407.0705}}.

\bibitem{Liu:2013una}
H.~Liu and M.~Mezei, ``{Probing renormalization group flows using entanglement
  entropy},'' {\em JHEP} {\bf 1401} (2014) 098,
  \href{http://xxx.lanl.gov/abs/1309.6935}{{\tt 1309.6935}}.

\bibitem{Hubeny:2013dea}
V.~E. Hubeny and H.~Maxfield, ``{Holographic probes of collapsing black
  holes},'' {\em JHEP} {\bf 03} (2014) 097,
  \href{http://xxx.lanl.gov/abs/1312.6887}{{\tt 1312.6887}}.

\bibitem{Emparan:2015rva}
R.~Emparan, R.~Suzuki, and K.~Tanabe, ``{Quasinormal modes of (Anti-)de Sitter
  black holes in the 1/D expansion},'' {\em JHEP} {\bf 04} (2015) 085,
  \href{http://xxx.lanl.gov/abs/1502.02820}{{\tt 1502.02820}}.

\bibitem{Bhattacharyya:2015dva}
S.~Bhattacharyya, A.~De, S.~Minwalla, R.~Mohan, and A.~Saha, ``{A membrane
  paradigm at large D},'' {\em JHEP} {\bf 04} (2016) 076,
  \href{http://xxx.lanl.gov/abs/1504.06613}{{\tt 1504.06613}}.

\bibitem{Liu:2012eea}
H.~Liu and M.~Mezei, ``{A Refinement of entanglement entropy and the number of
  degrees of freedom},'' {\em JHEP} {\bf 1304} (2013) 162,
  \href{http://xxx.lanl.gov/abs/1202.2070}{{\tt 1202.2070}}.

\bibitem{Casini:2012ei}
H.~Casini and M.~Huerta, ``{On the RG running of the entanglement entropy of a
  circle},'' {\em Phys.Rev.} {\bf D85} (2012) 125016,
  \href{http://xxx.lanl.gov/abs/1202.5650}{{\tt 1202.5650}}.

\end{thebibliography}\endgroup

\end{document}